\documentclass[aip,jmp,preprint]{revtex4-1}

\usepackage{amsfonts,amssymb,amsmath,graphicx}
\usepackage{verbatim,xcolor}
\def\ez{\mathbf e_z}
\def\ex{\mathbf e_x}
\def\ey{\mathbf e_y}
\def\dl{ {\rm d}\lambda}
\def\dlp{{\rm d}\lambda'}
\def\dxi{{\rm d}\xi}
\def\de{{\rm d}\eta}
\def\dph{{\rm d}\phi'}
\def\hate{\widehat E}
\def\hatb{\widehat B}
\def\php{\phi'}
\def\d{{\rm d}}

\begin{document}
%\preprint{To be submitted to the Journal of Mathematical Physics}

\title{On solutions of Maxwell's equations with dipole sources over a thin conducting film}

\author{Dionisios Margetis}
\email{dio@math.umd.edu}

\affiliation{Department of Mathematics, and Institute for Physical Science
and Technology, and Center for Scientific Computation and Mathematical Modeling, University of Maryland, College Park, Maryland 20742, USA}

\author{Mitchell Luskin}
\email{luskin@umn.edu}

\affiliation{School of Mathematics, University of Minnesota, Minneapolis, Minnesota 55455, USA}

\begin{abstract}
We derive and interpret solutions of time-harmonic Maxwell's equations with a vertical and a horizontal electric dipole near a planar, thin conducting film, e.g. graphene sheet, lying between two unbounded isotropic and non-magnetic media. Exact expressions for all field components are extracted in terms of rapidly convergent series
of known transcendental functions when the ambient media have equal permittivities and both the dipole and observation point lie on the plane of the film. These solutions are simplified for all distances from the source when the film surface resistivity is large in magnitude compared to the intrinsic impedance of the ambient space. The formulas reveal the analytical structure of two types of waves that can possibly be excited by the dipoles and propagate on the film. One of these waves is intimately related to the surface plasmon-polariton of transverse-magnetic (TM) polarization of plane waves.
\end{abstract}

%\begin{keyword}Crystal surface; Epitaxial growth; Morphological evolution;
%Burton-Cabrera-Frank (BCF) model;
%Terrace anisotropy; Step edge diffusion; Ehrlich-Schwoebel barrier; Surface
%flux; Continuum limit; Partial differential equation
%\vskip10pt

%\PACS 68.55.-a; 81.10.Aj; 81.15.Aa; 68.35.Md; 61.46.-w; 61.50.Ah

%\end{keyword}

\maketitle

%%%%%%%%%%%%%%%%%%%%%%%%%%%%
%%%%%%%%%%%%%%%%%%%%%%%%%%%%
\section{Introduction}
\label{sec:Intro}
%%%%%%%%%%%%%%%%%%%%%%%%%%%%
%%%%%%%%%%%%%%%%%%%%%%%%%%%%
In the last decade, rapid advances have been made in the design and fabrication of two-dimensional materials~\cite{Torres-book} which can be used to manipulate light at  small scales.~\cite{Zhang-book}
The  highly active field of plasmonics focuses on the interaction of electromagnetic radiation at the mid- and near-infrared spectrum with the collective motion
of electrons in conducting interfaces and nanostructures.~\cite{Maier-book} A goal is to generate electromagnetic waves that propagate with relatively small energy loss close to the interface between a conducting material, e.g. graphene, and a dielectric.~\cite{Vakil11,Ju11,Cheng13,Hanson08,Hanson11,Lovat15} In plasmonics such {\em lateral waves} should decay fast enough away from the interface; while, on the other hand, they should attenuate slowly enough in the direction of propagation along the interface. A wave that has attracted much attention in this context is the surface plasmon-polariton,~\cite{Raether-book,Maier-book,Bludov13} with a variety of reported applications~\cite{Zhang12,Huidobro10} including invisibility cloaking,~\cite{Renger10} photovoltaics~\cite{Green12} and nanolithography.~\cite{Luo04}

Most recently, direct experimental evidence was provided for generating surface plasmons by placing a receiving resonant antenna near a graphene sheet.~\cite{Gonzalez14}
Motivated by this advance, our  goal with this paper is to analytically study the generation of surface plasmons by current-carrying sources via a solvable model for a fundamental setting. To this end, we formulate and solve exactly boundary value problems for the time-harmonic Maxwell equations in the presence of vertical and horizontal electric (Hertzian) dipoles near an isotropic and homogeneous conducting sheet between two isotropic and non-magnetic unbounded media.~\cite{Maier-book,Cheng13} \looseness=-1

The underlying theme, wave propagation near boundaries, has been the subject of important studies for over a century;~\cite{Sommerfeld1899,Sommerfeld1909,Zenneck1907,vanderPol31,Fok33,Norton36,Banos,Wait,KOW} see particularly the systematic and extensive treatment of Ref.~\onlinecite{KOW}.
In these works, approximation techniques are developed for radiowave propagation; these  have offered valuable insights into the lateral electromagnetic waves traveling near the boundaries between media of very different indices of refraction.~\cite{KOW} In the  frequency band of plasmonics,~\cite{Zhang-book} however, additional considerations have emerged because of properties of novel two-dimensional materials used in microscale applications. For instance, a thin layer of graphene has a complex surface conductivity, with a positive or negative imaginary part depending on frequency and doping, and  introduces a jump in the tangential component of the magnetic field across the interface. The analytical consequences of this discontinuity are largely unexplored.~\cite{Hanson08,Hanson11,Nikitin11}
This view suggests that the associated lateral electromagnetic waves be studied in detail in the near- and mid-infrared spectrum.\looseness=-1

In this paper we analytically address aspects of the following question. What is the structure of the waves generated by dipole sources on a thin conducting film? Our goal is to single out fundamental attributes of the field which are intimately related to the film by analyzing a minimal solvable model. Our tasks can be summarized as follows.\looseness=-1

\begin{itemize}

\item
We explicitly represent all field components in terms of one-dimensional Fourier-Bessel (Sommerfeld-type) integrals, in the spirit of Ref.~\onlinecite{KOW}. Our derivations, focusing on the electromagnetic field itself, differ from the use of the (non-physical) Hertzian potential invoked, e.g. in Ref.~\onlinecite{Hanson08}, by which the field components are derived via successive differentiations.  \looseness=-1

\item
By a generalized Schwinger-Feynman representation for a class of integrals,~\cite{Margetis01} we compute all field components via fast convergent series of known functions such as the Fresnel integrals, when both the dipole and observation point lie on the plane of the film and the ambient media have equal permittivities. Our model is thus simplified, yet without obscuring the goal to analytically understand the role of the interface. \looseness=-1

\item
In accord with applications in plasmonics,~\cite{Cheng13} we further simplify the exact solutions when the surface resistivity (inverse of conductivity) of the thin layer is much larger in magnitude than the intrinsic impedance of the ambient space. Then, a few terms are retained in the series expansions for the fields yielding simple approximate formulas for all distances from the source.

\end{itemize}

Our approach is based on systematically solving Maxwell's equations in the spirit of Refs.~\onlinecite{KOW,Margetis01}. Thus, we avoid any a-priori plane-wave approximations. We recognize a specific type of lateral wave as intimately related to the surface plasmon-polariton of transverse-magnetic (TM) polarization of plane waves~\cite{Maier-book,Bludov13} via the contribution of a certain pole in the complex plane of the dual (Fourier) variable; see also the treatments of Refs.~\onlinecite{Hanson08,Hanson11,Nikitin11}. For a horizontal dipole on a thin film in free space, when contributions related to TM polarization in principle may co-exist with contributions of transverse-electric (TE) polarization, our analysis reveals that the TM surface plasmon-polariton, when present, is accompanied by a wave expressed by Fresnel integrals.~\cite{KOW}

The analysis presented here, with focus on explicit, physically transparent  expressions for the electromagnetic field in terms of known functions, differs in methodology from previous studies of waves in similar settings.~\cite{Bludov13,Hanson08,Hanson11,Nikitin11} For example, in Ref.~\onlinecite{Bludov13}, the authors review dispersion relations for plane waves in plasmonics for a variety of experimentally relevant geometries, without discussing effects of point sources. In Ref.~\onlinecite{Hanson08}, exact integrals are formulated for the Hertzian potential produced by dipoles in the presence of a graphene sheet; the electromagnetic field is then computed by numerical evaluation of integrals. In Ref.~\onlinecite{Hanson11}, a similar task is carried out more extensively, with numerical evidence that the surface plasmon-polariton of TM polarization, recognized as a discrete spectral contribution, may dominate wave propagation under certain conditions on the surface conductivity. In Ref.~\onlinecite{Nikitin11}, the authors numerically describe the field produced by dipoles near a graphene sheet, distinguishing a ``core region'', where the electric field can be much larger than its values in free space, from an ``outer region'', where the field approaches its values in free space. 

Our work expands previous numerical approaches~\cite{Hanson08,Hanson11,Nikitin11} in the following sense. By focusing on a simple yet nontrivial model with a conducting thin film, we are able to derive closed-form expressions which explicitly separate the primary field of the dipole, produced in the absence of the layer, from the scattered field which is sensitive to the film conductivity, for all distances from the point source. This approach singles out analytic aspects of the wave produced by the point source that are intimately connected to the surface conductivity of the film, thus showing {\em how} the surface plasmon related to TM polarization can dominate propagation in cases of physical and practical interest.  Our analysis lacks generality, since we restrict attention to the case where the source and the observation point both lie on the plane of the thin film; nonetheless, we  view our treatment as a step necessary for tackling the problem of radiation by a realistic current-carrying source (rather than an incident plane wave) placed on the material surface.~\cite{Gonzalez14} The character of the wave produced by the source of course becomes important at short distances or high enough frequencies. Our work aims at illuminating the complicated structure of the field in a simple nontrivial setting.

The present work illustrates analytic aspects of the tensor (or, dyadic) Green function for the geometry of a thin conducting film at fixed frequency. Naturally, the response to any imposed current-carrying source can then be derived by superposition. This task lies beyond our present scope. Furthermore, we do not pursue numerical computations of the fields. The explicit computation by asymptotic methods~\cite{KOW} of field components when the observation point or the dipole is away from the layer is left for future work.

The remainder of this paper is organized as follows. In Section~\ref{sec:BVP}, we describe the boundary value problem for Maxwell's equations. Section~\ref{sec:Fourier} focuses on the derivation of Fourier-Bessel representations, known as the Hankel transform,~\cite{Lebedev-book} for the electromagnetic field. In Section~\ref{sec:Exact}, we evaluate exactly the requisite integrals  when the dipole and the observation point are on the film, which is placed in a homogeneous space. Section~\ref{sec:Lateral} discusses the field resulting from our computations for  distances from the source that are comparable to the wavelength in free space. Section~\ref{sec:Conclusion} concludes our paper with an outline of open problems. \looseness=-1
\smallskip

{\bf Notation and terminology.} The $e^{-i\omega t}$ time dependence is assumed throughout, where $\omega$ is the angular frequency. $\mathbb{R}$ is the set of reals, $\mathbb{Z}$ is the set of integers, and boldface symbols denote vectors in $\mathbb{R}^3$. We write $f=\mathcal O(g)$ ($f=o(g)$) to mean that $|f/g|$ is bounded by a nonzero constant (approaches zero) in a prescribed limit. $f\sim g$ implies $f-g=o(g)$ in a prescribed limit. $\Re w$ ($\Im w$) denotes the real (imaginary) part of complex $w$. The term ``sheet'' has a two-fold meaning as either a material thin film or, as a ``Riemann sheet'', a particular branch of a multiple-value function of a complex variable; and the terms ``top Riemann sheet'' and ``first Riemann sheet'' are used interchangeably. We use the terms ``TM polarization'' and ``TE polarization'' in the context of waves produced by dipoles to indicate the presence of certain denominators, denoted by $\mathcal P$ and $\mathcal Q$ in the main text, respectively, in the Fourier representations of the corresponding fields; each denominator appears in the reflection coefficient for the TM- or TE-polarized plane wave incident upon the thin film. The terms ``surface plasmon-polariton''~\cite{Maier-book} and ``surface plasmon'' are used interchangeably.\looseness=-1

%%%%%%%%%%%%%%%%%%%%%%%%%%%%%%%%%%%%%%%%%%%%%
%%%%%%%%%%%%%%%%%%%%%%%%%%%%%%%%%%%%%%%%%%%%%
\section{Boundary value problem}
\label{sec:BVP}
%%%%%%%%%%%%%%%%%%%%%%%%%%%%%%%%%%%%%%%%%%%%%
%%%%%%%%%%%%%%%%%%%%%%%%%%%%%%%%%%%%%%%%%%%%%

\begin{figure}
\includegraphics*[scale=0.5, trim=0in 1in 0in 1.4in]{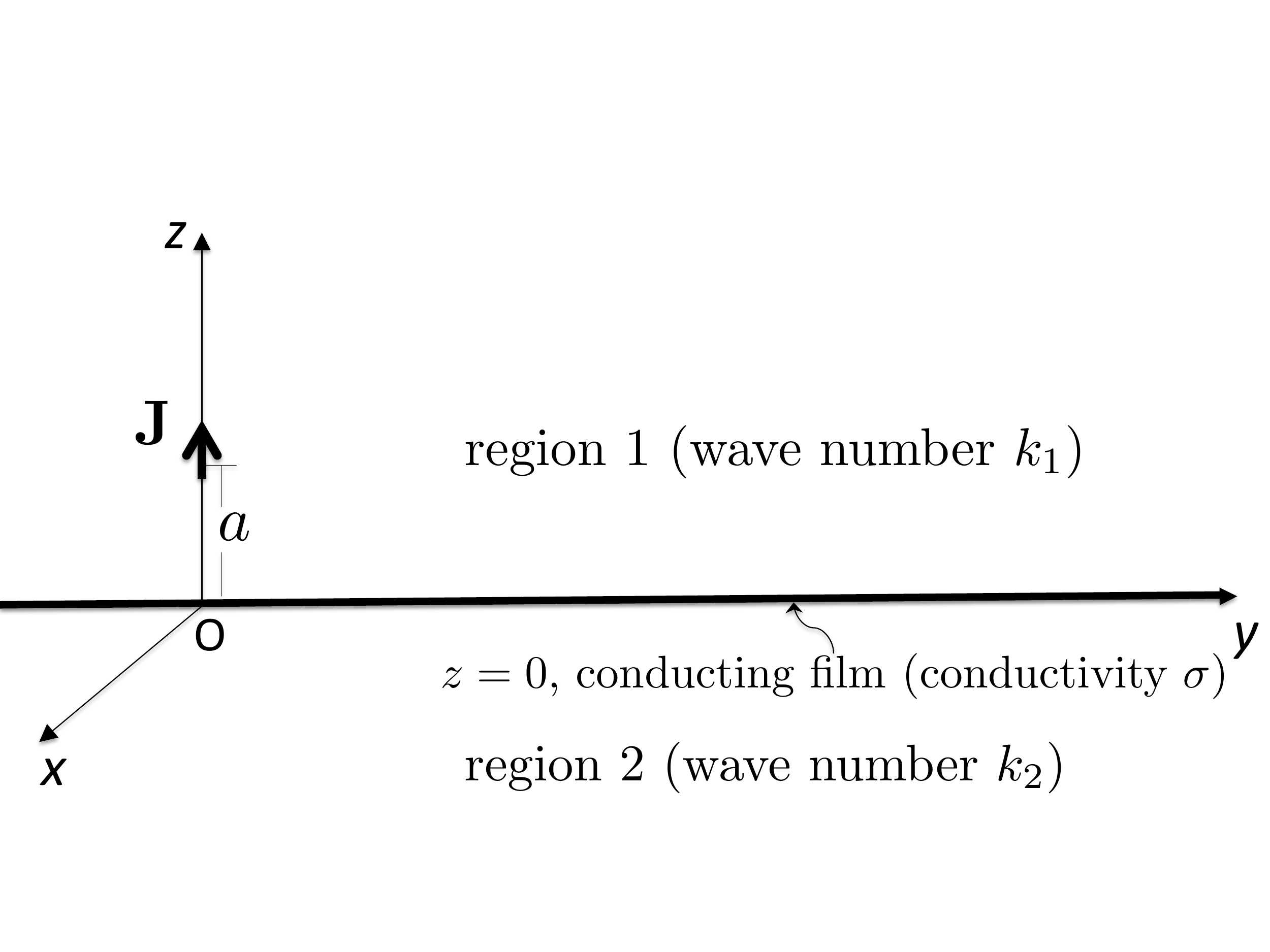}
%[width=3.6in,height=1in,trim=0in 2.6in 0 3.0in]{VerticalD-fig.pdf}
%\includegraphics*[width=3.7in,height=0.56in,trim=0 4.8in 0 4.9in]{Fig1.pdf}
%\includegraphics*[width=3.7in]{geom.eps}
    \caption{Vertical unit electric dipole at distance $a$ from planar thin conducting film. The infinitely thin film lies in the plane $z=0$, between region $1$ (half space $\{z>0\}$ with wave number $k_1$) and region $2$ ($\{z<0\}$ with wave number $k_2$); and has surface conductivity $\sigma$.}\label{fig:VD}
\end{figure}

\begin{figure}
\includegraphics*[scale=0.5, trim=0in 1in 0in 1.4in]{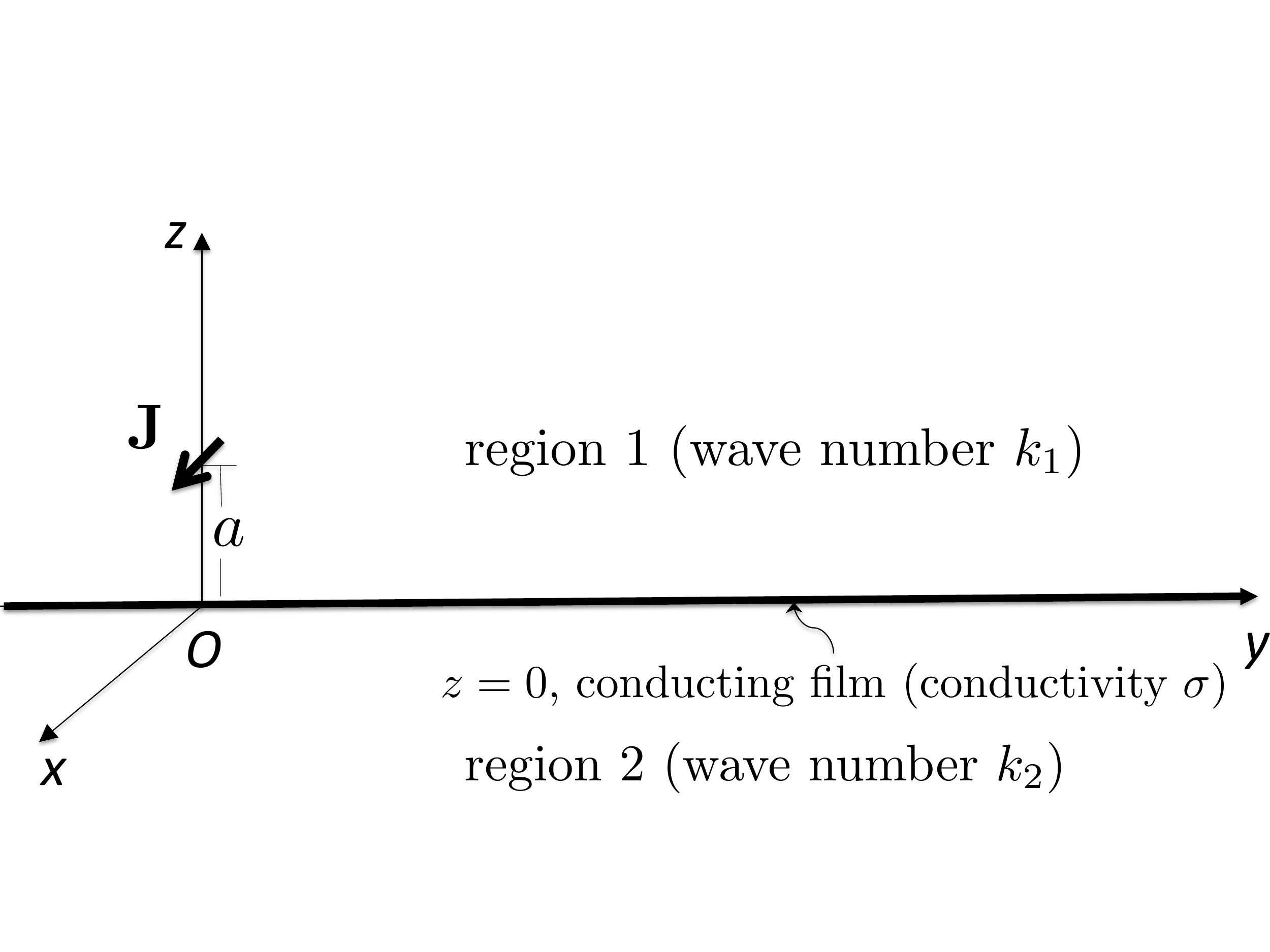}
    \caption{Horizontal unit electric dipole at distance $a$ from planar thin conducting film. The film lies in the plane $z=0$, separating region $1$ from region $2$; and has surface conductivity $\sigma$. }
\label{fig:HD}
\end{figure}

In this section, we formulate the boundary value problem for Maxwell's equations. The current density of the vertical unit electric dipole, shown in Fig.~\ref{fig:VD}, is\looseness=-1
\begin{equation}\label{eq:J-VD}
\mathbf J(x,y,z)=\mathbf J^v(x,y,z)=\ez\,\delta_{\mathbf r_0}~,\quad \mathbf r_0=(0,0,a)~,
\end{equation}
where $a$ is the distance of the dipole from the layer, $a>0$, $\ez$ is the unit vector along the $z$-axis, and $\delta_{\mathbf r}$ denotes the Dirac mass at point $\mathbf r$. For the horizontal unit electric dipole, shown in Fig.~\ref{fig:HD}, the current density reads
\begin{equation}\label{eq:J-HD}
\mathbf J(x,y,z)=\mathbf J^h(x,y,z)=\ex \,\delta_{\mathbf r_0}~.
\end{equation}
The film has infinitesimal thickness and scalar surface conductivity $\sigma$, which is in principle complex and $\omega$-dependent.~\cite{Hanson08} The film lies in the plane $z=0$ which separates region 1, the upper half space $\{z>0\}$ with wave number $k_1$, from region 2, the lower half space $\{z<0\}$ with wave number $k_2$ (Figs.~\ref{fig:VD} and~\ref{fig:HD}). We assume that $\Re k_j>0$ and $\Im k_j> 0$ ($j=1,\,2$), i.e., a lossy medium $j$, including the case with $\Im k_j\ll \Re k_j$. Note that $k_j^2=\omega^2\mu_0\tilde \epsilon_j$ where $\mu_0$ is the magnetic permeability of free space, since the media are assumed non-magnetic, and $\tilde\epsilon_j$ is the complex permittivity of medium $j$; in practically appealing situations, this $\tilde\epsilon_j$ has a small imaginary part.

The time-harmonic Maxwell equations for the field ($\mathbf E_j, \mathbf B_j$) in region $j$  dictate that
\begin{subequations}\label{eq:M-Eqs}
\begin{align}
\nabla \times \mathbf E_j&=i\omega \mathbf B_j~,\label{eq:M-Eqs-a}\\
\nabla\times \mathbf B_j&=-i(k_j^2/\omega) \mathbf E_j+\mu_0 \mathbf J~.\label{eq:M-Eqs-b}
\end{align}
\end{subequations}
By Gauss' law, $\nabla \cdot \mathbf B_j=0$ and $\nabla\cdot \mathbf E_j=0$ if $(x,y,z)\neq \mathbf r_0$; however, these equations are not independent from~\eqref{eq:M-Eqs} in the time-harmonic case and, therefore, are not utilized here. Equations~\eqref{eq:M-Eqs} are supplemented with boundary conditions for the tangential components, viz.,~\cite{Bludov13,Hanson08}\looseness=-1
\begin{subequations}\label{eq:BCs}
\begin{align}
&\ez\times (\mathbf E_1-\mathbf E_2)\big|_{z=0}=0~,\label{eq:BCs-a}\\
&\ez\times (\mathbf B_1-\mathbf B_2)\big|_{z=0}=\mu_0\sigma \mathbf E_\parallel~,\label{eq:BCs-b}
\end{align}
\end{subequations}
where $\mathbf E_\parallel:=\{\mathbf E_1-(\ez\cdot \mathbf E_1)\ez\}\big|_{z=0}=\{\mathbf E_2-(\ez\cdot \mathbf E_2)\ez\}\big|_{z=0}$ denotes the (continuous) tangential electric field at $z=0$. Notably, condition~\eqref{eq:BCs-b} expresses the physical property that the thin conducting film amounts to an effective surface current of density $\mathbf J_s=\sigma \mathbf E_\parallel$ at $z=0$; this $\mathbf J_s$ is viewed as a free current density for the field outside the film.
The boundary conditions for the normal components of $(\mathbf E, \mathbf B)$ are redundant for the derivation of a solution; see Appendix~\ref{app:normal}. In addition to~\eqref{eq:BCs}, we impose the Sommerfeld radiation condition, viz.,~\cite{Muller}
\begin{equation}\label{eq:Sommerfeld}
\biggl(\frac{\partial }{\partial r}-ik_j\biggr)\mathfrak F_{js}=o\biggl(\frac{1}{r}\biggr)\quad \mbox{as}\ r\to\infty\qquad (r=\sqrt{x^2+y^2+z^2})~,
\end{equation}
uniformly in $\mathbf r/r$ if $z\neq 0$, for each scalar component $\mathfrak F_{js}$ ($s=x,\,y,\,z$) of the vector-valued field $\boldsymbol{\mathfrak F}_j$ ($\boldsymbol{\mathfrak F}_j=\mathbf E_j, \mathbf B_j$); $\mathbf r=x\,\ex+y\,\ey+z\,\ez$ is the position vector in Cartesian coordinates.
Equations~\eqref{eq:M-Eqs}--\eqref{eq:Sommerfeld} with~\eqref{eq:J-VD} or~\eqref{eq:J-HD} constitute the desired boundary value problem. \looseness=-1

%%%%%%%%%%%%%%%%%%%%%%%%%%%%%%%%%%%%%%%%%%%%%
%%%%%%%%%%%%%%%%%%%%%%%%%%%%%%%%%%%%%%%%%%%%%
\section{Fourier representation of solution}
\label{sec:Fourier}
%%%%%%%%%%%%%%%%%%%%%%%%%%%%%%%%%%%%%%%%%%%%%
%%%%%%%%%%%%%%%%%%%%%%%%%%%%%%%%%%%%%%%%%%%%%
In this section, we derive one-dimensional integral representations for $(\mathbf E_j, \mathbf B_j)$ for a vertical and a horizontal electric dipole. The starting point is the Fourier transform with respect to $(x,y)$ of~\eqref{eq:M-Eqs} and~\eqref{eq:BCs}. Accordingly, let
\begin{equation}\label{eq:EB-FT}
\boldsymbol{\mathfrak F}_j(x,y,z)=\frac{1}{(2\pi)^2}\int_{\mathbb{R}^2}\de\,\dxi\ \widehat{\boldsymbol{\mathfrak F}}_j(\xi,\eta,z)\,
e^{i(\xi x+\eta y)}~,
\end{equation}
where $\boldsymbol{\mathfrak F}_j=\mathbf E_j,\,\mathbf B_j$ and $\widehat{\boldsymbol{\mathfrak F}}_j$
is the Fourier transform of $\boldsymbol{\mathfrak F}_j$ in $(x,y)$, assuming that the integral converges in an appropriate sense. Consequently, by~\eqref{eq:M-Eqs} the transformed variables obey
\begin{subequations}\label{eq:ME-FT}
\begin{align}
i\eta \hate_{jz}-\frac{\partial}{\partial z}\hate_{jy}&=i\omega \hatb_{jx}~,\nonumber\\
-i\xi \hate_{jz}+\frac{\partial}{\partial z}\hate_{jx}&=i\omega \hatb_{jy}~,\nonumber\\
i\xi \hate_{jy}-i\eta\hate_{jx}&=i\omega \hatb_{jz}~,\label{eq:ME-FT-Biot}
\end{align}
and
\begin{align}
i\eta \hatb_{jz}-\frac{\partial}{\partial z}\hatb_{jy}&=-i(k_j^2/\omega)\hate_{jx}+\mu_0 \hat J_x~,\nonumber\\
\frac{\partial}{\partial z}\hatb_{jx}-i\xi \hatb_{jz}&=-i(k_j^2/\omega)\hate_{jy}+\mu_0 \hat J_y~,\nonumber\\
i\xi \hatb_{jy}-i\eta \hatb_{jx}&=-i(k_j^2/\omega)\hate_{jz}+\mu_0 \hat J_z~,\label{eq:ME-FT-Ampere}
\end{align}
\end{subequations}
where $(\hat J_x,\hat J_y,\hat J_z)= (0, 0, \delta_a)$ or $(\delta_a, 0, 0)$ is the Fourier transform of $\mathbf J^v$ or $\mathbf J^h$, respectively; $\delta_a=\delta(z-a)$. Equations~\eqref{eq:ME-FT} are complemented with the transformation of boundary conditions~\eqref{eq:BCs} and radiation condition~\eqref{eq:Sommerfeld}.

%%%%%%%%%%%%%%%%%%%%%%%%%%%%%%%%%%%%%%%%%%%
\subsection{Vertical dipole}
\label{subsec:VD-FT}
%-%%%%%%%%%%%%%%%%%%%%%%%%%%%%%%%%%%%%%%%%%%

Consider Fig.~\ref{fig:VD}. In this case, by symmetry we have $B_{jz}\equiv 0$ for $j=1,\,2$. The remaining field components can be expressed in terms of $\hatb_{jy}$, as shown below. Equations~\eqref{eq:ME-FT} combined yield the differential equation~\cite{KOW}
\begin{equation*}
\biggl(\frac{\partial^2}{\partial z^2}+\beta_j^2\biggr)\hatb_{jy}=i\xi\mu_0 \,\delta(z-a)\qquad z\in \mathbb{R}\setminus \{0\}~,
\end{equation*}
where
\begin{equation*}
\beta_j:=(k_j^2-\xi^2-\eta^2)^{1/2}~,\qquad \Im\beta_j> 0\qquad (j=1,\,2)~.
\end{equation*}
In compliance with radiation condition~\eqref{eq:Sommerfeld}, we write
\begin{equation}\label{eq:By-vd}
\hatb_{1y}(\xi,\eta,z)=C_{>} e^{i\beta_1 z} +\frac{\xi\mu_0}{2\beta_1} e^{i\beta_1 |z-a|}\quad (z>0)~,\quad
\hatb_{2y}(\xi,\eta,z)=C_{<} e^{-i\beta_2 z}\quad (z<0)~,
\end{equation}
where $C_{>}$ and $C_<$ are integration constants to be determined.
In view of~\eqref{eq:ME-FT}, the remaining field components are given in terms of $\hatb_{jy}$ by the relations~\cite{KOW}
\begin{equation}\label{eq:Bx-vd}
\hatb_{jx}=-\frac{\eta}{\xi}\hatb_{jy}~,
\end{equation}
\begin{equation}\label{eq:E-vd}
\hate_{jx}=-\frac{i\omega}{k_j^2}\frac{\partial \hatb_{jy}}{\partial z}~,\quad
\hate_{jy}=-\frac{i\omega}{k_j^2}\frac{\eta}{\xi}\,\frac{\partial \hatb_{jy}}{\partial z}~,\quad
\hate_{jz}=-\frac{\omega}{k_j^2}\frac{\xi^2+\eta^2}{\xi}\,\hatb_{jy}\qquad (z\neq 0,\,a)~.
\end{equation}

To determine $C_>$ and $C_<$ we resort to conditions~\eqref{eq:BCs}, by which $\hate_{1x}=\hate_{2x}$, $\hate_{1y}=\hate_{2y}$, $\hatb_{1x}-\hatb_{2x}=\mu_0\sigma \hate_{2y}$, and $\hatb_{1y}-\hatb_{2y}=-\mu_0\sigma \hate_{2x}$ at $z=0$. In fact, we need only apply the first and third (or, second and fourth) of these conditions since we have set $\hatb_{jz}\equiv 0$ ab initio; the other two conditions are then satisfied. Thus, we obtain
\begin{align*}
C_>&=-\frac{\mu_0 \xi}{2\beta_1}\, \mathcal R_m\,e^{i\beta_1 a}~,\\
C_<&= \mu_0 k_2^2\xi\, \frac{e^{i\beta_1 a}}{\mathcal P}~,
\end{align*}
where the factor $\mathcal R_m$ is defined as
\begin{equation}\label{eq:RE-def}
\mathcal R_{m}=\frac{k_1^2\beta_2-k_2^2\beta_1-\omega\mu_0\sigma\beta_1\beta_2}{\mathcal P}
\end{equation}
which is associated with the reflection of TM-polarized plane waves from the thin layer (the subscript ``m'' stands for ``magnetic'', implying TM polarization); and the corresponding
denominator, $\mathcal P$, is~\cite{Maier-book,Hanson08}
\begin{equation}\label{eq:P-def}
\mathcal P=k_1^2\beta_2+k_2^2\beta_1+\omega\mu_0\sigma\beta_1\beta_2~.
\end{equation}
The two-dimensional Fourier integrals for $(\mathbf E, \mathbf B)$ then follow from~\eqref{eq:EB-FT} with \eqref{eq:By-vd}--\eqref{eq:E-vd}.

To reduce representation~\eqref{eq:EB-FT} to one-dimensional integrals, we resort to the cylindrical coordinates $(\rho, \phi, z)$ where $x=\rho \cos\phi$ and $y=\rho\sin\phi$ ($0\le\phi<2\pi$), following Ref.~\onlinecite{KOW}. The cylindrical components of the field $\boldsymbol{\mathfrak F}_j$ ($\boldsymbol{\mathfrak F}_j=\mathbf E_j, \mathbf B_j$) include $\mathfrak F_{j\rho}=\mathfrak F_{jx} \cos\phi+\mathfrak F_{jy} \sin\phi$ and $\mathfrak F_{j\phi}=-\mathfrak F_{jx}\sin\phi+\mathfrak F_{jy} \cos\phi$. Accordingly, let $(\xi,\eta)\mapsto (\lambda,\php)$ with $(\xi,\eta)=(\lambda\cos\php,\lambda\sin\php)$ where $\lambda\ge 0$ and $0\le\php< 2\pi$; thus, $\xi x+\eta y=\lambda\rho \cos(\phi-\phi')$ and $\dxi\,\de=\lambda\,\dph\,\dl$. By direct integration in $\phi'$, we find
\begin{equation*}
E_{j\phi}\equiv 0\quad \mbox{and}\quad B_{j\rho}\equiv 0~,
\end{equation*}
as expected by symmetry because of the dipole orientation.
For the remaining components, we invoke the known formula~\cite{Bateman-II}
\begin{equation}\label{eq:Jn-int}
e^{in\phi}J_n(w)=\frac{i^{-n}}{2\pi}\int_0^{2\pi}\dph\,e^{iw\cos(\phi-\phi')+in\phi'}
\end{equation}
for $n=0$ and $1$, where $J_n$ is the Bessel function of $n$th order, and $n\in \mathbb{Z}$.

Consequently, after some algebra, we find the following integral representations.

For $z>0$ (region $1$) with ($x,y,z)\neq (0,0,a)$,
\begin{subequations}\label{eq:E-reg1-vd}
\begin{align}
E_{1\rho}(\rho,\phi,z)&=\frac{i\omega\mu_0}{4\pi k_1^2}\int_0^\infty \dl\,\lambda^2 J_1(\lambda\rho)\bigl[ {\rm sgn}(z-a) e^{i\beta_1|z-a|}-\mathcal R_{m}\,e^{i\beta_1(z+a)}\bigr]~,\label{eq:E1r-vd}\\
E_{1z}(\rho,\phi,z)&=-\frac{\omega\mu_0}{4\pi k_1^2}\int_0^\infty \dl\,\frac{\lambda^3}{\beta_1}J_0(\lambda\rho)\bigl[e^{i\beta_1|z-a|}-\mathcal R_{m}\,e^{i\beta_1(z+a)}\bigr]~,\label{eq:E1z-vd}
\end{align}
\end{subequations}
\begin{equation}\label{eq:B-reg1-vd}
B_{1\phi}(\rho,\phi,z)=\frac{i\mu_0}{4\pi}\int_0^\infty \dl\,\frac{\lambda^2}{\beta_1}J_1(\lambda\rho)\bigl[e^{i\beta_1|z-a|}-\mathcal R_{m}\,e^{i\beta_1(z+a)}\bigr]~;
\end{equation}
and for $z<0$ (region $2$),
\begin{subequations}\label{eq:E-reg2-vd}
\begin{align}
E_{2\rho}(\rho,\phi,z)&=-\frac{i\omega\mu_0}{2\pi}\int_0^\infty \dl\,\lambda^2 J_1(\lambda\rho)\,\frac{\beta_2}{\mathcal P}\ e^{-i\beta_2 z+i\beta_1 a}~,\label{eq:E2r-vd}\\
E_{2z}(\rho,\phi,z)&=-\frac{\omega\mu_0}{2\pi}\int_0^\infty \dl\,\lambda^3 J_0(\lambda\rho) \frac{1}{\mathcal P}\ e^{-i\beta_2 z+i\beta_1a}~,\label{eq:E2z-vd}
\end{align}
\end{subequations}
\begin{equation}
B_{2\phi}(\rho,\phi,z)=\frac{i\mu_0k_2^2}{2\pi}\int_0^\infty \dl\,\lambda^2 J_1(\lambda\rho)\,\frac{1}{\mathcal P}\,e^{-i\beta_2 z+i\beta_1 a}~.\label{eq:B2ph-vd}
\end{equation}

In the above, ${\rm sgn}(z)=1$ if $z>0$ and ${\rm sgn}(z)=-1$ if $z<0$; and, by~\eqref{eq:RE-def} and~\eqref{eq:P-def}, $\mathcal R_m$ and $\mathcal P$ are functions of $\lambda$ with
\begin{subequations}\label{eq:betaj-def}
\begin{equation}\label{eq:betaj-def-a}
\beta_j(\lambda)=(k_j^2-\lambda^2)^{1/2}\qquad (j=1,\,2)~.
\end{equation}
In accord with the Sommerfeld radiation condition, the top (physical) Riemann sheet in the four-sheeted $\lambda$-Riemann surface for the field components is fixed by imposition of
\begin{equation}\label{eq:betaj-def-b}
\Im\beta_j(\lambda)> 0~,
\end{equation}
\end{subequations}
for each $j=1,\,2$; thus, $\beta_j(\lambda)=\beta_j(-\lambda)$ in this Riemann sheet. We note in passing that in the limit where $a\downarrow 0$ and $z\to 0$, integrals~\eqref{eq:E-reg1-vd}--\eqref{eq:B2ph-vd} approach expressions that are divergent in the conventional sense yet become meaningful as finite in the sense of Abel.~\cite{Hardy,Margetis01} This physically transparent interpretation permeates Section~\ref{sec:Exact}.

%%%%%%%%%%%%%%%%%%%%%%%%%%%%%%%%%%%%%%%%%%%
\subsection{Horizontal dipole}
\label{subsec:HD-FT}
%%%%%%%%%%%%%%%%%%%%%%%%%%%%%%%%%%%%%%%%%%%
Next, we focus on the geometry of Fig.~\ref{fig:HD}. In this case, all cylindrical field components are in principle nonzero. By~\eqref{eq:ME-FT}, $\widehat{E}_{jx}$ and $\widehat{B}_{jx}$  satisfy~\cite{KOW}
\begin{align*}
\biggl(\frac{\partial^2}{\partial z^2}+\beta_j^2\biggr)\hate_{jx}&=-i\frac{\omega\mu_0(k_1^2-\xi^2)}{k_1^2}\delta(z-a)~,\nonumber\\
\biggl(\frac{\partial^2}{\partial z^2}+\beta_j^2\biggr)\hatb_{jx}&=0~,
\end{align*}
where $z\neq 0$ and $\beta_j(\lambda)$ is defined in~\eqref{eq:betaj-def}, with admissible solutions
\begin{align}
\hate_{1x}&=-\frac{\omega\mu_0 (k_1^2-\xi^2)}{2\beta_1 k_1^2} e^{i\beta_1|z-a|}+K_> e^{i\beta_1z}\quad (z>0)~,\quad \hate_{2x}=K_< e^{-i\beta_2 z}\quad (z<0)~, \label{eq:Bx-hd}\\
\hatb_{1x}&=C_> e^{i\beta_1 z}\quad (z>0)~,\quad \hatb_{2x}=C_< e^{-i\beta_2 z}\quad (z<0)~,\label{eq:Ex-hd}
\end{align}
consistent with radiation condition~\eqref{eq:Sommerfeld}.
Note that $C_>\neq C_<$ here because, in view of condition~\eqref{eq:BCs-b}, the tangential component of the $\mathbf B$ field is not continuous across $z=0$, in contrast to the formulation of Ref.~\onlinecite{KOW}. The integration constants $C_<$, $C_>$, $K_<$ and $K_>$ are determined through boundary conditions~\eqref{eq:BCs}.
By transformed Maxwell equations~\eqref{eq:ME-FT}, the remaining field components can be expressed in terms of $\hate_{jx}$ and $\hatb_{jx}$ as
\begin{equation}\label{eq:B-yz}
\hatb_{jy}=-\frac{1}{k_j^2-\xi^2}\biggl(i\frac{k_j^2}{\omega} \frac{\partial \hate_{jx}}{\partial z}+\eta\xi \hatb_{jx}\biggr)~,\quad \hatb_{jz}=\frac{1}{k_j^2-\xi^2}\biggl( -\eta\frac{k_j^2}{\omega}\hate_{jx}+i\xi \frac{\partial\hatb_{jx}}{\partial z}\biggr)~,
\end{equation}
\begin{equation}\label{eq:E-yz}
\hate_{jy}=\frac{i}{k_j^2-\xi^2}\biggl(i\eta\xi \hate_{jx}+\omega \frac{\partial\hatb_{jx}}{\partial z}\biggr)~,\quad \hate_{jz}=\frac{i}{k_j^2-\xi^2}\biggl( \xi \frac{\partial\hate_{jx}}{\partial z}-i\eta\omega \hatb_{jx}\biggr)~\quad (z\neq 0,\,a)~.
\end{equation}

We proceed to compute the fields. By~\eqref{eq:BCs-a} we impose $\hate_{1x}=\hate_{2x}$ at $z=0$; thus,
\begin{equation*}
K_>=K_<+\frac{\omega\mu_0(k_1^2-\xi^2)}{2\beta_1 k_1^2}~.
\end{equation*}
We henceforth set $K=K_<$ for ease of notation. By~\eqref{eq:BCs-b}, we enforce $\hatb_{1x}-\hatb_{2x}=\mu_0\sigma \hate_{2y}$ at $z=0$ which yields
\begin{equation*}
C_>=-\frac{\mu_0\sigma\,\eta\xi}{k_2^2-\xi^2}K+\biggl(1+\frac{\omega\mu_0\sigma\beta_2}{k_2^2-\xi^2}\biggr)C_<~.
\end{equation*}
Now introduce $C=C_<$. Thus, it suffices to determine the integration constants $K$ and $C$.

By~\eqref{eq:BCs}, the continuity condition $\hate_{1y}=\hate_{2y}$ and the jump condition $\hatb_{1y}-\hatb_{2y}=-\mu_0\sigma \hate_{2x}$ at $z=0$ lead to the system
\begin{equation*}
\eta\xi\biggl[ \frac{1}{k_1^2-\xi^2}\biggl(1-\frac{\omega\mu_0\sigma\beta_1}{k_2^2-\xi^2}\biggr)-\frac{1}{k_2^2-\xi^2}\biggr]K+\omega\biggl[\frac{\beta_1}{k_1^2-\xi^2}\biggl(1+\frac{\omega\mu_0\sigma\beta_2}{k_2^2-\xi^2}\biggr)+\frac{\beta_2}{k_2^2-\xi^2}\biggr]C=0~,
\end{equation*}
\begin{align*}
&-\biggl\{\frac{k_1^2\beta_1}{k_1^2-\xi^2}+\frac{k_2^2\beta_2}{k_2^2-\xi^2}+\omega\mu_0\sigma\biggl[1+\frac{\eta^2\xi^2}{(k_1^2-\xi^2)(k_2^2-\xi^2)}\biggr]\biggr\}K\nonumber\\
&\mbox{}\qquad +\omega\eta\xi\biggl[\frac{1}{k_1^2-\xi^2}\biggl(1+\frac{\omega\mu_0\sigma\beta_2}{k_2^2-\xi^2}\biggr)-\frac{1}{k_2^2-\xi^2}\biggr]C=\omega\mu_0 e^{i\beta_1 a}~.
\end{align*}
After some algebra, the determinant for this system is found to be
\begin{equation*}
\mathcal D=\frac{\mathcal P\,\mathcal Q}{(k_1^2-\xi^2)(k_2^2-\xi^2)}~,
\end{equation*}
where $\mathcal P$ is defined by~\eqref{eq:P-def}, and
\begin{equation}\label{eq:Q-def}
\mathcal Q=\beta_1+\beta_2+\omega\mu_0\sigma
\end{equation}
corresponds to TE polarization.~\cite{Hanson08} It follows that
\begin{equation*}
C=\mu_0\eta\xi\,\frac{k_2^2-k_1^2-\omega\mu_0\sigma\beta_1}{\mathcal P\,\mathcal Q}\,e^{i\beta_1 a}~,
\end{equation*}
\begin{equation*}
K=-\omega\mu_0 \,\frac{\beta_1(k_2^2-\xi^2)+\beta_2(k_1^2-\xi^2)+\omega\mu_0\sigma\beta_1\beta_2}{\mathcal P\,\mathcal Q}\,e^{i\beta_1a}~.
\end{equation*}

Accordingly, all field components are now obtained via~\eqref{eq:Bx-hd}--\eqref{eq:E-yz} by use of Fourier integral~\eqref{eq:EB-FT}. By representing the resulting two-dimensional integrals in cylindrical coordinates $(\rho,\phi, z)$ and using~\eqref{eq:Jn-int},~\cite{KOW} we obtain the following expressions.

For $z>0$ (region 1):
\begin{subequations}\label{eq:E-rg1-hd}
\begin{align}
E_{1\rho}(\rho,\phi,z)&=-\frac{\omega\mu_0}{4\pi}\cos\phi\Biggl(\frac{1}{2}\int_0^\infty \dl\,\lambda\biggl\{[J_0(\lambda\rho)-J_2(\lambda\rho)]\, \frac{\beta_1}{k_1^2}\,\mathcal R_{m}\nonumber\\
&\mbox{}\hphantom{ =-\frac{\omega\mu_0}{4\pi}\cos\phi\Biggl(\frac{1}{2}\int_0^\infty \dl\, }-[J_0(\lambda\rho)+J_2(\lambda\rho)]\,\frac{1}{\beta_1}\,\mathcal R_{e}\biggr\}e^{i\beta_1(z+a)}\nonumber\\
&\mbox{}\hphantom{=\frac{\omega\mu_0}{4\pi}\cos\phi}+\int_0^\infty\dl\,\lambda\biggl\{J_0(\lambda\rho)-\frac{\lambda^2}{2k_1^2}[J_0(\lambda\rho)-J_2(\lambda\rho)]\biggr\}\frac{1}{\beta_1}\,e^{i\beta_1|z-a|}\Biggr)~,\label{eq:E1r-hd}
\end{align}
\begin{align}
E_{1\phi}(\rho,\phi,z)&=\frac{\omega\mu_0}{8\pi k_1^2}\sin\phi\Biggl(\int_0^\infty \dl\,\lambda\biggl\{-[J_0(\lambda\rho)-J_2(\lambda\rho)]\,\frac{k_1^2}{\beta_1}\,\mathcal R_{e}\nonumber\\
&\mbox{}\hphantom{=\frac{\omega\mu_0}{4\pi k_1^2}\sin\phi\Biggl(\frac{1}{2}\int_0^\infty \dl\,}+[J_0(\lambda\rho)+J_2(\lambda\rho)]\,\beta_1\,\mathcal R_{m}\biggr\}e^{i\beta_1(z+a)}\nonumber\\
&\mbox{}\hphantom{=\frac{\omega\mu_0}{4\pi k_1^2}}+\int_0^\infty\dl\,\lambda\biggl\{k_1^2 J_0(\lambda\rho)-\frac{\lambda^2}{2}[J_0(\lambda\rho)+J_2(\lambda\rho)]\biggr\}\frac{2}{\beta_1}\,e^{i\beta_1|z-a|}\Biggr)~,\label{eq:E1ph-hd}
\end{align}
\begin{align}
E_{1z}(\rho,\phi,z)&= \frac{i\omega \mu_0}{4\pi k_1^2}\cos\phi \int_0^\infty \dl\,\lambda^2 J_1(\lambda\rho) \bigl[{\rm sgn}(z-a) e^{i\beta_1|z-a|}+ \mathcal R_{m}\,e^{i\beta_1(z+a)}\bigr]~,\label{eq:E1z-hd}
\end{align}
\end{subequations}
\begin{subequations}\label{eq:B-rg1-hd}
\begin{align}
B_{1\rho}(\rho,\phi,z)&=-\frac{\mu_0}{8\pi}\sin\phi \Biggl(\int_0^\infty \dl\,\lambda \bigl\{ [J_0(\lambda\rho)+J_2(\lambda\rho)]\,\mathcal R_{m}\nonumber\\
&\mbox{}\hphantom{=-\frac{\mu_0}{4\pi}\sin\phi \Biggl(\frac{1}{2}\int_0^\infty \dl\,}-[J_0(\lambda\rho)-J_2(\lambda\rho)]\,\mathcal R_{e}\bigr\} e^{i\beta_1(z+a)}\nonumber\\
&\mbox{}\hphantom{ =-\frac{\mu_0}{4\pi}\sin\phi \Biggl(\frac{1}{2} }+2\,{\rm sgn}(z-a)\int_0^\infty \dl\,\lambda J_0(\lambda\rho)\,e^{i\beta_1|z-a|}\Biggr)~,\label{eq:B1r-hd}
\end{align}
\begin{align}
B_{1\phi}(\rho,\phi,z)&=-\frac{\mu_0}{8\pi}\cos\phi \Biggl(\int_0^\infty \dl\,\lambda \bigl\{ [J_0(\lambda\rho)-J_2(\lambda\rho)]\,\mathcal R_{m}\nonumber\\
&\mbox{} \hphantom{=-\frac{\mu_0}{4\pi}\cos\phi \Biggl(\frac{1}{2}\int_0^\infty \dl\,  }-[J_0(\lambda\rho)+J_2(\lambda\rho)]\,\mathcal R_{e}\bigr\} e^{i\beta_1(z+a)}\nonumber\\
&\mbox{}\hphantom{ =-\frac{\mu_0}{4\pi}\cos\phi }+2\,{\rm sgn}(z-a)\int_0^\infty \dl\,\lambda J_0(\lambda\rho)\,e^{i\beta_1|z-a|}\Biggr)~,\label{eq:B1ph-hd}
\end{align}
\begin{align}
B_{1z}(\rho,\phi,z)&=-\frac{i\mu_0}{4\pi}\sin\phi\int_0^\infty \dl\,\lambda^2 J_1(\lambda\rho)\,\frac{1}{\beta_1}\bigl[ \mathcal R_{e}\,e^{i\beta_1(z+a)}-e^{i\beta_1|z-a|}\bigr]~;\label{eq:B1z-hd}
\end{align}
\end{subequations}
and for $z<0$ (region $2$),
\begin{subequations}\label{eq:E-rg2-hd}
\begin{align}
E_{2\rho}(\rho,\phi,z)&=-\frac{\omega\mu_0}{4\pi}\cos\phi\int_0^\infty \dl\,\lambda \biggl\{ [J_0(\lambda\rho)+J_2(\lambda\rho)]\,\frac{1}{\mathcal Q}\nonumber\\
&\mbox{}\hphantom{ -\frac{\omega\mu_0}{4\pi}\cos\phi\int_0^\infty \dl\,} +[J_0(\lambda\rho)-J_2(\lambda\rho)]\,\frac{\beta_1\beta_2}{\mathcal P}\biggr\}e^{i\beta_1a-i\beta_2z}~,\label{eq:E2r-hd}
\end{align}
\begin{align}
E_{2\phi}(\rho,\phi,z)&=\frac{\omega\mu_0}{4\pi}\sin\phi \int_0^\infty \dl\,\lambda \biggl\{[J_0(\lambda\rho)-J_2(\lambda\rho)]\,\frac{1}{\mathcal Q}\nonumber\\
&\mbox{}\hphantom{ =\frac{\omega\mu_0}{4\pi}\sin\phi \int_0^\infty \dl\,}+[J_0(\lambda\rho)+J_2(\lambda\rho)]\,\frac{\beta_1\beta_2}{\mathcal P}\biggr\}e^{i\beta_1a-i\beta_2z}~,\label{eq:E2ph-hd}
\end{align}
\begin{align}
E_{2z}(\rho,\phi,z)&=-\frac{i\omega\mu_0}{2\pi}\cos\phi \int_0^\infty \dl\,\lambda^2 J_1(\lambda\rho)\,\frac{\beta_1}{\mathcal P}\,e^{i\beta_1a-i\beta_2 z}~,\label{eq:E2z-hd}
\end{align}
\end{subequations}
\begin{subequations}\label{eq:B-rg2-hd}
\begin{align}
B_{2\rho}(\rho,\phi,z)&=\frac{\mu_0}{4\pi}\sin\phi \int_0^\infty \dl\,\lambda \biggl\{[J_0(\lambda\rho)+J_2(\lambda\rho)]\,\frac{k_2^2\beta_1}{\mathcal P}\nonumber\\
&\mbox{}\hphantom{\frac{\mu_0}{4\pi}\sin\phi \int_0^\infty \dl\, }+[J_0(\lambda\rho)-J_2(\lambda\rho)]\,\frac{\beta_2}{\mathcal Q}\biggr\} e^{i\beta_1 a-i\beta_2 z}~,\label{eq:B2r-hd}
\end{align}
\begin{align}
B_{2\phi}(\rho,\phi,z)&=\frac{\mu_0}{4\pi}\cos\phi\int_0^\infty \dl\,\lambda \biggl\{[J_0(\lambda\rho)-J_2(\lambda\rho)]\,\frac{k_2^2\beta_1}{\mathcal P}\nonumber\\
&\mbox{} \hphantom{ \frac{\mu_0}{4\pi}\cos\phi\int_0^\infty \dl\,}+[J_0(\lambda\rho)+J_2(\lambda\rho)]\,\frac{\beta_2}{\mathcal Q}\biggr\}e^{i\beta_1 a-i\beta_2 z}~,\label{eq:B2ph-hd}
\end{align}
\begin{equation}
B_{2z}(\rho,\phi,z)=\frac{i\mu_0}{2\pi}\sin\phi\int_0^\infty \dl\,\lambda^2 J_1(\lambda\rho)\,\frac{1}{\mathcal Q}\,e^{i\beta_1 a-i\beta_2 z}~.\label{eq:B2z-hd}
\end{equation}
\end{subequations}
In the above, the factor
\begin{equation}\label{eq:RH-def}
\mathcal R_{e}=\frac{\beta_2-\beta_1+\omega\mu_0\sigma}{\mathcal Q}
\end{equation}
is associated with the reflection of plane waves with TE polarization from a thin layer (the subscript ``e'' stands for ``electric'', implying TE polarization); also, recall~\eqref{eq:RE-def}, \eqref{eq:P-def} and~\eqref{eq:Q-def} for $\mathcal R_m$, $\mathcal P$ and $\mathcal Q$, respectively. Similarly to the vertical-dipole case (Section~\ref{subsec:VD-FT}), the top Riemann sheet is defined by $\Im\beta_j(\lambda)> 0$, as in~\eqref{eq:betaj-def-b}.

%%%%%%%%%%%%%%%%%%%%%%%%%%%%%%%%%%%%%%%%%%%%%%
\subsection{Poles in $\lambda$-Riemann surface}
\label{subsec:poles}
%%%%%%%%%%%%%%%%%%%%%%%%%%%%%%%%%%%%%%%%%%%%%%
We now discuss the role of singularities in evaluating integrals~\eqref{eq:E-reg1-vd}--\eqref{eq:B2ph-vd} and~\eqref{eq:E-rg1-hd}--\eqref{eq:B-rg2-hd}. The singularities of the integrands include: branch points at $\lambda=k_1,\,k_2$ because of the multiple-valued $\beta_j(\lambda)$ of~\eqref{eq:betaj-def-a}; and simple poles identified with zeros of $\mathcal P(\lambda)$ and $\mathcal Q(\lambda)$ by~\eqref{eq:P-def} and~\eqref{eq:Q-def}. The $\lambda$-Riemann surface associated with each integrand consists of four Riemann sheets if $k_1\neq k_2$; the physically relevant (top) one is fixed by $\Im \beta_j(\lambda)>0$.
The location of the poles in this Riemann surface is closely related to the complex surface conductivity, $\sigma$.~\cite{Hanson08}

%--------------------------------------------------%
\subsubsection{Denominator $\mathcal P(\lambda)$}
%--------------------------------------------------%

By~\eqref{eq:P-def}, the zeros of $\mathcal P(\lambda)$ satisfy
\begin{equation}\label{eq:P-zeros}
\frac{k_1^2}{(k_1^2-\lambda^2)^{1/2}}+\frac{k_2^2}{(k_2^2-\lambda^2)^{1/2}}+\omega\mu_0\sigma=0~.
\end{equation}
Equation~\eqref{eq:P-zeros} is identified with the dispersion relation for the surface plasmon in the context of TM polarization for plane waves~\cite{Maier-book} if $\lambda$ is viewed as the component of the vector wave number tangential to the interface. For nearly lossless ambient media ($\Im k_j\simeq 0$), roots of~\eqref{eq:P-zeros} lie in the first Riemann sheet if
\begin{equation}\label{eq:sigma-cond}
\Im \sigma >0~.
\end{equation}
For a graphene layer, this is consistent with the Kubo formula in the far-infrared regime.~\cite{Hanson08,Falkovsky07,Gan12,Cheng13}

Consider the special case with $k_1=k_2=:k$ (Section~\ref{sec:Exact}) in which the $\lambda$-Riemann surface for the fields has two sheets. Then, $\mathcal P(\lambda)=(2k^2+\omega\mu_0\sigma \sqrt{k^2-\lambda^2})\sqrt{k^2-\lambda^2}$ and  \eqref{eq:P-zeros} reduces to $\sqrt{k^2-\lambda^2}=-2k^2/(\omega\mu_0\sigma)$, which is solved in the first Riemann sheet if $-\pi< \arg k^2-\arg\sigma<0$. Equation \eqref{eq:P-zeros} has two solutions, $\lambda=\pm k_m$,  where
\begin{equation}\label{eq:kp-iso}
k_m=\sqrt{k^2-\frac{4k^4}{(\omega\mu_0\sigma)^2}}~;
\end{equation}
the branch of the square root is chosen so that $\Im k_m > 0$.  These $\pm k_m$ are simple poles of the corresponding integrands for the fields. In the ``nonretarded frequency regime'',~\cite{Cheng13} one imposes $|\omega\mu_0 \sigma|\ll |k|$ by which $1/|\sigma|\gg \sqrt{\mu_0/|\tilde\epsilon|}$, where  $\sqrt{\mu_0/|\tilde\epsilon|}$ is the magnitude of the intrinsic impedance of the adjacent homogeneous space; thus,
we obtain $k_m\sim i\,2k^2/(\omega\mu_0\sigma)$. For  $\arg\sigma=\pi/2-\delta$, $0<\delta<\pi$, the analytically continued square-root function yields $(k^2-k_m^2)^{1/2}=ie^{i\delta}2k^2/(\omega\mu_0|\sigma|)$ and $k_m$ lies in the first Riemann sheet for the appropriate range of $\delta$ and phase of $k$, as outlined above; this $k_m$ reads
\begin{equation}\label{eq:km-nonr}
k_m\sim \frac{2k^2}{\omega\mu_0|\sigma|}\,e^{i\delta}~,\qquad |\omega\mu_0\sigma|\ll |k|~.
\end{equation}
If $\Im \sqrt{k^2-k_m^2}<0$, then $k_m$ is in the second Riemann sheet.

We adhere to the following definition for the purposes of the analysis in Section~\ref{sec:Exact}.
\smallskip

{\em Definition 1 (TM surface plasmon). The \textbf {TM surface plasmon} corresponds to the residue contribution to the electromagnetic field from the pole $\lambda=k_m$, provided this pole lies in the first Riemann sheet.}

We note in passing that in the more general setting with $k_1\neq k_2$, one finds the simple poles $\pm k_m\sim \pm i(k_1^2+k_2^2)/(\omega\mu_0\sigma)$ if $|(k_1^2+k_2^2)/(\omega\mu_0\sigma)|\gg \max\{|k_1|, |k_2|\}$.

%--------------------------------------------------%
\subsubsection{Denominator $\mathcal Q(\lambda)$}
%--------------------------------------------------%

By~\eqref{eq:Q-def}, the zeros of $\mathcal Q(\lambda)$ obey
\begin{equation}\label{eq:Q-zeros}
(k_1^2-\lambda^2)^{1/2}+(k_2^2-\lambda^2)^{1/2}+\omega\mu_0\sigma=0~,
\end{equation}
which has two roots, $\pm k_e$, in the $\lambda$-Riemann surface, where~\cite{Hanson08,Hanson11}
\begin{equation*}
k_e=\frac{1}{2}\sqrt{-\frac{(k_1^2-k_2^2)^2}{(\omega\mu_0\sigma)^2}+2(k_1^2+k_2^2)-(\omega\mu_0\sigma)^2}~.
\end{equation*}
For definiteness, the branch of the square root is chosen so that $k_e\to k$ if one sets $k_1=k_2=k$ and then lets $\omega\mu_0\sigma\to 0$.
The $\pm k_e$ are simple poles of the corresponding integrands for the fields. By~\eqref{eq:Q-zeros} and for nearly lossless ambient media ($\Im k_j\simeq 0$), $\pm k_e$ lie in the top Riemann sheet if
\begin{equation}\label{eq:sigma-cond-n}
\Im \sigma <0~,
\end{equation}
in contrast to~\eqref{eq:sigma-cond}. Thus, for the appealing case with nonzero $\Im \sigma$ and an ambient lossless dielectric,~\cite{Hanson08,Cheng13} only one of $k_m$ and $k_e$ lies in the top Riemann sheet.~\cite{Hanson08} If $|\omega\mu_0\sigma|^2\ll |k_1^2+k_2^2|$, then $\pm k_e\sim \pm (i/2)(k_1^2-k_2^2)/(\omega\mu_0\sigma)$ for $k_1\neq k_2$; hence, $|k_e|$ may take a wide range of values.\looseness=-1

In the special case with $k_1=k_2=k$ (Section~\ref{sec:Exact}),
\eqref{eq:Q-zeros} reduces to $\sqrt{k^2-\lambda^2}=-\omega\mu_0\sigma/2$, which is solved in the first Riemann sheet if $-\pi< \arg\sigma< 0$.
We find
\begin{equation}\label{eq:kst-iso}
k_e=\sqrt{k^2-\frac{(\omega\mu_0\sigma)^2}{4}}~.
\end{equation}
If $|\omega\mu_0\sigma|\ll |k|$ then $k_e\sim k -(\omega\mu_0\sigma)^2/(8k)\sim k$. By setting $\arg\sigma=\pi/2-\delta$, we find
\begin{equation}\label{eq:ke-nonr}
k_e- k\sim \frac{(\omega\mu_0|\sigma|)^2}{8k}\,e^{-i2\delta}~,\qquad |\omega\mu_0\sigma|\ll |k|\qquad (0<\delta<\pi)~.
\end{equation}

The above discussion leads to the following remark.
\smallskip

{\em Remark 1 (On the co-existence of poles $k_m$ and $k_e$ in top Riemann sheet). Consider the case with $k_1=k_2=k$. We henceforth assume that $0\le \arg k^2<\pi/2$ and $\Re\sigma>0$ with $\arg\sigma=\pi/2-\delta$ ($0<\delta<\pi$). Thus, at most one of the poles at $k_m$ and $k_e$ is present in the top $\lambda$-Riemann sheet.

Specifically:  If $\Im\sigma >0$ with $0<\delta<\pi/2-\arg k^2$, then $k_m$ lies in the top Riemann sheet; and if $\Im \sigma<0$ ($\pi/2<\delta<\pi$), then $k_e$ lies in the top Riemann sheet.

}

%%%%%%%%%%%%%%%%%%%%%%%%%%%%%%%%%%%%%%%%%%%%%%%%%%%
%%%%%%%%%%%%%%%%%%%%%%%%%%%%%%%%%%%%%%%%%%%%%%%%%%%
\section{Exact Evaluation of Field on Plane of Film}
\label{sec:Exact}
%%%%%%%%%%%%%%%%%%%%%%%%%%%%%%%%%%%%%%%%%%%%%%%%%%%
%%%%%%%%%%%%%%%%%%%%%%%%%%%%%%%%%%%%%%%%%%%%%%%%%%%
In this section, we analytically evaluate the integrals for the electromagnetic field (Sections~\ref{subsec:VD-FT} and~\ref{subsec:HD-FT}) on one side of the layer by setting $k_1=k_2=k$ and allowing the dipole and observation point ($x, y, z$) to approach  $z=0$.

%%%%%%%%%%%%%%%%%%%%%%%%%%%%%%%%%%%%%%%%
\subsection{Formalism}
\label{subsec:components}
%%%%%%%%%%%%%%%%%%%%%%%%%%%%%%%%%%%%%%%%

For the vertical dipole (Fig.~\ref{fig:VD}), let $a\downarrow 0$ with $0<z<a$, so that $(\rho,\phi,z)$ approaches the film from region 1. Equations~\eqref{eq:E-reg1-vd} and \eqref{eq:B-reg1-vd} reduce to the following expressions.\looseness=-1
\begin{align}
E_{1\rho}^{\rm (vd)}&=-\frac{i\omega\mu_0}{2\pi}\frac{1}{(\omega\mu_0\sigma)^2}\biggl[\frac{2k^2}{\rho}\, I_1-\frac{\omega\mu_0\sigma}{\rho^2} \,I_2+2k^2k_m^2\rho\, I_3(k_m\rho)-\omega\mu_0\sigma k_m^2 \,I_4(k_m\rho)\biggr],\label{eq:E1r-vd-0}\\
B_{1\phi}^{\rm (vd)}&=\frac{i\mu_0}{2\pi}\biggl\{ \frac{k^2}{\omega\mu_0\sigma}\frac{1}{\rho}\,I_1-\frac{1}{\rho^2}I_2+\biggl[k^2-\frac{2k^4}{(\omega\mu_0\sigma)^2}\biggr]\,\breve I_2+\frac{k^2k_m^2\rho}{\omega\mu_0\sigma}\,I_3(k_m\rho)\nonumber\\
&\mbox{}\qquad -\frac{2k^4k_m^2\rho^2}{(\omega\mu_0\sigma)^2}\,\breve I_4(k_m\rho)\biggr\}~,\label{eq:B1ph-vd-0}\\
E_{1z}^{\rm (vd)}&=\frac{i\omega}{k^2}\frac{1}{\rho}\frac{\partial}{\partial\rho}(\rho B_{1\phi}^{({\rm vd})})\nonumber\\
&=-\frac{\omega\mu_0}{2\pi k^2}\frac{1}{\rho}\frac{\partial}{\partial\rho}\rho
\biggl\{ \frac{k^2}{\omega\mu_0\sigma}\frac{1}{\rho}\,I_1-\frac{1}{\rho^2}\,I_2+\biggl[k^2-\frac{2k^4}{(\omega\mu_0\sigma)^2}\biggr]\,\breve I_2+\frac{k^2k_m^2\rho}{\omega\mu_0\sigma}\,I_3(k_m\rho)\nonumber\\
&\mbox{}\qquad -\frac{2k^4k_m^2\rho^2}{(\omega\mu_0\sigma)^2}\,\breve I_4(k_m\rho)\biggr\}
~,\label{eq:E1z-vd-0}
\end{align}
where the requisite integrals read
\begin{subequations}\label{eq:I-int}
\begin{align}
I_1&=\int_0^\infty \dlp\,J_1(\lambda')=-\int_0^\infty \dlp\ \frac{d}{d\lambda'}J_0(\lambda')=J_0(0)=1~,\label{eq:I-int1}\\
I_2&=\int_0^\infty \dlp\,\sqrt{(k\rho)^2-\lambda'^2}\,J_1(\lambda')~,\quad
\breve I_2=\int_0^\infty \dlp\,\frac{J_1(\lambda')}{\sqrt{(k\rho)^2-\lambda'^2}}~,\label{eq:I-int-2}\\
I_3(\varpi)&=\int_0^\infty \dlp\ \frac{J_1(\lambda')}{\lambda'^2-\varpi^2}~,\label{eq:I-int-3}\\
I_4(\varpi)&=\int_0^\infty \dlp\ \frac{\sqrt{(k\rho)^2-\lambda'^2}}{\lambda'^2-\varpi^2} J_1(\lambda')~,\quad
\breve I_4(\varpi)=\int_0^\infty \dlp\ \frac{1}{\sqrt{(k\rho)^2-\lambda'^2}}
\frac{J_1(\lambda')}{\lambda'^2-\varpi^2}~,\label{eq:I-int-4}
\end{align}
\end{subequations}
with $\breve I_4(\varpi)=-[\varpi^2-(k\rho)^2]^{-1}[\breve I_2+I_4(\varpi)]$. Thus, we only need to compute $I_2,\,\breve I_2,\,I_3$ and $I_4$.\looseness=-1

For a horizontal dipole (Fig.~\ref{fig:HD}), we choose to evaluate the field on the side of the layer that faces the source-free region: Let $a\downarrow 0$ and $z\uparrow 0$ so that the observation point approaches the plane $z=0$ from region 2. By~\eqref{eq:E-rg2-hd} and~\eqref{eq:B-rg2-hd}, the ensuing representations for the field components at $z=0$ in region 2 read
\begin{align}
E_{2z}^{\rm (hd)}&=-\frac{i\omega\mu_0}{2\pi}\frac{1}{(\omega\mu_0\sigma)^2}\cos\phi
\biggl[\frac{2k^2}{\rho}\, I_1-\frac{\omega\mu_0\sigma}{\rho^2}\,I_2+2k^2k_m^2\rho\, I_3(k_m\rho)\nonumber\\
&\mbox{} \hphantom{=-\frac{i\omega\mu_0}{2\pi}\frac{1}{(\omega\mu_0\sigma)^2}\cos\phi
\biggl[} -\omega\mu_0\sigma k_m^2\, I_4(k_m\rho)\biggr]~,\label{eq:E2z-hd-0}
\end{align}
\begin{align}
E_{2\rho}^{\rm (hd)}&=-\frac{\omega\mu_0}{4\pi}\cos\phi \biggl\{ \frac{1}{2\rho}[\omega\mu_0\sigma\rho\, I_3(k_e\rho)-2\,I_4(k_e\rho)]+\frac{2}{(\omega\mu_0\sigma)^2}\frac{d}{d\rho}\biggl[\frac{\omega\mu_0\sigma}{\rho}\,I_1\nonumber\\
& \mbox{} \qquad -\frac{4k^4\rho}{\omega\mu_0\sigma}\,I_3(k_m\rho)+2k^2 \,I_4(k_m\rho)\biggr]\biggr\}~,\label{eq:E2r-hd-0}\\
E_{2\phi}^{\rm (hd)}&=\frac{\omega\mu_0}{4\pi}\,\sin\phi \biggl\{ \frac{1}{2}\frac{d}{d\rho}[\omega\mu_0\sigma\rho\, I_3(k_e\rho)-2\,I_4(k_e\rho)]+\frac{2}{(\omega\mu_0\sigma)^2}\frac{1}{\rho}\biggl[\frac{\omega\mu_0\sigma}{\rho}\,I_1\nonumber\\
&\mbox{}\qquad -\frac{4k^4\rho}{\omega\mu_0\sigma}\,I_3(k_m\rho)+2k^2\,I_4(k_m\rho)\biggr]\biggr\}~,\label{eq:E2ph-hd-0}
\end{align}
\begin{align}
B_{2z}^{\rm (hd)}&=\frac{i\mu_0}{8\pi}\,\sin\phi \biggl[\frac{\omega\mu_0\sigma}{\rho}\,I_1-\frac{2}{\rho^2}\,I_2+\omega\mu_0\sigma k_e^2\rho \,I_3(k_e\rho)-2k_e^2\, I_4(k_e\rho)\biggr]~,
\label{eq:B2z-hd-0}
\end{align}
\begin{align}
B_{2\rho}^{\rm (hd)}&=\frac{\mu_0}{4\pi}\,\sin\phi \biggl\{\frac{2k^2}{(\omega\mu_0\sigma)^2}\frac{1}{\rho}\, [2k^2\rho\,I_3(k_m\rho)-\omega\mu_0\sigma \,I_4(k_m\rho)]+\frac{1}{2}\frac{d}{d\rho}\biggl[\frac{2}{\rho}\,I_1 \nonumber\\
&\mbox{}\qquad -\frac{(\omega\mu_0\sigma)^2}{2}\rho\,I_3(k_e\rho)+\omega\mu_0\sigma\,I_4(k_e\rho)\biggr]\biggr\}~,\label{eq:B2r-hd-0}\\
B_{2\phi}^{\rm (hd)}&=\frac{\mu_0}{4\pi}\,\cos\phi \biggl\{\frac{2k^2}{(\omega\mu_0\sigma)^2}\,\frac{d}{d\rho}[2k^2\rho\,I_3(k_m\rho)-\omega\mu_0\sigma\,I_4(k_m\rho)]+\frac{1}{2\rho} \biggl[\frac{2}{\rho}\,I_1\nonumber\\
&\mbox{}\qquad -\frac{(\omega\mu_0\sigma)^2}{2}\rho\,I_3(k_e\rho)+\omega\mu_0\sigma\,I_4(k_e\rho)\biggr]\biggr\}~.\label{eq:B2ph-hd-0}
\end{align}

In the above, we provide expressions for the fields on one side of the layer for each source; the fields on the other side can be found through the imposed boundary conditions (Section~\ref{sec:BVP} and Appendix~\ref{app:normal}). The integrals $I_3(\varpi),\,I_4(\varpi)$ and $\breve I_4(\varpi)$ are evaluated in Section~\ref{subsec:integrals}. (The reader who wishes to skip technical derivations may proceed directly to Section~\ref{subsec:Eval-asym}.)

The field of the vertical dipole bears TM polarization, as indicated by~\eqref{eq:E1r-vd-0}--\eqref{eq:E1z-vd-0} with~\eqref{eq:I-int}. On the other hand, the field of the horizontal dipole carries the signatures of TM and TE polarization.

%%%%%%%%%%%%%%%%%%%%%%%%%%%%%%%%%%%%%%%%%%%%%%%%%%%%%%%%%%%%%%%%%%%%%%%%%%%
\subsection{Key integrals and generalized Schwinger-Feynman representation}
\label{subsec:integrals}
%%%%%%%%%%%%%%%%%%%%%%%%%%%%%%%%%%%%%%%%%%%%%%%%%%%%%%%%%%%%%%%%%%%%%%%%%%%

Next, we provide derivations for requisite integrals \eqref{eq:I-int-2}--\eqref{eq:I-int-4} in terms of known transcendental functions.
This section forms a crucial part of our formal analysis.
We follow the formalism of Ref.~\onlinecite{Margetis01}, elements of which we repeat here for the sake of completeness. This procedure should clarify the role of certain singularities related to material properties of the film.\looseness=-1

For later algebraic convenience, set $k\rho=:iq$ and $\varpi=:i\ss$ recalling that $\varpi=k_m\rho$ or $k_e\rho$; and treat $q$ and $\ss$ as positive for the purpose of computing the integrals. The extension of the requisite integrals to the actual, physically relevant values of these parameters will be carried out once the integrals are evaluated.
Accordingly, set $\sqrt{(k\rho)^2-\lambda'^2}=i\sqrt{\lambda'^2+q^2}$ and treat  $\sqrt{\lambda'^2+q^2}$ as positive if $\lambda'$ is real. Firstly, we directly find that~\cite{Watson,Margetis01}
\begin{equation}\label{eq:I2}
I_2=i(q+e^{-q})=k\rho+ie^{ik\rho}~,\quad \breve I_2=-iq^{-1}(1-e^{-q})=\frac{1-e^{ik\rho}}{k\rho}~.
\end{equation}

%-------------------------------------%
\subsubsection{Integral $I_3(\varpi)$}
%-------------------------------------%

\begin{figure}
\includegraphics*[scale=0.45, trim=0in 0.2in 0in 0in]{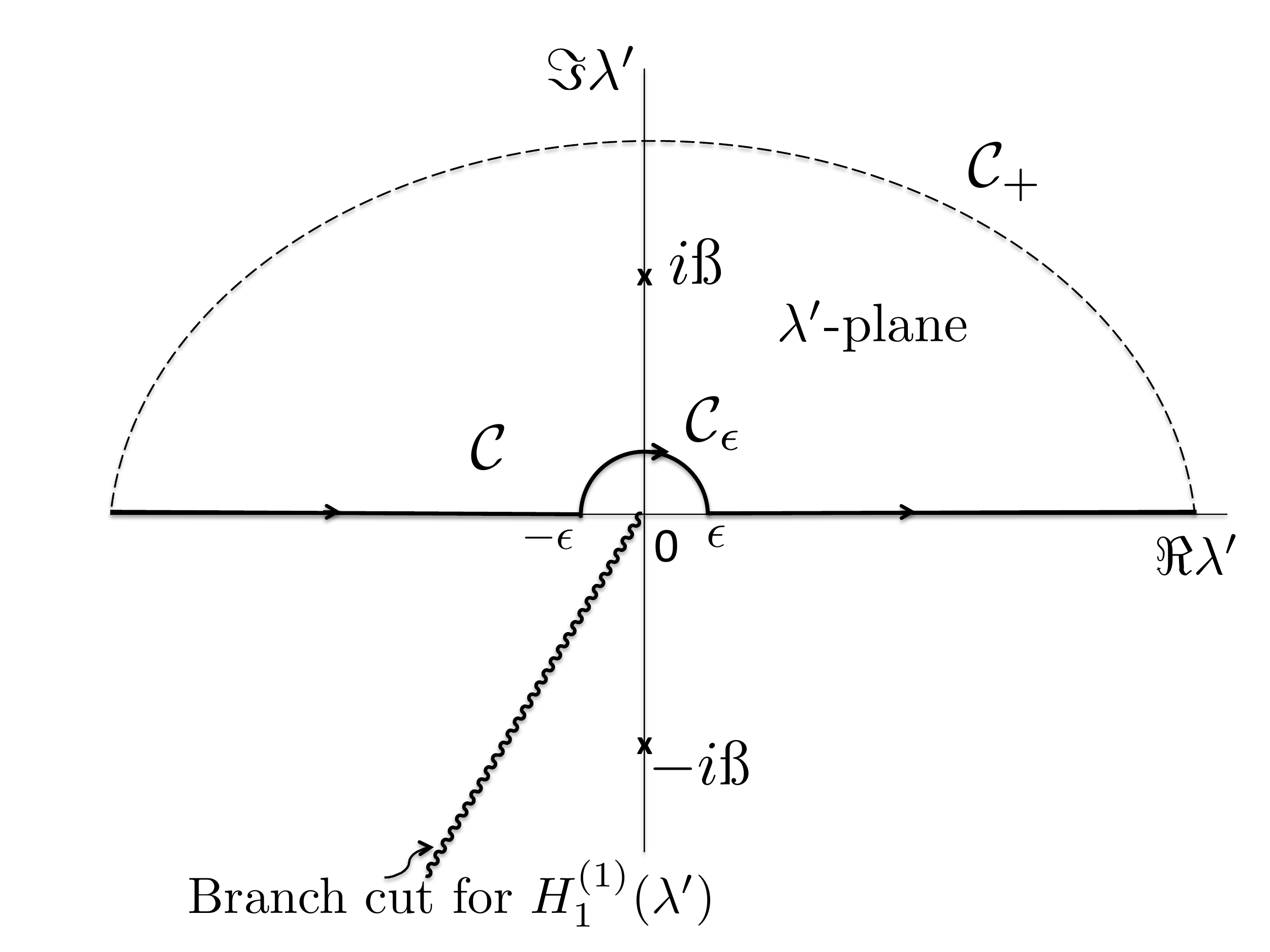}
    \caption{Branch-cut configuration and contours in the $\lambda'$-plane for the integral $I_3(\varpi)$; see~\eqref{eq:I3-cmpx}. The integrand has simple poles at $\pm i\ss$ ($\ss=-i\varpi$ is treated as positive). Contour $\mathcal C=(-\infty, -\epsilon)\cup \mathcal C_\epsilon\cup (\epsilon, \infty)$ (solid line with arrows) is identified with the real axis except for a small indentation, $\mathcal C_\epsilon$, of radius $\epsilon$ around $0$ in the upper half-plane; $\epsilon\downarrow 0$.
Contour  $\mathcal C_+$ (dashed line) in the upper half-plane closes the path; the contour $\mathcal C\cup \mathcal C_+$  picks up the residue at $\lambda=i \ss$. }\label{fig:I3-cont}
\end{figure}

By~\eqref{eq:I-int-3}, we resort to writing $J_1(\lambda)=(1/2)[H_1^{(1)}(\lambda)+H_1^{(2)}(\lambda)]$ where $H_\nu^{(1,2)}$ is the (first or second) Hankel function of order $\nu$.~\cite{Bateman-II}  By the identity  $H_1^{(2)}(\lambda e^{-i\pi})=H_1^{(1)}(\lambda)$, valid for $\lambda\neq 0$, \eqref{eq:I-int-3} becomes \looseness=-1
\begin{align}\label{eq:I3-cmpx}
I_3(\varpi)&=\frac{1}{2}\int_{\mathcal C}\dlp\ \frac{H_1^{(1)}(\lambda')}{\lambda'^2+\ss^2}-\frac{1}{2}\lim_{\epsilon\downarrow 0}\int_{\mathcal C_\epsilon}\dlp\ \frac{H_1^{(1)}(\lambda')}{\lambda'^2+\ss^2}=\frac{\pi}{2\ss}\,H_1^{(1)}(i\ss)+\ss^{-2}\nonumber\\
&=\frac{i\pi}{2\varpi} H_1^{(1)}(\varpi)-\frac{1}{\varpi^2}~,
\end{align}
by contour integration, where the contours $\mathcal C$ and $\mathcal C_\epsilon$ are depicted in Fig.~\ref{fig:I3-cont}; cf. Ref.~\onlinecite{Fikioris-book}. The integrals in the first line of~\eqref{eq:I3-cmpx} provide the Cauchy principal value of $\frac{1}{2}\int_{-\infty}^{\infty}\dlp\,H_1^{(1)}(\lambda') \,(\lambda^{'2}+\ss^2)^{-1}$ which accounts for the singularity of $H_1^{(1)}(\lambda')$ at $\lambda'=0$.

%---------------------------------------------------------------%
\subsubsection{Integrals $I_4(\varpi)$ and $\breve I_4(\varpi)$}
%---------------------------------------------------------------%

We seek exact formulas for~\eqref{eq:I-int-4} with $\varpi=k_m\rho$ and $\varpi=k_e\rho$. Such formulas should be particularly useful for $|\omega\mu_0\sigma|\ll |k|$, by which $|k_e|\sim |k|\ll |k_m|$; cf.~\eqref{eq:kp-iso} and~\eqref{eq:kst-iso}. Thus, we view as {\em small} the parameters: $|k/ k_m|$, with $\varpi=k_m\rho$; and $|(k_e-k)/k_e|$, with $\varpi = k_e\rho$. Our derivation is tailored to the case where $k_m$ lies in the first Riemann sheet (see Remark~1); the extension to other values of the physical parameters is discussed in the end of this section.

\begin{figure}
\includegraphics*[scale=0.5, trim=0in 2in 0in 0in]{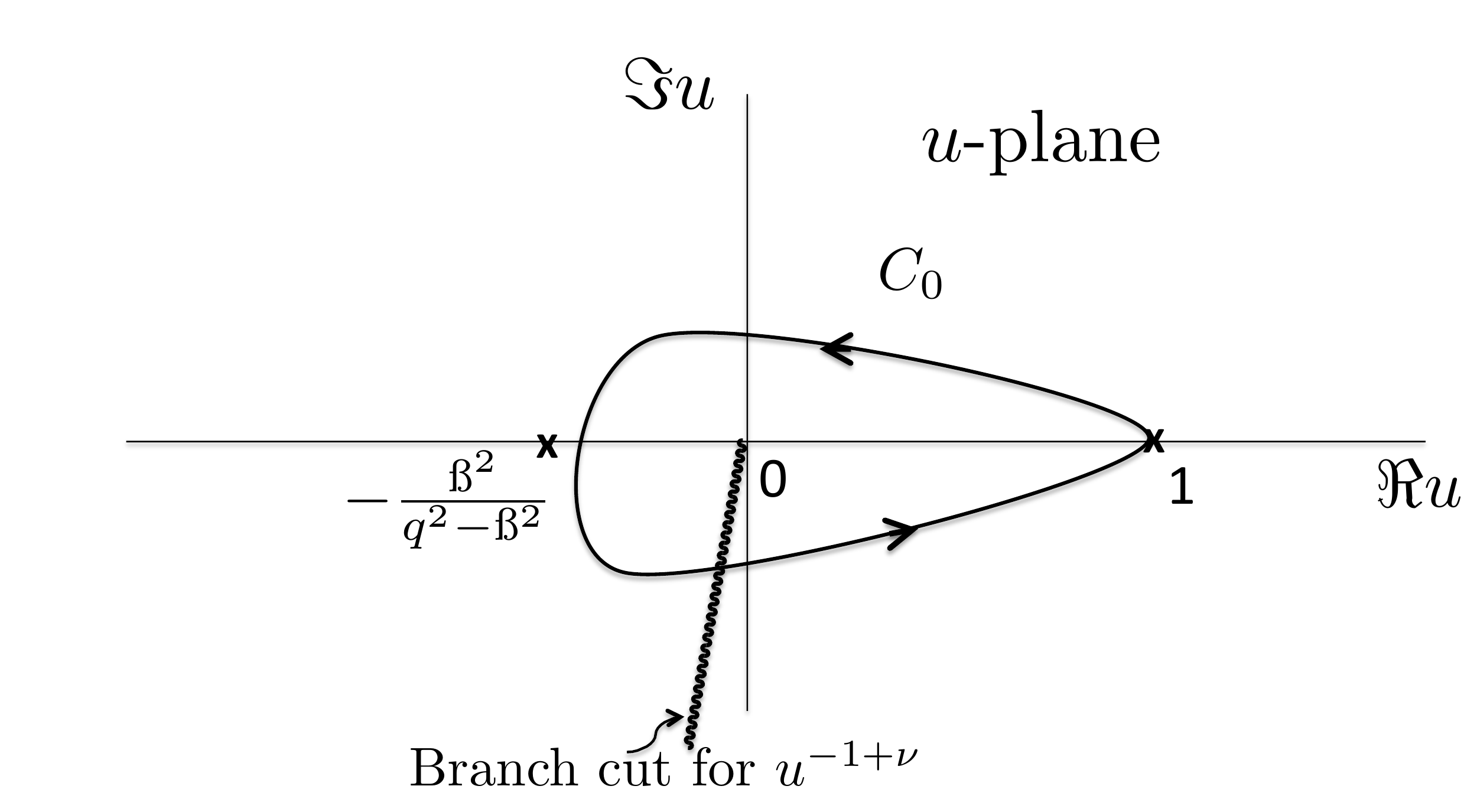}
    \caption{Contour $C_0$ of integration for generalized integral on right-hand side of~\eqref{eq:u-int} pertaining to $I_4(\varpi)$; $0<\ss^2<q^2$. The integrand has branch points at $u=0,\,-\ss^2/(q^2-\ss^2)$; only the cut emanating from $0$ is shown here, indicating the passage of the integrand from the first Riemann sheet for $u^{-1+\nu}$, where $u^{-1+\nu}>0$ if $u>0$, to an adjacent Riemann sheet.}\label{fig:I4-cont}
\end{figure}

The first step is to convert $I_4(\varpi)$ in~\eqref{eq:I-int-4} to an integral of an elementary function. In the spirit of the Schwinger-Feynman approach, consider the generalized representation~\cite{Margetis01}
\begin{displaymath}
\mathfrak{A}^{-\nu}\mathfrak{C}^{-1}=\frac{1}{\Gamma(\nu)}\int_0^\infty\int_0^\infty \d s_1\,\d s_2\ s_1^{\nu-1} e^{-(\mathfrak A s_1+\mathfrak C s_2)}=
\nu \int_0^1 du\,u^{-1+\nu}\,[(\mathfrak A-\mathfrak C)u+\mathfrak C]^{-1-\nu}~,\ \Re \nu>0~,
\end{displaymath}
via the change of variable $(s_1, s_2)\mapsto (u,v)$ with $(s_1, s_2)=(uv, (1-u)v)$ and integration in $v$; $\Gamma(z)$ is the Gamma function. We assume that $0<\mathfrak C< \mathfrak A$, a condition to be relaxed later; here, $\mathfrak A=\lambda^{\prime 2}+q^2$, $\mathfrak C=\lambda^{\prime 2}+\ss^2$.
By using the above representation for $\mathfrak{A}^{-\nu}\mathfrak{C}^{-1}$, we obtain
\begin{align}\label{eq:u-int}
\int_0^{\infty}\dlp\ \frac{(\lambda'^2+q^2)^{-\nu}}{\lambda'^2+\ss^2}\,J_1(\lambda')&= \frac{\nu}{1-e^{i2\pi \nu}}\oint_{C_0}\d u\ u^{-1+\nu}
[q^2 u+\ss^2 (1-u)]^{-(1+\nu)}\biggl\{ 1-\frac{2^{-\nu}}{\Gamma(1+\nu)} \nonumber\\
&\mbox{}  \times
[q^2 u+\ss^2 (1-u)]^{\frac{1+\nu}{2}}\, K_{1+\nu}\big((q^2 u+\ss^2(1-u))^{1/2}\big)\biggr\}~.
\end{align}
The contour $C_0$ serves the analytic continuation of the integral on the left-hand side to values of $\nu$ with $\Re \nu <0$ (see Fig.~\ref{fig:I4-cont}). To derive~\eqref{eq:u-int}, we interchanged  the order of integration (in $\lambda'$ and $u$), and made use of a result from Ref.~\onlinecite{Watson} to carry out the integration in $\lambda'$; $K_{\nu}$ is the modified Bessel function of order $\nu$, with $K_{1/2}(\zeta)=\sqrt{\pi/(2\zeta)}\, e^{-\zeta}$.~\cite{Bateman-II} By the change of variable $u\mapsto \zeta$ with $\zeta^2=q^2 u+ \ss^2 (1-u)$, integration by parts via the identity $\frac{d}{d\zeta}[\zeta^{-2\nu}(\zeta^2-\ss^2)^\nu]=2\ss^2\nu \zeta^{-1-2\nu}(\zeta^2-\ss^2)^{\nu-1}$, and the subsequent substitution $\nu=-1/2$, we find
\begin{align}\label{eq:I4-elem}
I_4(\varpi)&=i\int_0^\infty \dlp\ \frac{\sqrt{\lambda^{\prime 2}+q^2}}{\lambda^{\prime 2}+\ss^2}\ J_1(\lambda')\nonumber\\
&=\frac{iq}{\ss^2}\,\big(1-e^{-q}\big)-i\frac{\sqrt{q^2-\ss^2}}{\ss^2}\int_{\ss}^q \d\zeta\ \frac{\zeta}{\sqrt{\zeta^2-\ss^2}}\,e^{-\zeta}\nonumber\\
&=-\frac{k\rho}{\varpi^2}\,\big(1-e^{ik\rho}\big)+i\frac{\sqrt{\varpi^2-(k\rho)^2}}{\varpi}\int_{k\rho/\varpi}^1\d\eta\ \frac{\eta}{\sqrt{1-\eta^2}}\,e^{i\varpi\eta}~,
\end{align}
which provides $I_4(\varpi)$ in terms of an integral of an elementary function.

We proceed to simplify~\eqref{eq:I4-elem}. For this purpose, define
\begin{equation}\label{eq:W-def}
W(\ell, \varsigma; \varpi):=\int_{\ell}^\varsigma\d\eta\ \frac{\eta}{\sqrt{1-\eta^2}}\, e^{i\varpi\eta}~.
\end{equation}
We seek appropriate expansions of this $W$ for $\varsigma=1$ with: (i) $\ell=k/k_m$ and $\varpi=k_m\rho$ where $|\ell|< 1$; and (ii) $\ell=k/k_e$ and $\varpi=k_e\rho$ where $|1-\ell|< 1$. $|\varpi|$ should be unrestricted in each case.
\smallskip

\noindent {\bf (i) Case $\ell=k/ k_m$, $\varpi=k_m\rho$.} By writing
\begin{equation}\label{eq:W-km}
W(k/k_m, 1; k_m\rho)=W(0, 1; k_m\rho)-W(0, k/k_m; k_m\rho)
\end{equation}
and resorting to Ref.~\onlinecite{Bateman-II}, we obtain the formula
\begin{equation}\label{eq:W0}
W(0, 1; k_m\rho)=1-\frac{\pi}{2}\big[\mathbb{H}_1(k_m\rho)-Y_1(k_m\rho)\big]+\frac{i\pi}{2}\,H_1^{(1)}(k_m\rho)~,
\end{equation}
where $\mathbb{H}_\nu$ ($Y_\nu$) is the Struve (Neumann) function of order $\nu$. In~\eqref{eq:W0}, the last term corresponds to the contribution to $I_4(k_m\rho)$ from the residue at $\lambda'=k_m\rho$. For the remaining terms, note the large-$|\zeta|$ expansion~\cite{Bateman-II}
\begin{equation}\label{eq:HY-asympt}
\mathbb{H}_\nu(\zeta)-Y_\nu(\zeta)=\pi^{-1}\sum_{n=0}^{M-1}\frac{\Gamma(\frac{1}{2}+n)}{\Gamma(\nu+\frac{1}{2}-n)}\,\biggl(\frac{2}{\zeta}\biggr)^{2n-\nu+1}+\mathcal O(|\zeta|^{-2M+\nu-1})~,\quad |\arg \zeta|<\pi~.
\end{equation}

Now consider the term $W(0, k/k_m; k_m\rho)$ in~\eqref{eq:W-km} by inspection of~\eqref{eq:W-def}.
By expanding $\eta (1-\eta^2)^{-1/2}$ at $\eta=0$ with $|\eta|<1$ and integrating term by term, we derive the geometrically convergent series\looseness=-1
\begin{equation}\label{eq:W1}
W(0, k/k_m; k_m\rho)=\sum_{l=0}^\infty \biggl(\frac{1}{2}\biggr)_l\ \frac{(-1)^{l+1}}{l!}\,\biggl(\frac{k^2}{k_m^2}\biggr)^{l+1}\,
\frac{d^{2l+1}}{d\zeta^{2l+1}}\biggl(\frac{1-e^{i\zeta}}{\zeta}\biggr)\biggl|_{\zeta=k\rho}~,\quad \biggl|\frac{k^2}{k_m^2}\biggr|<1~,
\end{equation}
where $(b)_l$ is Pochhammer's symbol; $(b)_l=\Gamma(l+b)/\Gamma(b)$.
Formula~\eqref{eq:W1} concludes the evaluation of $W(k/k_m,1; k_m\rho)$ by~\eqref{eq:W-km}.
\smallskip

\noindent {\bf (ii) Case $\ell=k/k_e$, $\varpi=k_e\rho$.}
By changing integration variable from $\eta$ to $v=1-\eta$ and expanding $\eta(1-\eta^2)^{-1/2}=(1-v)v^{-1/2}(2-v)^{-1/2}$ at $v=0$, we recast~\eqref{eq:W-def} to
\begin{equation*}
W(k/k_e, 1; k_e\rho)=\frac{e^{ik\rho}}{\sqrt{2k_e\rho}}\sum_{l=0}^\infty \biggl(\frac{1}{2}\biggr)_l\,\frac{1+2l}{1-2l}\frac{(2k_e\rho)^{-l}}{l!}\,f_l((k_e-k)\rho)~,
\end{equation*}
where
\begin{equation*}
f_l(z)=\int_0^z \d v\ v^{-1/2+l}\,e^{i(z-v)}~.
\end{equation*}
This $f_l(z)$ has generating function
\begin{equation*}
\mathfrak G(\zeta; z):=\sum_{l=0}^\infty \frac{f_l(z)}{l!} (-i\zeta)^l=\sqrt{2\pi}\,e^{iz}\,(1+\zeta)^{-1/2}\mathcal F_0((1+\zeta)z)~,
\end{equation*}
where $\mathcal F_0$ is given in terms of the Fresnel integrals $C(z)$ and $S(z)$,~\cite{Bateman-I} viz.,
\begin{subequations}\label{eq:F0-main}
\begin{align}\label{eq:F0}
\mathcal F_0(z)&=\int_0^z \d\tau\ \frac{e^{-i\tau}}{\sqrt{2\pi\tau}}=C(z)-iS(z)\nonumber\\
&= \frac{1}{\sqrt{2}}e^{-i\pi/4}-e^{-iz}\mathcal F(z)~;\quad \mathcal F(z):=e^{iz}\int_z^\infty \d\tau\ \frac{e^{-i\tau}}{\sqrt{2\pi\tau}}~.
\end{align}
In particular, note the asymptotic expansion
\begin{equation}\label{eq:F-asymp}
\mathcal F(z)=-\frac{i}{\sqrt{2\pi z}}\biggl[ 1+\frac{i}{2z}+\mathcal O(z^{-2})\biggr]\qquad\mbox{as}\ |z|\to \infty~.
\end{equation}
\end{subequations}
Thus, by use of the generating function $\mathfrak G$ we write
\begin{displaymath}
f_l(z)= \sqrt{2\pi}\,i^l z^{l+1/2}\,e^{iz}\,\frac{d^l}{dz^l}\big[z^{-1/2}\mathcal F_0(z)\big]~,
\end{displaymath}
and derive the desired expansion for $W$:
\begin{equation}\label{eq:W-ke}
W(k/k_e, 1; k_e\rho)=\sqrt{\pi}\,e^{ik\rho}\sum_{l=0}^\infty \biggl(\frac{1}{2}\biggr)_l\,\frac{1+2l}{1-2l}\,\frac{(i/2)^l}{l!}\biggl(\frac{k_e-k}{k_e}\biggr)^{l+1/2}e^{i\wp}\,\frac{d^l}{d z^l}\big[z^{-1/2}\mathcal F_0(z)\big]\biggl|_{z=\wp}~.
\end{equation}
This series exhibits geometric rate of convergence for $|(k_e-k)/k_e|< 1$ and all $\wp$ where
\begin{equation}\label{eq:Som-dist}
\wp:=(k_e-k)\rho \sim e^{-i2\delta}\,\frac{(\omega\mu_0|\sigma|)^2\rho}{8k}~,\qquad\ |\omega\mu_0\sigma|\ll |k|~,
\end{equation}
with $\arg\sigma=\pi/2-\delta$ ($0<\delta<\pi$). This $\wp$ is  analogous to Sommerfeld's ``numerical distance'';~\cite{Sommerfeld-pde} cf.~\eqref{eq:ke-nonr}.

Thus, by~\eqref{eq:I4-elem}, integral $I_4(\varpi)$ reads
\begin{equation}\label{eq:I4-fin}
I_4(\varpi)=-\frac{k\rho}{\varpi^2}\,\big(1-e^{ik\rho}\big)+i\frac{\sqrt{\varpi^2-(k\rho)^2}}{\varpi}\,W(k\rho/\varpi, 1; \varpi)~,
\end{equation}
where $W$ is provided by~\eqref{eq:W-km} with~\eqref{eq:W0} and~\eqref{eq:W1} for $\varpi=k_m\rho$; and by~\eqref{eq:W-ke} for $\varpi=k_e\rho$. By~\eqref{eq:I-int-4}, $\breve I_4(\varpi)$ equals
\begin{equation}\label{eq:brevI4-fin}
\breve I_4(\varpi)=\frac{-1}{\varpi^2-(k\rho)^2}\biggl[\biggl(\frac{1}{k\rho}-\frac{k\rho}{\varpi^2}\biggr)(1-e^{ik\rho}) +i\,\frac{\sqrt{\varpi^2-(k\rho)^2}}{\varpi}\,W(k\rho/\varpi, 1; \varpi)\biggr]~.
\end{equation}
In the above, $\Re \sqrt{\varpi^2-(k\rho)^2}=\Im \sqrt{(k\rho)^2-\varpi^2}>0$, in accord with the definition of the top $\lambda$-Riemann sheet, and $k_m$ is assumed to lie in this Riemann sheet; see~\eqref{eq:km-nonr}.

We now discuss formulas~\eqref{eq:I4-fin} and~\eqref{eq:brevI4-fin} for a wider range of parameters.
\smallskip

{\em Remark 2 (On the values of $I_4$ and $\breve I_4$). Equations~\eqref{eq:I4-fin} and~\eqref{eq:brevI4-fin} can be extended to all physically relevant values of $k_m$ (and $k_e$) with $\Re \sigma>0$ (Remark~1), by computing $(1/\varpi)\sqrt{\varpi^2-(k\rho)^2}\,W(k\rho/\varpi, 1; \varpi)$ according to the following rules. These rules are consistent with the starting formulation of the Sommerfeld-type integrals (Section~\ref{sec:Fourier}).

\begin{itemize}

\item 
For $\varpi=k_m\rho$: In regard to the residue contribution in $W(0,1;k_m\rho)$, set
\begin{subequations}\label{eq:varpi-fac-gen}
\begin{equation}\label{eq:varpi-fac}
\frac{\sqrt{(k_m\rho)^2-(k\rho)^2}}{k_m\rho}H_1^{(1)}(k_m\rho)=\frac{\sqrt{k_m^2-k^2}}{k_m}H_1^{(1)}(k_m\rho)=\pm \frac{i\,2k^2}{\omega\mu_0\sigma\,k_m}\, H_1^{(1)}(k_m\rho)~,
\end{equation}
where $\Im k_m>0$ and the upper (lower) sign holds if the pole at $k_m$ lies in the first (second) Riemann sheet; see~\eqref{eq:km-nonr}.
For the remaining terms in $W,$ set $\sqrt{(k_m\rho)^2-(k\rho)^2}/(k_m\rho)=\sqrt{1-k^2/k_m^2}$ and choose the branch so that $\lim_{k_m\to\infty}\sqrt{1-k^2/k_m^2}=1$ for fixed $k$; thus,
\begin{equation}\label{eq:alpha-fac}
\alpha_m:=\sqrt{1-k^2/k_m^2}=\sum_{n=0}^\infty \biggl(-\frac{1}{2}\biggr)_n\,\frac{(k^2/k_m^2)^n}{n!}~,\qquad \mbox{if}\quad |k^2/k_m^2|< 1~.
\end{equation}

\item
For $\varpi=k_e\rho$: Set 
\begin{equation}\label{eq:varpi-fac-ke}
\sqrt{(k_e\rho)^2-(k\rho)^2}=-i\sqrt{(k\rho)^2-(k_e\rho)^2} =
\pm i\,\omega\mu_0\sigma\rho/2~;
\end{equation}
\end{subequations}
the upper (lower) sign holds if the pole at $k_e$ lies in the first (second) $\lambda$-Riemann sheet.

\end{itemize}

The sign in~\eqref{eq:varpi-fac} controls the appearance or absence of the TM surface plasmon (Definition~1). If the pole at $\lambda=k_m$ does not lie on the top Riemann sheet, e.g., when $k>0$ and $\Im\sigma<0$, contribution \eqref{eq:varpi-fac} causes cancellation of the residue from $I_3(k_m\rho)$ in each relevant field component; then, the TM surface plasmon is absent. An analogous effect comes from~\eqref{eq:varpi-fac-ke} in regard to TE polarization. Equation~\eqref{eq:alpha-fac} warrants that the primary field (in the absence of the layer) of the dipole is recovered as $\sigma\to 0$ regardless of the location of the pole at $k_m$. See Section~\ref{subsec:Eval-asym} for consequences of this solution.}

%%%%%%%%%%%%%%%%%%%%%%%%%%%%%%%%%%%%%%%%%%%%%%%%%%%%%%%%%%%%%%%%%%%%%%%%%%
\subsection{Field components: Evaluation}
\label{subsec:Eval-asym}
%%%%%%%%%%%%%%%%%%%%%%%%%%%%%%%%%%%%%%%%%%%%%%%%%%%%%%%%%%%%%%%%%%%%%%%%%

The field components are computed by substitution of the formulas for integrals $I_n$ ($n=1,\,2,\,3,\,4$) and $\breve I_n$ ($n=1,\,2$) from Section~\ref{subsec:integrals} into~\eqref{eq:E1r-vd-0}--\eqref{eq:E1z-vd-0} and \eqref{eq:E2z-hd-0}--\eqref{eq:B2ph-hd-0}, bearing in mind Remarks~1 and~2. To render the resulting expressions physically transparent, we subsequently apply the condition $|\omega\mu_0\sigma|\ll |k|$ which warrants that $|k/k_m|\ll 1$ and $|(k_e-k)/k_e|\ll 1$; the requisite series expansions can be approximated by their first few terms for all distances $\rho$ from the source. In Appendix~\ref{app:asymp}, we provide alternate evaluations of the field components for sufficiently small and large distance, $\rho$, from the source.
\smallskip

{\em Remark 3 (On the pole at $k_m$). We assume that the parameter $\sigma$ is chosen so that the pole at $\lambda=k_m$ lies in the top $\lambda$-Riemann sheet (see Remark~1), unless we state otherwise. }

%----------------------------------------------------
\subsubsection{Vertical dipole}
%----------------------------------------------------
The $\rho$-component of the electric field deserves special attention, since it vanishes at $z=0$ in the absence of the layer.~\cite{KOW} By using this component, we illustrate the role of the location of the pole at $k_m$ in the Riemann surface associated with the dual ($\lambda$) variable. By~\eqref{eq:E1r-vd-0}, we write\looseness=-1
\begin{subequations}\label{eq:E1r-vd-ex}
\begin{align}\label{eq:E1r-vd-ex-a}
E_{1\rho}^{({\rm vd})}&=\frac{i}{2\pi\sigma}\Biggl\{\Biggl(\frac{i}{\rho^2}+\frac{k}{\rho}\Biggr)e^{ik\rho}-\frac{i\pi}{2} k_m\Biggl(\frac{2k^2}{\omega\mu_0\sigma}-i\sqrt{k_m^2-k^2}\Biggr)H_1^{(1)}(k_m\rho)\nonumber\\
&\mbox{} +ik_m^2\alpha_m\Biggl[1-\frac{\pi}{2}\bigl(\mathbb{H}_1(k_m\rho)-Y_1(k_m\rho)\bigr)\nonumber\\
&\mbox{} +\sum_{l=0}^\infty \biggl(\frac{1}{2}\biggr)_l\,\frac{(-1)^{l+1}}{l!}\biggl(\frac{k^2}{k_m^2}\biggr)^{l+1}\,\frac{d^{2l+1}}{d\zeta^{2l+1}}\biggr(\frac{1-e^{i\zeta}}{\zeta}\biggr)\biggr|_{\zeta=k\rho}\Biggr]\Biggr\}~.
\end{align}
Notice the term proportional to $[2k^2/(\omega\mu_0\sigma)-i\sqrt{k_m^2-k^2}]H_1^{(1)}(k_m\rho)$, which expresses the TM surface plasmon; see Remark~2. By~\eqref{eq:varpi-fac}, in this term $\sqrt{k_m^2-k^2}=\pm i 2k^2/(\omega\mu_0\sigma)$, depending on the location of the associated pole. On the other hand, in regard to the second line of~\eqref{eq:E1r-vd-ex-a}, $k_m^2\alpha_m \sim k_m^2[1-k^2/(2k_m^2)]\sim k_m^2$ if $|k_m|^2\gg |k|^2$. Suppose that the pole at $k_m$ is present in the top $\lambda$-Riemann sheet, if, say, $\Im\sigma >0$ and $k>0$ as outlined in Section~\ref{subsec:poles}. By retaining {\em two terms} in the series expansion of~\eqref{eq:E1r-vd-ex-a} and enforcing $|\omega\mu_0\sigma|\ll |k|$, we obtain
\begin{align}\label{eq:E1r-vd-ex-b}
E_{1\rho}^{({\rm vd})}&\sim \frac{\omega\mu_0}{2\pi}\Biggl\{
\frac{\omega\mu_0\sigma}{4}\Biggl(\frac{3}{k^4\rho^4}-\frac{3i}{k^3\rho^3}-\frac{1}{k^2\rho^2}\Biggr)e^{ik\rho}
+i\pi \frac{4k^4}{(\omega\mu_0\sigma)^3}H_1^{(1)}(k_m\rho)+\frac{4k^4}{(\omega\mu_0\sigma)^3}\nonumber\\
&\mbox{}\qquad \times \Biggl(1+\frac{(\omega\mu_0\sigma)^2}{8k^2}\Biggr)\Biggl[1-\frac{\pi}{2}\bigl(\mathbb{H}_1(k_m\rho)-Y_1(k_m\rho)\bigr)+\frac{1}{(k_m\rho)^2}-\frac{3}{(k_m\rho)^4}\Biggr]\Biggr\}~.
\end{align}
The TM surface plasmon contribution has {\em survived} here as the term proportional to $H_1^{(1)}(k_m\rho)$; see also Appendix~\ref{app:asymp}.
In contrast, if $k_m$ lies in the second Riemann sheet, if, for example, $\Im \sigma<0$ and $k>0$, the above residue contribution disappears (see Remark~2), viz.,
\begin{align}\label{eq:E1r-vd-ex-c}
E_{1\rho}^{({\rm vd})}&\sim \frac{\omega\mu_0}{2\pi}\Biggl\{
\frac{\omega\mu_0\sigma}{4}\Biggl(\frac{3}{k^4\rho^4}-\frac{3i}{k^3\rho^3}-\frac{1}{k^2\rho^2}\Biggr)e^{ik\rho}
+\frac{4k^4}{(\omega\mu_0\sigma)^3}\Biggl(1+\frac{(\omega\mu_0\sigma)^2}{8k^2}\Biggr)\nonumber\\
&\mbox{}\qquad \times \Biggl[1-\frac{\pi}{2}\bigl(\mathbb{H}_1(k_m\rho)-Y_1(k_m\rho)\bigr)+\frac{1}{(k_m\rho)^2}-\frac{3}{(k_m\rho)^4}\Biggr]\Biggr\}~,\quad |\omega\mu_0\sigma|\ll |k|~.
\end{align}
\end{subequations}

A similar calculation can be carried out for the other field components of the vertical dipole. We henceforth emphasize mainly consequences of Remark~3.

Next, we turn our attention to the magnetic field. By~\eqref{eq:B1ph-vd-0} and Remark~3, we compute
\begin{align}\label{eq:B1ph-vd-ex}
B_{1\phi}^{({\rm vd})}&=-\frac{i\mu_0}{4\pi}\Biggl\{2\Biggl(\frac{i}{\rho^2}+\frac{k}{\rho}\Biggr)e^{ik\rho}
-i\pi\frac{k^2k_m}{\omega\mu_0\sigma}\Biggl(1+\frac{2ik^2}{\omega\mu_0\sigma}\frac{1}{\sqrt{k_m^2-k^2}}\Biggr) H^{(1)}_1(k_m\rho)\nonumber\\
&\mbox{}\quad -\frac{4i k^4}{(\omega\mu_0\sigma)^2}\frac{1}{\alpha_m}
\Biggl[1-\frac{\pi}{2}\bigl(\mathbb{H}_1(k_m\rho) -Y_1(k_m\rho)\bigr)\nonumber\\
&\mbox{} \quad +\sum_{l=0}^\infty \biggl(\frac{1}{2}\biggr)_l\ \frac{(-1)^{l+1}}{l!}\,\biggl(\frac{k^2}{k_m^2}\biggr)^{l+1}\,
\frac{d^{2l+1}}{d\zeta^{2l+1}}\biggl(\frac{1-e^{i\zeta}}{\zeta}\biggr)\biggl|_{\zeta=k\rho}\Biggr]\Biggr\}\nonumber\\
&\sim -\frac{i\mu_0}{4\pi}\Biggl\{\Biggl(\frac{i}{\rho^2}+\frac{k}{\rho}\Biggr)e^{ik\rho}+4\pi\frac{k^4}{(\omega\mu_0\sigma)^2}H_1^{(1)}(k_m\rho)-i\frac{4k^4}{(\omega\mu_o\sigma)^2}\Biggl[1-\frac{\pi}{2}\bigl(\mathbb{H}_1(k_m\rho)\nonumber\\
&\mbox{} \qquad -Y_1(k_m\rho)\bigr)+\frac{1}{k_m^2\rho^2}\Biggr]\Biggr\}~,\qquad |\omega\mu_0\sigma|\ll |k|~.
\end{align}
The terms with the $e^{ik\rho}$ factor in the fourth line (but not on the first line) of~\eqref{eq:B1ph-vd-ex} correspond to the primary field of a $z$-directed dipole in {\em free space};~\cite{KOW} the remaining terms in the second equation amount to the scattered field due to the presence of the thin layer for $|\omega\mu_0\sigma|\ll |k|$. Notice the surviving residue contribution expressed by the term containing $H_1^{(1)}(k_m\rho)$; this is identified with the TM surface plasmon via Definition~1. In Appendix~B, we provide alternate calculations for $B_{1\phi}^{({\rm vd})}$ in the disparate cases with $|k\rho|\gg 1$ and $|k_m\rho|\ll 1$.

The $z$-component of the electric field is likewise computed by~\eqref{eq:E1z-vd-0} under Remark~3:
\begin{align}\label{eq:E1z-vd-ex}
E_{1z}^{({\rm vd})}&=-\frac{\omega\mu_0}{4\pi k^2}\Biggl\{2\Biggl(\frac{i}{\rho^3}+\frac{k}{\rho^2}-\frac{ik^2}{\rho}\Biggr)e^{ik\rho}+i\pi\frac{k^2k_m^2}{\omega\mu_0\sigma}\biggl(1+i\frac{2k^2}{\omega\mu_0\sigma}\frac{1}{\sqrt{k_m^2-k^2}}\biggr)H_0^{(1)}(k_m\rho)\nonumber\\
&\mbox{} \qquad +i\frac{4k^4}{(\omega\mu_0\sigma)^2}\frac{1}{\alpha_m}\frac{1}{\rho}\Biggl[ 1-\frac{\pi}{2}k_m\rho \bigl(\mathbb{H}_0(k_m\rho)-Y_0(k_m\rho)\bigr)\nonumber\\
&\mbox{}\qquad +\sum_{l=0}^\infty \biggl(\frac{1}{2}\biggr)_l\ \frac{(-1)^{l+1}}{l!}\biggl(\frac{k^2}{k_m^2}\biggr)^{l+1}\,\frac{d}{d\zeta}\biggl\{\zeta\frac{d^{2l+1}}{d\zeta^{2l+1}}\biggr(\frac{1-e^{i\zeta}}{\zeta}\biggr)\biggr\}\biggr|_{\zeta=k\rho}\Biggr]\Biggr\}\nonumber\\
&\sim -\frac{\omega\mu_0}{4\pi}\Biggl\{ \Biggl(\frac{i}{k^2\rho^3}+\frac{1}{k\rho^2}-\frac{i}{\rho}\Biggr)e^{ik\rho}-8\pi i\, \frac{k^4}{(\omega\mu_0\sigma)^3}\,H_0^{(1)}(k_m\rho)+i\frac{4k^2}{(\omega\mu_0\sigma)^2}\frac{1}{\rho}\Biggl[1-\frac{\pi}{2}k_m\rho\nonumber\\
&\mbox{}\qquad \times \bigl(\mathbb{H}_0(k_m\rho)-Y_0(k_m\rho)\bigr)-\frac{1}{k_m^2\rho^2}\Biggr]\Biggr\}~,\qquad |\omega\mu_0\sigma|\ll |k|~.
\end{align}
The terms proportional to $e^{ik\rho}$ in the fourth line of~\eqref{eq:E1z-vd-ex} amount to the primary field of the dipole.~\cite{KOW} The TM surface plasmon contribution is expressed by the term proportional to $H_0^{(1)}(k_m\rho)$. If $|\omega\mu_0\sigma|\ll |k|$ and $|k_m\rho|= \mathcal O(1)$, the primary field is subdominant to this contribution provided $|k\rho| \gg |(\omega\mu_0\sigma)/k|$ (in addition to having $|k\rho|\ll 1$). See Appendix~\ref{app:asymp} for asymptotic evaluations of $E_{1z}^{({\rm vd})}$ in the far- and near-field regimes.

%----------------------------------------------------
\subsubsection{Horizontal dipole}
%----------------------------------------------------
Next, by~\eqref{eq:E2z-hd-0}--\eqref{eq:B2ph-hd-0} we compute the field of the horizontal dipole. We address the $z$-components first, since each of these fields involves only one type of polarization, as indicated by the presence of only one denominator ($\mathcal P$ or $\mathcal Q$) in their Fourier representations for $k_1\neq k_2$.

By~\eqref{eq:E2z-hd-0}, $E_{2z}^{({\rm hd})}$ involves only TM polarization and is computed as
\begin{subequations}\label{eq:E2z-hd-ex}
\begin{align}\label{eq:E2z-hd-ex-a}
E_{2z}^{({\rm hd})}&=-\frac{i\omega\mu_0}{2\pi}\frac{1}{(\omega\mu_0\sigma)^2}\cos\phi \Biggl\{-i\omega\mu_0\sigma\Biggl(\frac{1}{\rho^2}-\frac{ik}{\rho}\Biggr)e^{ik\rho}+\frac{i\pi k_m}{2}\biggl(2k^2-i\omega\mu_0\sigma \sqrt{k_m^2-k^2}\biggr)   \nonumber\\
&\mbox{} \hphantom{=-\frac{i\omega\mu_0}{2\pi}\frac{1}{(\omega\mu_0\sigma)^2}} \times H_1^{(1)}(k_m\rho)-i\omega\mu_0\sigma k_m^2\alpha_m\Biggl[1-\frac{\pi}{2}\bigl(\mathbb{H}_1(k_m\rho)-Y_1(k_m\rho)\bigr)\nonumber\\
&\mbox{} \hphantom{=-\frac{i\omega\mu_0}{2\pi}\frac{1}{(\omega\mu_0\sigma)^2}}
+\sum_{l=0}^\infty \biggl(\frac{1}{2}\biggr)_l \frac{(-1)^{l+1}}{l!}\biggl(\frac{k^2}{k_m^2}\biggr)^{l+1}\,\frac{d^{2l+1}}{d\zeta^{2l+1}}\biggl(\frac{1-e^{i\zeta}}{\zeta}\biggr)\biggl|_{\zeta=k\rho}\Biggr]\Biggr\}~.
\end{align}
Notice the term $(2k^2-i\omega\mu_0\sigma \sqrt{k_m^2-k^2}) H_1^{(1)}(k_m\rho)$, in conjunction with Remark~2. By Remark~3 and $|\omega\mu_0\sigma|\ll |k|$ (i.e., $|k_m|^2\gg |k|^2$), we keep two terms in the series expansion and readily compute (see also Appendix~\ref{app:asymp})
\begin{align}\label{eq:E2z-hd-ex-b}
E_{2z}^{({\rm hd})}&\sim -\frac{i\omega\mu_0}{2\pi}\,\cos\phi \Biggl\{ -\frac{i\omega\mu_0\sigma}{4}\Biggl(\frac{1}{k^2\rho^2}+\frac{3i}{k^3\rho^3}-\frac{3}{k^4\rho^4}\Biggr)e^{ik\rho}-4\pi  \frac{k^4}{(\omega\mu_0\sigma)^3}H_1^{(1)}(k_m\rho)\nonumber\\
&\mbox{} \hphantom{\sim -\frac{i\omega\mu_0}{2\pi}\,\cos\phi}
+\frac{2k^2 k_m}{(\omega\mu_0\sigma)^2}\Biggl[1-\frac{\pi}{2}\bigl(\mathbb{H}_1(k_m\rho)-Y_1(k_m\rho)\bigr)+\frac{1}{k_m^2\rho^2}-\frac{3}{k_m^4\rho^4}\Biggr]\Biggr\}~.
\end{align}
In the last formula, the TM surface plasmon is evident as the term proportional to $H_1^{(1)}(k_m\rho)$.
In contrast, if $k_m$ lies in the second Riemann sheet (see Remark~2), then
\begin{align}\label{eq:E2z-hd-ex-c}
E_{2z}^{({\rm hd})}&\sim -\frac{i\omega\mu_0}{2\pi}\,\cos\phi \Biggl\{-\frac{i\omega\mu_0\sigma}{4}\Biggl(\frac{1}{k^2\rho^2}+\frac{3i}{k^3\rho^3}-\frac{3}{k^4\rho^4}\Biggr)e^{ik\rho}\nonumber\\
&\mbox{} \hphantom{\sim -\frac{i\omega\mu_0}{2\pi}\,\cos\phi}
+\frac{2k^2 k_m}{(\omega\mu_0\sigma)^2}\Biggl[1-\frac{\pi}{2}\bigl(\mathbb{H}_1(k_m\rho)-Y_1(k_m\rho)\bigr)+\frac{1}{k_m^2\rho^2}-\frac{3}{k_m^4\rho^4}\Biggr]\Biggr\}~.
\end{align}
\end{subequations}

We next compute the $z$-component of the magnetic field by resorting to Remark~2. Equation~\eqref{eq:B2z-hd-0} yields
\begin{subequations}\label{eq:B2z-hd-ex}
\begin{align}\label{eq:B2z-hd-ex-a}
B_{2z}^{({\rm hd})}&=\frac{i\mu_0}{8\pi}\,\sin\phi \Biggl\{-2i\Biggr(\frac{1}{\rho^2}-\frac{ik}{\rho}\Biggr)e^{ik\rho}+\frac{i\pi}{2}\,\omega\mu_0\sigma k_e\,H_1^{(1)}(k_e\rho)-2ik_e\sqrt{k_e^2-k^2}\nonumber\\
&\mbox{} \qquad \times\sqrt{\pi}e^{ik\rho}\sum_{l=0}^\infty \biggl(\frac{1}{2}\biggr)_l\,\frac{1+2l}{1-2l}\frac{(i/2)^l}{l!}\biggl(\frac{k_e-k}{k_e}\biggr)^{l+1/2}\,e^{i\wp}\frac{d^l}{d z^l}\bigl[z^{-1/2}\mathcal F_0(z)\bigr]\biggl|_{z=\wp}\Biggr\}~,
\end{align}
which involves TE polarization; $\mathcal F_0$ and $\wp$ are defined by~\eqref{eq:F0} and~\eqref{eq:Som-dist}. The term proportional to $e^{ik\rho}$ in the first line of~\eqref{eq:B2z-hd-ex-a} corresponds to the field of the $x$-directed dipole in free space. Beware of the factor $\sqrt{k_e^2-k^2}$ in~\eqref{eq:B2z-hd-ex-a}; see also Remark~2 with $\varpi=k_e\rho$. To simplify formula~\eqref{eq:B2z-hd-ex-a}, consider $|\omega\mu_0\sigma|\ll |k|$, by which $k_e\sim k$. Thus,~\eqref{eq:B2z-hd-ex-a} is reduced to
\begin{align}\label{eq:B2z-hd-ex-b}
B_{2z}^{({\rm hd})}&\sim \frac{i\mu_0}{8\pi}\,\sin\phi \Biggl\{-2i\Biggr(\frac{1}{\rho^2}-\frac{ik}{\rho}\Biggr)e^{ik\rho}+\frac{i\pi}{2}\omega\mu_0\sigma k\Biggl[H_1^{(1)}(k_e\rho)-2\frac{\sqrt{k_e^2-k^2}}{\omega\mu_0\sigma}\sqrt{\frac{2}{\pi k_e\rho}} e^{i(k_e\rho-\pi/4)}\Biggr]\nonumber\\
&\mbox{} \hphantom{=\frac{i\mu_0}{8\pi}\,\sin\phi \Biggl\{}+2ik\sqrt{k_e^2-k^2}\,e^{ik\rho}\,\sqrt{\frac{\pi}{k\rho}}\ \mathcal F(\wp)\Biggr\}~,
\end{align}
where $\mathcal F$ is defined in~\eqref{eq:F0}.
If we adhere to Remark~3, we expect that \eqref{eq:B2z-hd-ex-b} should not manifest the contribution, $H_1^{(1)}(k_e\rho)$, of the residue at the pole $\lambda=k_e$.~\cite{KOW} To demonstrate its elimination for large enough distances in this component, we use the asymptotic expansion~\cite{Bateman-II}
$H_\nu^{(1)}(z)\sim \sqrt{2/(\pi z)}\, e^{i(z-\nu\pi/2-\pi/4)}$, which provides a reasonable approximation for $H_\nu^{(1)}(k_e\rho)$ for all but small values of $|k_e\rho|$. Recall that $\sqrt{k_e^2-k^2}=-i\omega\mu_0\sigma/2$ if $k_e$ does not lie on the top Riemann sheet of $\sqrt{k^2-\lambda^2}$. Thus, for sufficiently large $|k_e\rho|\sim |k\rho|$, \eqref{eq:B2z-hd-ex-b} becomes
\begin{equation}\label{eq:B2z-hd-ex-c}
B_{2z}^{({\rm hd})}=\frac{i\mu_0}{8\pi}\,\sin\phi \Biggl\{-2i\Biggr(\frac{1}{\rho^2}-\frac{ik}{\rho}\Biggr)e^{ik\rho}+\omega\mu_0\sigma k\,\sqrt{\frac{\pi}{k\rho}}\,e^{ik\rho}\,\mathcal F(\wp)\Biggr\}~,
\end{equation}
which manifests the $e^{ik\rho}$-wave behavior in view of~\eqref{eq:F-asymp}; see Appendix~\ref{app:asymp}. In contrast, by the hypothetical scenario when $k_e$ lies in the physical (top) $\lambda$-Riemann sheet, \eqref{eq:B2z-hd-ex-b} yields
\begin{align}\label{eq:B2z-hd-ex-d}
B_{2z}^{({\rm hd})}&=\frac{i\mu_0}{8\pi}\,\sin\phi \Biggl\{-2i\Biggr(\frac{1}{\rho^2}-\frac{ik}{\rho}\Biggr)e^{ik\rho}+\omega\mu_0\sigma k \sqrt{\frac{2\pi }{ k\rho}} e^{i(k_e\rho-\pi/4)}\nonumber\\
&\mbox{} \hphantom{=\frac{i\mu_0}{8\pi}\,\sin\phi \Biggl\{}-\omega\mu_0\sigma k\,\sqrt{\frac{\pi}{k\rho}}\,e^{ik\rho}\,\mathcal F(\wp)\Biggr\}~,
\end{align}
\end{subequations}
in view of Remark~2.

In regard to the  $\rho$-component of the electric field, formula \eqref{eq:E2r-hd-0} entails
\begin{subequations}\label{eq:E2r-hd-ex}
\begin{align}\label{eq:E2r-hd-ex-a}
E_{2\rho}^{({\rm hd})}&=-\frac{\omega\mu_0}{4\pi}\cos\phi \Biggl\{ \Biggl[-i\biggl(1-\frac{k^2}{k_m^2}\biggr)\frac{1}{\rho}+\biggl(1-\frac{k^2}{k_e^2}-\frac{k^2}{k_m^2}\biggr)\frac{1}{k\rho^2}\Biggr]e^{ik\rho}\nonumber\\
&\mbox{}\quad  +i\pi \frac{\omega\mu_0\sigma}{4k_e\rho} H_1^{(1)}(k_e\rho)-i\frac{\sqrt{k_e^2-k^2}}{k_e\rho}\sqrt{\pi}\,e^{ik\rho}\nonumber\\
&\mbox{} \quad \times \sum_{l=0}^\infty\biggl(\frac{1}{2}\biggr)_l \frac{1+2l}{1-2l}\frac{(i/2)^l}{l!}\biggl(\frac{k_e-k}{k_e}\biggr)^{l+1/2} e^{i \wp}\frac{d^l}{d z^l}\big[z^{-1/2} \mathcal F_0(z)\big]\biggl|_{z=\wp}\nonumber\\
&\mbox{}\quad -2\pi i \frac{k^2}{(\omega\mu_0\sigma)^2}\biggl[\frac{2k^2}{\omega\mu_0\sigma}-i\sqrt{k_m^2-k^2}\biggr] H_1^{(1)\prime}(k_m\rho) +i\frac{4k^2\,k_m}{(\omega\mu_0\sigma)^2}\alpha_m\Biggl[ -\frac{\pi}{2}\big(\mathbb{H}_1'(k_m\rho)\nonumber\\
&\mbox{}\quad -Y_1'(k_m\rho)\big)
+\frac{k}{k_m}\sum_{l=0}^\infty \biggl(\frac{1}{2}\biggr)_l \frac{(-1)^{l+1}}{l!}\biggl(\frac{k^2}{k_m^2}\biggr)^{l+1}\frac{d^{2l+2}}{d \zeta^{2l+2}}\biggl(\frac{1-e^{i\zeta}}{\zeta}\biggr)\biggl|_{\zeta=k\rho}\Biggr]\Biggr\}~,
\end{align}
where the prime denotes derivative with respect to the argument and $\mathbb{H}'_1(z)=\frac{1}{2}[\mathbb{H}_0(z)-\mathbb{H}_2(z)]+\Gamma(5/2)^{-1}(4\sqrt{\pi})^{-1}z$. By imposing $|\omega\mu_0\sigma|\ll |k|$ under Remark~3, we keep one term from each of the above series expansion and find
\begin{align}\label{eq:E2r-hd-ex-b}
E_{2\rho}^{({\rm hd})}&\sim -\frac{\omega\mu_0}{4\pi}\cos\phi \,\Biggl\{ -2 \Biggl(\frac{1}{k\rho^2}+\frac{i}{k^2\rho^3}\Biggr)e^{ik\rho}\nonumber\\
&\mbox{}\quad + i\pi \frac{\omega\mu_0\sigma}{4k\rho}\,\biggl[H_1^{(1)}(k_e\rho)-\sqrt{\frac{2}{\pi k_e\rho}}\,e^{i(k_e\rho-3\pi/4)}\biggr]+\frac{\omega\mu_0\sigma}{2k \rho}\sqrt{\frac{\pi}{k\rho}}\,e^{ik\rho} \mathcal F(\wp)\nonumber\\
&\mbox{}\quad -i\pi \frac{8k^4}{(\omega\mu_0\sigma)^3}\,H_1^{(1)\prime}(k_m\rho)+\frac{4k^4}{(\omega\mu_0\sigma)^3}\biggl[\pi\bigl(\mathbb{H}_1'(k_m\rho)-Y_1'(k_m\rho)\bigr)+\frac{4}{(k_m\rho)^3}\biggr]\Biggr\}~.
\end{align}
\end{subequations}
The primary field of the dipole is singled out in the first line of formula~\eqref{eq:E2r-hd-ex-b}, while the remaining terms of this formula express the scattered wave; see Appendix~\ref{app:asymp}. In the scattered wave, there are two noteworthy contributions. The TM surface plasmon contribution is expressed by the term proportional to $H_1^{(1)\prime}(k_m\rho)$. The other contribution, $H_1^{(1)}(k_e\rho)$, is cancelled out for large enough $|k_e\rho|$ in a fashion similar to that shown for $B_{2z}^{({\rm hd})}$ above.

By~\eqref{eq:E2ph-hd-0}, we likewise obtain the exact formula
\begin{subequations}\label{eq:E2ph-hd-ex}
\begin{align}\label{eq:E2ph-hd-ex-a}
E_{2\phi}^{({\rm hd})}&=\frac{\omega\mu_0}{4\pi}\sin\phi\,\Biggl\{ \biggl[\frac{k}{k_e^2}\biggl(1+\frac{4k^2k_e^2}{k_m^2(\omega\mu_0\sigma)^2}\Biggr)\frac{1}{\rho^2}-\frac{ik^2}{k_e^2}\frac{1}{\rho}\biggr] e^{ik\rho}\nonumber\\
&\mbox{} \quad +\frac{i\pi}{4}\omega\mu_0\sigma H_1^{(1)\prime}(k_e\rho)+\sqrt{k_e^2-k^2}\sqrt{\pi}\,e^{ik\rho}\nonumber\\
&\mbox{}\quad \times \sum_{l=0}^\infty \biggl(\frac{1}{2}\biggr)_l\frac{3+4l^2}{(1-2l)(3-2l)}\frac{(i/2)^l}{l!}\biggl(\frac{k_e-k}{k_e}\biggr)^{l+1/2}e^{i\wp}\frac{d^l}{d z^l}\big[z^{-1/2}\mathcal F_0(z)\big]\biggl|_{z=\wp}\nonumber\\
&\mbox{}\quad -\frac{2\pi i k^2}{(\omega\mu_0\sigma)^2}\biggl(\frac{2k^2}{\omega\mu_0\sigma}-i\sqrt{k_m^2-k^2}\biggr)\frac{H_1^{(1)}(k_m\rho)}{k_m\rho}+\frac{4i k^2}{(\omega\mu_0\sigma)^2}\frac{\alpha_m}{\rho}\biggl[1-\frac{\pi}{2}\bigl(\mathbb{H}_1(k_m\rho)\nonumber\\
&\mbox{}\quad -Y_1(k_m\rho)\bigr)+\sum_{l=0}^\infty \biggl(\frac{1}{2}\biggr)_l \frac{(-1)^{l+1}}{l!}\biggl(\frac{k^2}{k_m^2}\biggr)^{l+1}\frac{d^{2l+1}}{d\zeta^{2l+1}}\biggl(\frac{1-e^{i\zeta}}{\zeta}\biggr)\biggl|_{\zeta=k\rho}\biggr]\Biggr\}~.
\end{align}
By Remark~3 and $|\omega\mu_0\sigma|\ll |k|$, this formula is simplified to
\begin{align}\label{eq:E2ph-hd-ex-b}
E_{2\phi}^{({\rm hd})}&\sim \frac{\omega\mu_0}{4\pi} \sin\phi \Biggl\{ \Biggl(-\frac{i}{\rho}+\frac{1}{k\rho^2}+\frac{i}{k^2\rho^3}\Biggr)e^{ik\rho}\nonumber\\
&\mbox{}\quad +\frac{i\pi}{4}\omega\mu_0\sigma \biggl[ H_1^{(1)\prime}(k_e\rho)-i\sqrt{\frac{2}{\pi k_e\rho}} e^{i(k_e\rho-3\pi/4)}\biggr]+\frac{i\omega\mu_0\sigma}{2}e^{ik\rho}\sqrt{\frac{\pi}{k\rho}}\,\mathcal F(\wp)\nonumber\\
&\mbox{}\quad -i\pi \frac{8k^4}{(\omega\mu_0\sigma)^3}\frac{1}{k_m\rho} H_1^{(1)}(k_m\rho)\nonumber\\
&\mbox{}\quad -\frac{8k^4}{(\omega\mu_0\sigma)^3}\frac{1}{k_m\rho}\biggl[1-\frac{\pi}{2}\bigl(\mathbb{H}_1(k_m\rho)-Y_1(k_m\rho)\bigr)+\frac{1}{(k_m\rho)^2}\biggr]\Biggr\}~.
\end{align}
\end{subequations}
In the first line of approximate formula~\eqref{eq:E2ph-hd-ex-b}, we display the radiation field in the absence of the layer. The additional terms of this formula account for the scattered field. See Appendix~\ref{app:asymp} for alternate computations based on asymptotics.

By~\eqref{eq:B2r-hd-0}, the  $\rho$-component of the magnetic field reads
\begin{subequations}\label{eq:B2r-hd-ex}
\begin{align}\label{eq:B2r-hd-ex-a}
B_{2\rho}^{({\rm hd})}&=\frac{\mu_0}{4\pi}\sin\phi\Biggl\{ \Biggl[-\Biggl(\frac{\omega\mu_0\sigma}{2}\frac{k}{k_e^2}+\frac{2k^3}{\omega\mu_0\sigma}\frac{1}{k_m^2}\Biggr)\frac{1}{\rho^2}+\frac{i\omega\mu_0\sigma}{2}\frac{k^2}{k_e^2}\frac{1}{\rho}\Biggr] e^{ik\rho}\nonumber\\
&\mbox{} \quad +i\pi \frac{2k^4}{(\omega\mu_0\sigma)^2}\frac{1}{k_m\rho}\biggl[1-\frac{i\omega\mu_o\sigma}{2k^2}\sqrt{k_m^2-k^2}\biggr] H_1^{(1)}(k_m\rho)-\frac{2ik^2}{\omega\mu_0\sigma}\frac{\alpha_m}{\rho}\nonumber\\
&\mbox{} \quad \times \Biggl[1-\frac{\pi}{2}\bigl(\mathbb{H}_1(k_m\rho)-Y_1(k_m\rho)\bigr)+\sum_{l=0}^\infty \biggl(\frac{1}{2}\biggr)_l\frac{(-1)^{l+1}}{l!}\biggl(\frac{k^2}{k_m^2}\biggr)^{l+1}\frac{d^{2l+1}}{d\zeta^{2l+1}}\biggl(\frac{1-e^{i\zeta}}{\zeta}\biggr)\biggl|_{\zeta=k\rho}\Biggr]\nonumber\\
&\mbox{}\quad -\frac{i\pi}{8}(\omega\mu_0\sigma)^2\,H_1^{(1)\prime}(k_e\rho)-\frac{\omega\mu_0\sigma}{2}\sqrt{k_e^2-k^2}\sqrt{\pi}\,e^{ik\rho}\nonumber\\
&\mbox{}\quad \times \sum_{l=0}^\infty \biggl(\frac{1}{2}\biggr)_l\ \frac{3+4l^2}{(1-2l)(3-2l)}
\frac{(i/2)^l}{l!}\biggl(\frac{k_e-k}{k_e}\biggr)^{l+1/2}e^{i\wp}\frac{d^l}{dz^l}\big[z^{-1/2}\mathcal F_0(z)\big]\biggl|_{z=\wp}\Biggr\}~.
\end{align}
By Remark~3 and condition $|\omega\mu_0\sigma|\ll |k|$, the last expression becomes
\begin{align}\label{eq:B2r-hd-ex-b}
B_{2\rho}^{({\rm hd})}&\sim \frac{\mu_0}{4\pi}\sin\phi\, \Biggl\{ \frac{\omega\mu_0\sigma}{2}\Biggl(\frac{i}{\rho}-\frac{1}{k\rho^2}-\frac{i}{k^2\rho^3}\Biggr)e^{ik\rho}\nonumber\\
&\mbox{}\quad +i\pi \frac{4k^4}{(\omega\mu_0\sigma)^2}\frac{1}{k_m\rho}H_1^{(1)}(k_m\rho)-i\frac{2k^2}{\omega\mu_0\sigma}\frac{1}{\rho}\Biggl[1-\frac{\pi}{2}\bigl(\mathbb{H}_1(k_m\rho)-Y_1(k_m\rho)\bigr)+\frac{1}{(k_m\rho)^2}\Biggr]\nonumber\\
&\mbox{}\quad -\frac{i\pi}{8} (\omega\mu_0\sigma)^2\,\biggl[H_1^{(1)\prime}(k_e\rho)-\sqrt{\frac{2}{\pi k_e\rho}}\,e^{i(k_e\rho-\pi/4)}\biggr]-\frac{i (\omega\mu_0\sigma)^2}{4}\,\sqrt{\frac{\pi}{k\rho}}\,e^{ik\rho}\mathcal F(\wp)\Biggr\}~;
\end{align}
\end{subequations}
see Appendix~\ref{app:asymp} for alternate derivations of the far and near field.

We conclude this section with the computation of the $\phi$-component of the magnetic field. By~\eqref{eq:B2ph-hd-0}, we obtain
\begin{subequations}\label{eq:B2ph-hd-ex}
\begin{align}\label{eq:B2ph-hd-ex-a}
B_{2\phi}^{({\rm hd})}&=\frac{\mu_0}{4\pi}\cos\phi\,\Biggl\{\Biggl[\Biggl(\frac{\omega\mu_0\sigma}{2}\frac{k}{k_e^2}+\frac{2k^3}{\omega\mu_0\sigma}\frac{1}{k_m^2}\Biggr)\frac{1}{\rho^2}-i\frac{2k^4}{\omega\mu_0\sigma}\frac{1}{k_m^2\rho}\Biggr]e^{ik\rho}\nonumber\\
&\mbox{}\quad +i\pi \frac{k^2}{\omega\mu_0\sigma}\biggl( \frac{2k^2}{\omega\mu_0\sigma}-i\sqrt{k_m^2-k^2}\biggr)H_1^{(1)\prime}(k_m\rho)-i\frac{2k^2k_m}{\omega\mu_0\sigma}\alpha_m\Biggl[-\frac{\pi}{2}\bigl(\mathbb{H}_1'(k_m\rho)\nonumber\\
&\mbox{}\quad -Y_1'(k_m\rho)\bigr)+\frac{k}{k_m}\sum_{l=0}^\infty \biggl(\frac{1}{2}\biggr)_l\ \frac{(-1)^{l+1}}{l!}\biggl(\frac{k^2}{k_m^2}\biggr)^{l+1}\frac{d^{2l+2}}{d\zeta^{2l+2}}\biggl(\frac{1-e^{i\zeta}}{\zeta}\biggr)\biggl|_{\zeta=k\rho}\Biggr]\nonumber\\
&\mbox{}\quad -\frac{i\pi}{8}\frac{(\omega\mu_0\sigma)^2}{k_e\rho}H_1^{(1)}(k_e\rho)+\frac{i\omega\mu_0\sigma}{2}\frac{\sqrt{k_e^2-k^2}}{k_e\rho}\nonumber\\
&\mbox{}\quad \times \sqrt{\pi}\,e^{ik\rho}\sum_{l=0}^\infty \biggl(\frac{1}{2}\biggr)_l\ \frac{1+2l}{1-2l}\frac{(i/2)^l}{l!}\biggl(\frac{k_e-k}{k_e}\biggr)^{l+1/2}e^{i\wp}\frac{d^l}{dz^l}\bigl[z^{-1/2}\mathcal F_0(z)\bigr]\biggl|_{z=\wp}\Biggr\}~.
\end{align}
Suppose that $|\omega\mu_0\sigma|\ll |k|$. By Remark~3, the preceding formula is reduced to
\begin{align}\label{eq:B2ph-hd-ex-b}
B_{2\phi}^{({\rm hd})}&\sim \frac{\mu_0}{4\pi}\cos\phi\Biggl\{ \omega\mu_0\sigma\Biggl(\frac{1}{k\rho^2}+\frac{i}{k^2\rho^3}\Biggr)e^{ik\rho}\nonumber\\
&\mbox{}\quad +i\pi \frac{4k^4}{(\omega\mu_0\sigma)^2}H_1^{(1)\prime}(k_m\rho)-\frac{4k^4}{(\omega\mu_0\sigma)^2}\Biggl[\frac{\pi}{2}\bigl(\mathbb{H}_1'(k_m\rho)-Y_1'(k_m\rho)\bigr)+\frac{2}{(k_m\rho)^3}\Biggr]\nonumber\\
&\mbox{}\quad -\frac{i\pi}{8}\frac{(\omega\mu_0\sigma)^2}{k_e\rho}\Biggl[H_1^{(1)}(k_e\rho)-\sqrt{\frac{2}{\pi k_e\rho}}e^{i(k\rho-3\pi/4)}\Biggr]-\frac{(\omega\mu_0\sigma)^2}{4}\frac{1}{k\rho}\sqrt{\frac{\pi}{k\rho}}\,e^{ik\rho}\,\mathcal F(\wp)\Biggr\}~.
\end{align}
\end{subequations}
In Appendix~\ref{app:asymp}, we compute this field component by asymptotic methods, in the near- ($|k_m\rho|\ll 1$) and far-field ($|k\rho|\gg 1$) regimes for $|\omega\mu_0\sigma|\ll |k|$; the results are in agreement with~\eqref{eq:B2ph-hd-ex-b}.

%%%%%%%%%%%%%%%%%%%%%%%%%%%%%%%%%%%%%%%%%%%%%
%%%%%%%%%%%%%%%%%%%%%%%%%%%%%%%%%%%%%%%%%%%%%
\section{Discussion}
\label{sec:Lateral}
%%%%%%%%%%%%%%%%%%%%%%%%%%%%%%%%%%%%%%%%%%%%%
%%%%%%%%%%%%%%%%%%%%%%%%%%%%%%%%%%%%%%%%%%%%%
In this section, we discuss the derived formulas of Section~\ref{sec:Exact} under Remark 3 in the case with $|k_m\rho|\gg 1$ and $k\rho= \mathcal O(1)$, by assuming that $|\omega\mu_0\sigma|\ll |k|$. An issue that we address is the relative strength of the TM surface plasmon contribution to the electromagnetic field. A practical consideration that motivates this study is that layers of two-dimensional materials such as graphene may sustain a TM surface plasmon wave number, $k_m$, with $0< \Im k_m\ll \Re k_m$ and $\Re k_m\sim |k_m|\gg |k|$ at a certain accessible frequency range;~\cite{Gonzalez14} thus, the wavelength of the generated surface plasmon can be much smaller than the wavelength of the incident electromagnetic field in free space. It is of interest to investigate how the TM surface plasmon can prevail over other field contributions on the layer at distances of the order of the free-space wavelength.

%%%%%%%%%%%%%%%%%%%%%%%%%%%%%%%%%%%%%%%%%%%%%
\subsection{Field of vertical  dipole, $|k_m\rho|\gg 1$ and $k\rho=\mathcal O(1)$}
\label{subsec:Lateral-vert}
%%%%%%%%%%%%%%%%%%%%%%%%%%%%%%%%%%%%%%%%%%%%%
By~\eqref{eq:E1r-vd-ex-b}, the $\rho$-component of the electric field reduces to
\begin{equation}\label{eq:E1r-vd-app}
E_{1\rho}^{({\rm vd})}\sim \frac{\omega\mu_0}{2\pi}\Biggl\{
\frac{\omega\mu_0\sigma}{4}\Biggl(\frac{3}{k^4\rho^4}-\frac{3i}{k^3\rho^3}-\frac{1}{k^2\rho^2}\Biggr)e^{ik\rho}
+\frac{4k^4}{(\omega\mu_0\sigma)^3}\sqrt{\frac{2\pi}{k_m\rho}}e^{i(k_m\rho-\pi/4)}\Biggr\}~,
\end{equation}
if $|k_m\rho|\gg 1$, by use of asymptotic expansions~\eqref{eq:HY-asympt} and $H_\nu^{(1)}(z)\sim \sqrt{2/(\pi z)}e^{i(z-\nu \pi/2-\pi/4)}$ as $z\to\infty$ with $\nu=1$. Hence, the TM surface plasmon dominates in this regime if
\begin{equation}\label{eq:plasm-cond-1}
\biggl|\frac{\omega\mu_0\sigma}{k}\biggr|\ll e^{-(2/7)\Im k_m/|k|}~.
\end{equation}

In the same vein, by~\eqref{eq:B1ph-vd-ex} we compute the $\phi$-component of the magnetic field:
\begin{equation}\label{eq:E1r-vd-app}
B_{1\phi}^{({\rm vd})}\sim -\frac{i\mu_0 k^2}{4\pi}\Biggl\{ \Biggl(\frac{i}{k^2\rho^2}+\frac{1}{k\rho}\Biggr)e^{ik\rho}-\frac{8k^2}{(\omega\mu_0\sigma)^2}\sqrt{\frac{\pi}{2k_m\rho}}e^{i(k_m\rho+\pi/4)}\Biggr\}~.
\end{equation}
Evidently, the TM surface plasmon contribution dominates in this component if
\begin{equation}\label{eq:plasm-cond-2}
\biggl|\frac{\omega\mu_0\sigma}{k}\biggr|\ll e^{-(2/3)\Im k_m/|k|}~.
\end{equation}

Equation~\eqref{eq:E1z-vd-ex} for the $z$-component of the electric field yields
\begin{equation}\label{eq:E1z-vd-app}
E_{1z}^{({\rm vd})}\sim -\frac{\omega\mu_0 k}{4\pi}\Biggl\{ \Biggl( \frac{i}{k^3\rho^3}+\frac{1}{k^2\rho^2}-\frac{i}{k\rho}\Biggr) e^{ik\rho}-\frac{8k^3}{(\omega\mu_0\sigma)^3}\sqrt{\frac{2\pi}{k_m\rho}} e^{i(k_m\rho+\pi/4)}\Biggr\}~,
\end{equation}
which implies that the TM surface plasmon is dominant  if
\begin{equation}\label{eq:plasm-cond-3}
\biggl|\frac{\omega\mu_0\sigma}{k}\biggr|\ll e^{-(2/5)\Im k_m/|k|}~.
\end{equation}
Thus, for $k\rho=\mathcal O(1)$ the TM surface plasmon dominates in all field components of the vertical dipole  if~\eqref{eq:plasm-cond-2} holds.

%%%%%%%%%%%%%%%%%%%%%%%%%%%%%%%%%%%%%%%%%%%%%
\subsection{Field of horizontal dipole, $|k_m\rho|\gg 1$ and $k\rho=\mathcal O(1)$}
\label{subsec:Lateral-horiz}
%%%%%%%%%%%%%%%%%%%%%%%%%%%%%%%%%%%%%%%%%%%%%
First, we address the $z$-components. By~\eqref{eq:E2z-hd-ex-b}, the $z$-component of the electric field reads
\begin{equation}\label{eq:E2z-hd-app}
E_{2z}^{({\rm hd})}\sim \frac{\omega\mu_0}{2\pi}\cos\phi \Biggl\{- \frac{\omega\mu_0\sigma}{4}\Biggl(\frac{1}{k^2\rho^2}+\frac{3i}{k^3\rho^3}-\frac{3}{k^4\rho^4}\Biggr)e^{ik\rho}+\frac{4k^4}{(\omega\mu_0\sigma)^3}\,\sqrt{\frac{2\pi}{k_m\rho}}\,e^{i(k_m\rho-\pi/4)}\Biggr\}~.
\end{equation}
For $k\rho=\mathcal O(1)$, the pole contribution dominates under condition~\eqref{eq:plasm-cond-1}. In contrast, by~\eqref{eq:B2z-hd-ex} the $z$-component of the magnetic field does not carry any TM surface plasmon contribution; thus, we omit its evaluation from this discussion.

The remaining components are more richly structured. By~\eqref{eq:E2r-hd-ex-b}, we have
\begin{align}\label{eq:E2r-hd-app}
E_{2\rho}^{({\rm hd})}&\sim -\frac{\omega\mu_0 k}{4\pi}\cos\phi \,\Biggl\{ -2 \Biggl(\frac{1}{k^2\rho^2}+\frac{i}{k^3\rho^3}\Biggr)e^{ik\rho}
-\frac{8k^3}{(\omega\mu_0\sigma)^3}\sqrt{\frac{2\pi}{k_m\rho}}\,e^{i(k_m\rho+\pi/4)}
\nonumber\\
&\mbox{}\quad + \frac{i\pi}{4} \frac{\omega\mu_0\sigma}{k}\frac{1}{k\rho}\,\biggl[H_1^{(1)}(k_e\rho)-\sqrt{\frac{2}{\pi k_e\rho}}\,e^{i(k_e\rho-3\pi/4)}\biggr]+\frac{\omega\mu_0\sigma}{2k}\frac{1}{k \rho}\sqrt{\frac{\pi}{k\rho}}\,e^{ik\rho} \mathcal F(\wp)\Biggr\}~.
\end{align}
If $k\rho=\mathcal O(1)$, all terms in the second line are relatively small. The residue contribution of the pole at $k_m$ dominates over the primary field if~\eqref{eq:plasm-cond-3} holds.

Equation~\eqref{eq:E2ph-hd-ex-b} for the $\phi$-component of the electric field gives
\begin{align}\label{eq:E2ph-hd-app}
E_{2\phi}^{({\rm hd})}&\sim \frac{\omega\mu_0 k}{4\pi} \sin\phi \Biggl\{ \Biggl(-\frac{i}{k\rho}+\frac{1}{k^2\rho^2}+\frac{i}{k^3\rho^3}\Biggr)e^{ik\rho}
-\frac{8k^3}{(\omega\mu_0\sigma)^3}\frac{1}{k_m\rho} \sqrt{\frac{2\pi}{k_m\rho}}
e^{i(k_m\rho-\pi/4)}
\nonumber\\
&\mbox{}\quad +\frac{i\pi}{4}\frac{\omega\mu_0\sigma}{k} \biggl[ H_1^{(1)\prime}(k_e\rho)-i\sqrt{\frac{2}{\pi k_e\rho}} e^{i(k_e\rho-3\pi/4)}\biggr]+\frac{i\omega\mu_0\sigma}{2k}e^{ik\rho}\sqrt{\frac{\pi}{k\rho}}\,\mathcal F(\wp)\Biggr\}~,
\end{align}
which is dominated by the TM surface plasmon at distances with $k\rho=\mathcal O(1)$ provided~\eqref{eq:plasm-cond-2} holds.

We now focus on the components of the magnetic field. By~\eqref{eq:B2r-hd-ex-b}, we compute
\begin{align}\label{eq:B2r-hd-app}
B_{2\rho}^{({\rm hd})}&\sim \frac{\mu_0 k^2}{4\pi}\sin\phi\, \Biggl\{ \frac{\omega\mu_0\sigma}{2k}\Biggl(\frac{i}{k\rho}-\frac{1}{k^2\rho^2}-\frac{i}{k^3\rho^3}\Biggr)e^{ik\rho}+\frac{4k^2}{(\omega\mu_0\sigma)^2}\frac{1}{k_m\rho}\sqrt{\frac{2\pi}{k_m\rho}}\,e^{i(k_m\rho-\pi/4)}\nonumber\\
&\mbox{}\quad -\frac{i\pi}{8} \frac{(\omega\mu_0\sigma)^2}{k^2}\biggl[H_1^{(1)\prime}(k_e\rho)-\sqrt{\frac{2}{\pi k_e\rho}}\,e^{i(k_e\rho-\pi/4)}\biggr]-i\frac{(\omega\mu_0\sigma)^2}{4k^2}\sqrt{\frac{\pi}{k\rho}}\,e^{ik\rho}\mathcal F(\wp)\Biggr\}~.
\end{align}
All terms of the second line are negligible; in contrast, the TM surface plasmon can dominate over the $e^{ik\rho}$-wave of the first line under~\eqref{eq:plasm-cond-2}.
Similarly, formula~\eqref{eq:B2ph-hd-ex-b} is reduced to
\begin{align}\label{eq:B2ph-hd-app}
B_{2\phi}^{({\rm hd})}&\sim \frac{\mu_0 k^2}{4\pi}\cos\phi\Biggl\{ \frac{\omega\mu_0\sigma}{k}\,\Biggl(\frac{1}{k^2\rho^2}+\frac{i}{k^3\rho^3}\Biggr)e^{ik\rho}
+\frac{4k^2}{(\omega\mu_0\sigma)^2}\sqrt{\frac{2\pi}{k_m\rho}}\,e^{i(k_m\rho+\pi/4)}
\nonumber\\
&\mbox{}\quad -\frac{i\pi}{8}\frac{(\omega\mu_0\sigma)^2}{k^2}\frac{1}{k\rho}\Biggl[H_1^{(1)}(k_e\rho)-\sqrt{\frac{2}{\pi k_e\rho}}e^{i(k_e\rho-3\pi/4)}\Biggr]-\frac{(\omega\mu_0\sigma)^2}{4k^2}\frac{e^{ik\rho}}{k\rho}\sqrt{\frac{\pi}{k\rho}}\,\mathcal F(\wp)\Biggr\}~,
\end{align}
which suggests the dominance of the TM surface plasmon if~\eqref{eq:plasm-cond-3} holds.

Therefore, on the basis of the approximate formulas of Sections~\ref{subsec:Lateral-vert} and~\ref{subsec:Lateral-horiz}, we reach the following conclusion.
\smallskip

{\em Proposition 1 (On dominance of TM surface plasmon contribution). Suppose that the pole at $\lambda=k_m$ is present in the top $\lambda$-Riemann sheet; and $|\omega\mu_0\sigma|\ll |k|$ which amounts to $\breve\zeta:=|\sigma|\sqrt{\mu_0/|\tilde\epsilon|}\ll 1$, where $k=\omega\sqrt{\mu_0\tilde\epsilon}$. Then, the TM surface plasmon is dominant at distances $\rho$ of the order of the wavelength in free space if condition~\eqref{eq:plasm-cond-2} holds. In particular, if $k>0$ ($\tilde\epsilon >0$) and $\arg\sigma=\pi/2-\delta$, $0<\delta<\pi/2$,  condition~\eqref{eq:plasm-cond-2} reads
\begin{equation}
\breve \zeta \ll \exp\biggl(-\frac{4}{3\breve\zeta}\,\sin\delta\biggr)~.
\end{equation}

}

%%%%%%%%%%%%%%%%%%%%%%%%%%%%%%%%%%%%%%%%%%%%%
%%%%%%%%%%%%%%%%%%%%%%%%%%%%%%%%%%%%%%%%%%%%%
\section{Conclusion}
\label{sec:Conclusion}
%%%%%%%%%%%%%%%%%%%%%%%%%%%%%%%%%%%%%%%%%%%%%
%%%%%%%%%%%%%%%%%%%%%%%%%%%%%%%%%%%%%%%%%%%%%
In this paper, we computed by analytical means the electromagnetic field of a vertical and a horizontal dipole located near a thin film with scalar, complex conductivity, $\sigma$, inside an unbounded medium of wave number $k$. This setting provides a minimal yet nontrivial model of wave propagation on a two-dimensional material such as a graphene sheet.

Our starting point was the Fourier-Bessel representation, or Sommerfeld-type integrals, for all field components. In the particular case when both the dipole and the observation point lie on the layer, we showed that the fields can be expressed in terms of geometrically convergent series and known transcendental functions such as the Fresnel integrals, and the Bessel and Struve functions. We simplified this result considerably when $|\omega\mu_0\sigma|\ll |k|$ and derived analytic formulas that hold for practically all distances, $\rho$, from the source. These formulas connect smoothly the near field with the far field of the dipole, as we verified via independent asymptotic evaluation of the requisite integral representations. Our treatment provides an analytical description of a particular case of the dyadic Green function associated with the geometry of the conducting film; and can expand insights obtained from previous numerical or semi-analytical approaches for related problems.~\cite{Hanson08,Hanson11,Nikitin11}

Depending on the frequency $\omega$ and the parameters $\sigma$ and $k$, the electromagnetic field of the dipole may manifest a contribution from a pole analogous to the TM surface plasmon of plane waves.~\cite{Maier-book} We demonstrated explicitly how this contribution may be manifest or suppressed in each relevant component of the electromagnetic field for all distances from the dipole source.

The present work admits several extensions and also points to pending issues. For example, we have not computed the field components at points away from the film; this task would require an elaborate asymptotic evaluation of the requisite Fourier-Bessel integrals by imposing some restriction on $|z|$. The possible anisotropy of the thin-film material, e.g., black phosphorus,~\cite{Low14} where $\sigma$ is a tensor, deserves attention; this anisotropy dramatically affects the dispersion relations for surface plasmons and the analytic structure of the electromagnetic field produced by current-carrying sources. Another problem is the excitation of surface plasmons by a resonant receiving antenna placed on the material sheet,~\cite{Gonzalez14} which requires solving an integral equation for the associated current distribution; this task is left for future work.

\section*{ACKNOWLEDGMENTS}
This paper is dedicated to the memory of the late Professor Ronold Wyeth Percival King, who  taught the first author (DM) electromagnetic wave propagation. The first author is also greatly indebted to Professor Tai Tsun Wu for his tremendous insights into lateral waves.  The authors wish to thank  Professor George Fikioris for valuable remarks on the manuscript; and Professor Mo Li and Professor Tony Low for illuminating discussions on surface plasmonics.

DM's work has been partially supported by Grant No. 1412769 of the Division of Mathematical Sciences (DMS) of the NSF. ML's work has been partially supported by the Army Research Office (ARO) Multidisciplinary University Research Initiative (MURI)  Award No. W911NF-14-1-0247.

\appendix

%%%%%%%%%%%%%%%%%%%%%%%%%%%%%%%%%%%%%%%%%%%%%
%%%%%%%%%%%%%%%%%%%%%%%%%%%%%%%%%%%%%%%%%%%%%
\section{On the normal field components }
\label{app:normal}
%%%%%%%%%%%%%%%%%%%%%%%%%%%%%%%%%%%%%%%%%%%%%
%%%%%%%%%%%%%%%%%%%%%%%%%%%%%%%%%%%%%%%%%%%%%
In this appendix, we discuss the conditions on the field components normal to the plane, $z=0$, of the film. Such boundary conditions were not explicitly invoked in the derivation of the Fourier-Bessel integral representations (Section~\ref{sec:Fourier}).

First, we describe the (effective) surface charge density of the film which produces a jump of the normal component, $\tilde\epsilon_j E_{jz}$, of the electric displacement field $\mathbf D_j=\tilde\epsilon_j \mathbf E_j$.
By~\eqref{eq:BCs-b}, the thin layer bears the surface {\em current} density $\mathbf J_s=\sigma \mathbf E_\parallel=\sigma \{\mathbf E_j-(\mathbf E_j\cdot \mathbf e_z)\mathbf e_z\}$ at $z=0$, which by the continuity equation is associated with the surface {\em charge} density
$$\varrho_s(x,y)=-(i/\omega)\nabla_s\cdot \mathbf J_s\bigl|_{z=0}=-(i\sigma/\omega)\nabla_s\cdot \mathbf E_\parallel\bigl|_{z=0}~;\quad  \nabla_s=\nabla-(\nabla\cdot \mathbf e_z)\mathbf e_z~.$$
By Gauss' law in region $j$, $\nabla_s\cdot \mathbf E_\parallel=- \partial E_{jz}/\partial z$ for $(x,y,z)\neq (0,0,a)$; allowing $z$ to approach $0$ from region $j$, we conclude that the normal derivative, $\partial E_{jz}/\partial z$, of $E_{jz}$ is continuous across $z=0$. Thus,
\begin{equation}\label{eq:rho-s}
\varrho_s=\frac{i\sigma}{\omega}\,\frac{\partial E_{1z}}{\partial z}\biggl|_{z=0^+}=\frac{i\sigma}{\omega}\,\frac{\partial E_{2z}}{\partial z}\biggl|_{z=0^-}~,
\end{equation}
where writing $z=0^+$ ($z=0^-$) implies that $z=0$ is approached from above (below).

The boundary conditions for the normal field components consist of: (i) the jump of the $z$-component of the electric displacement field due to $\varrho_s$; and (ii) the continuity of the $z$-component of the magnetic field.  Accordingly, we write
\begin{subequations}\label{eq:BCs-normal}
\begin{align}
[k_j^2 E_{jz}]:=&(k_1^2 E_{1z}-k_2^2 E_{2z})|_{z=0}= \omega^2\mu_0 \,\varrho_s=i\omega\mu_0\sigma \,\frac{\partial E_{jz}}{\partial z}\biggl|_{z=0}~,\label{eq:BCs-normal-a}\\
[B_{jz}]:=&(B_{1z}-B_{2z})|_{z=0}=0~.\label{eq:BCs-normal-b}
\end{align}
\end{subequations}

Next, we proceed to demonstrate that~\eqref{eq:BCs-normal} are satisfied by the integral representations of  Section~\ref{sec:Fourier} for an elevated dipole ($a>0$), as expected.

%%%%%%%%%%%%%%%%%%%%%%%%%%%%%%%%%%%%%%%%%%%%%
\subsection{Vertical dipole}
%%%%%%%%%%%%%%%%%%%%%%%%%%%%%%%%%%%%%%%%%%%%%
By differentiation in $z$ of~\eqref{eq:E1z-vd} for $0<z<a$ and~\eqref{eq:E2z-vd} for $z<0$, we obtain
\begin{equation}\label{eq:Ez-vd-norm-deriv}
\frac{\partial E_{1z}}{\partial z}\biggl|_{z=0^+}=\frac{i\omega\mu_0}{2\pi}\int_0^\infty \dl\,\lambda^3\,J_0(\lambda\rho)\ \frac{\beta_2(\lambda)}{\mathcal P(\lambda)}\,e^{i\beta_1 a}=\frac{\partial E_{2z}}{\partial z}\biggl|_{z=0^-}~,
\end{equation}
which shows the continuity of the normal derivative of $E_{jz}$.
On the other hand, we compute
\begin{align*}
[k_j^2 E_{jz}]&=-\frac{\omega\mu_0}{4\pi}\int_0^\infty \dl\ \frac{\lambda^3}{\beta_1}\,J_0(\lambda\rho)\,\Biggl(1+\frac{k_2^2\beta_1-k_1^2\beta_2+\omega\mu_0\sigma\beta_1\beta_2}{k_2^2\beta_1+k_1^2\beta_2+\omega\mu_0\sigma\beta_1\beta_2}\Biggr)\,e^{i\beta_1a}\nonumber\\
&\mbox{} \qquad +\frac{\omega\mu_0}{2\pi} \int_0^\infty \dl\,\lambda^3\,J_0(\lambda\rho)\ \frac{k_2^2}{k_2^2\beta_1+k_1^2\beta_2+\omega\mu_0\sigma\beta_1\beta_2}\,e^{i\beta_1 a}\nonumber\\
&= -\frac{\omega\mu_0}{2\pi}\,(\omega\mu_0\sigma) \int_0^\infty \dl\,\lambda^3\,J_0(\lambda\rho)\ \frac{\beta_2}{\mathcal P}\,e^{i\beta_1 a}~,
\end{align*}
which, by~\eqref{eq:Ez-vd-norm-deriv}, is equal to $i\omega\mu_0\sigma (\partial E_{jz}/\partial z)$ at $z=0$. Thus, condition~\eqref{eq:BCs-normal-a} is satisfied. Note that~\eqref{eq:BCs-normal-b} is trivially satisfied in this case because $B_{jz}\equiv 0$.

%%%%%%%%%%%%%%%%%%%%%%%%%%%%%%%%%%%%%%%%%%%%%
\subsection{Horizontal dipole}
%%%%%%%%%%%%%%%%%%%%%%%%%%%%%%%%%%%%%%%%%%%%%
By differentiating~\eqref{eq:E1z-hd} and~\eqref{eq:E2z-hd}, we verify that $\partial E_{jz}/\partial z$ is continuous across $z=0$:
\begin{equation}\label{eq:Ez-hd-norm-deriv}
\frac{\partial E_{1z}}{\partial z}\biggl|_{z=0^+}=-\frac{\omega\mu_0}{2\pi}\cos\phi \int_0^\infty \dl\,\lambda^2\,J_1(\lambda\rho)\ \frac{\beta_1\beta_2}{\mathcal P}\,e^{i\beta_1 a}=\frac{\partial E_{2z}}{\partial z}\biggl|_{z=0^-}~.
\end{equation}
On the other hand, the jump of $k_j^2 E_{jz}$ is equal to
\begin{align*}
[k_j^2 E_{jz}]&= \frac{i\omega\mu_0}{4\pi}\cos\phi \int_0^\infty \dl\,\lambda^2\,J_1(\lambda\rho)\,\Biggl(\frac{k_1^2\beta_2-k_2^2\beta_1-\omega\mu_0\sigma\beta_1\beta_2}{k_1^2\beta_2+k_2^2\beta_1+\omega\mu_0\sigma\beta_1\beta_2}-1\Biggr)\,e^{i\beta_1 a}\nonumber\\
& + \frac{i\omega\mu_0}{2\pi}\,\cos\phi \int_0^\infty \dl\,\lambda^2\,J_1(\lambda\rho)\ \frac{k_2^2\beta_1}{k_1^2\beta_2+k_2^2\beta_1+\omega\mu_0\sigma\beta_1\beta_2}\,e^{i\beta_1 a}~,
\end{align*}
which, after elementary algebra, is equal to $i\omega\mu_0\sigma  (\partial E_{jz}/\partial z)$ at $z=0$; cf.~\eqref{eq:Ez-hd-norm-deriv}.

We now turn our attention to $B_{jz}$. By~\eqref{eq:B1z-hd}, as $z\downarrow 0$ (with  $0<z<a$) we obtain
\begin{equation*}
B_{1z}|_{z=0^+}=\frac{i\mu_0}{4\pi}\sin\phi\int_0^\infty \dl\,\lambda^2\,J_1(\lambda\rho)\frac{1}{\beta_1}\Biggl(1-\frac{\beta_2-\beta_1+\omega\mu_0\sigma}{\beta_2+\beta_1+\omega\mu_0\sigma}\Biggr) e^{i\beta_1 a}~,
\end{equation*}
which is equal to $B_{2z}$ at $z=0^-$ by~\eqref{eq:B2z-hd}. This shows the expected validity of~\eqref{eq:BCs-normal-b}.

%%%%%%%%%%%%%%%%%%%%%%%%%%%%%%%%%%%%%%%%%%%%%
%%%%%%%%%%%%%%%%%%%%%%%%%%%%%%%%%%%%%%%%%%%%%
\section{Asymptotic evaluation of Sommerfeld-type integrals}
\label{app:asymp}
%%%%%%%%%%%%%%%%%%%%%%%%%%%%%%%%%%%%%%%%%%%%%
%%%%%%%%%%%%%%%%%%%%%%%%%%%%%%%%%%%%%%%%%%%%%
In this appendix, we derive the near- and far-field of each dipole by asymptotic evaluation of the Sommerfeld-type integrals under $|\omega\mu_0\sigma|\ll |k|$. The results of these calculations are compared to the exact formulas of Section~\ref{subsec:Eval-asym}. We assume that the pole at $\lambda=k_m$ is present in the physical Riemann sheet throughout this appendix (see Remarks~1--3).

In the following analysis, we will need the relations~\cite{Bateman-II}
\begin{equation}\label{eq:J}
J_0(z)+J_2(z)=\frac{2}{z}\,J_1(z)~,\quad J_0(z)-J_2(z)=2\,J_1'(z)~,
\end{equation}
along with the leading-order asymptotic formula~\cite{Bateman-II}
\begin{equation}\label{eq:J-asymp}
J_\nu (z)\sim \sqrt{\frac{2}{\pi z}}\,\cos(z-\nu\pi/2-\pi/4)\quad\mbox{as}\ |z|\to\infty~,\quad |\arg z|< \pi~.
\end{equation}

In addition, we will make use of the following result from Ref.~\onlinecite{KOW}.

{\em Lemma B.1 (On the Fresnel integrals). The integral
\begin{equation}
\mathfrak I:=\int_{-\infty}^\infty d\tau \ \frac{e^{i(k_e-k)\rho\tau}}{\sqrt{\tau} (\omega\mu_0\sigma+ i2\sqrt{2k}\sqrt{k_e-k}\sqrt{\tau})}=\frac{\pi}{\sqrt{k}\sqrt{k_e-k}} e^{i\pi/4}\mathcal F(\wp_0)~,
\end{equation}
where $\sqrt{\tau}>0$ for $\tau>0$, $\mathcal F(z)$ is defined in~\eqref{eq:F0}, and $\wp_0=-(\omega\mu_0\sigma)^2\rho /(8k)$ [see~\eqref{eq:ke-nonr} and~\eqref{eq:Som-dist}].}
\smallskip

{\em Proof.} By Ref.~\onlinecite{KOW}, first write
\begin{equation*}
\mathfrak I=\int_0^\infty d\tau \ \frac{e^{i(k_e-k)\rho\tau}}{\sqrt{\tau}(\omega\mu_0\sigma +i2\sqrt{2k}\sqrt{k_e-k}\sqrt{\tau})}+\int_0^\infty d\tau \ \frac{e^{-i(k_e-k)\rho\tau}}{-i\sqrt{\tau}(\omega\mu_0\sigma +2\sqrt{2k}\sqrt{k_e-k}\sqrt{\tau})}~.
\end{equation*}
Then, apply $\tau\mapsto \varsigma$ with $\sqrt{\varsigma}=2i\sqrt{2k}\sqrt{k_e-k}(\omega\mu_0\sigma)^{-1}\sqrt{\tau}$ to find
\begin{equation*}
\mathfrak I=\frac{i}{2\sqrt{2k}\sqrt{k_e-k}}\mathfrak I_0(\wp_0)~,
\end{equation*}
where
\begin{equation*}
\mathfrak I_0(z)=-\int_0^\infty d\varsigma\ \frac{e^{i z\varsigma}}{\sqrt{\varsigma}(\sqrt{\varsigma}+1)}+\int_0^\infty d\varsigma \ \frac{e^{-i z \varsigma}}{\sqrt{\varsigma}(\sqrt{\varsigma}+i)}~.
\end{equation*}

This $\mathfrak I_0(z)$ satisfies
\begin{equation}
\frac{d\mathfrak I_0}{d z}-i\mathfrak I_0(z)=-2e^{-i\pi/4}\sqrt{\frac{\pi}{z}}~,\quad \lim_{z\to +\infty}\mathfrak I_0(z)=0~,
\end{equation}
with solution $\mathfrak I_0(z)=2\pi\sqrt{2}\,e^{-i\pi/4} \mathcal F(z)$, which concludes the proof.  $\square$

\subsection{Far field: $|k\rho|\gg 1$ with $|\omega\mu_0\sigma|\ll |k|$}
In this regime, the major contribution to integration in the Sommerfeld-type integrals comes from the singularities at $\lambda=k$ and $\lambda=k_m$. The Bessel functions in all integrands are expressed in terms of $J_\nu(\lambda\rho)$ for $\nu=0,\,1$ by use of~\eqref{eq:J}; $J_\nu (\lambda\rho)$ is in turn replaced by its large-argument approximation according to~\eqref{eq:J-asymp}.

\subsubsection{Field of vertical dipole on film}
Taking the limit as $a\downarrow 0$ with $0<z<a$ of~\eqref{eq:E1r-vd} and then invoking~\eqref{eq:J-asymp} readily yield
\begin{align}
E_{1\rho}&=-\frac{i\omega \mu_0}{2\pi}\int_0^\infty \dl\ \lambda^2\,J_1(\lambda\rho)\ \frac{1}{2k^2+\omega\mu_0\sigma\sqrt{k^2-\lambda^2}}\nonumber\\
&\stackrel{|k\rho|\gg 1}{\sim} -\frac{i\omega\mu_0}{4\pi}\int_{-\infty}^\infty \dl\ \lambda^2\ \sqrt{\frac{2}{\pi\lambda\rho}}\ \frac{1}{2k^2+\omega\mu_0\sigma \sqrt{k^2-\lambda^2}}\,e^{i(\lambda\rho-3\pi/4)}~.
\end{align}
In  this integral, we single out the contributions from the simple pole at $\lambda=k_m$ and the branch point at $\lambda=k$, viz.,
\begin{align*}
E_{1\rho}^{({\rm vd})}&\sim -\frac{i\omega\mu_0}{4\pi}\Biggl\{2\pi i\,k_m^2 \sqrt{\frac{2}{\pi k_m\rho}}\, e^{i(k_m\rho-3\pi/4)} \lim_{\lambda\to k_m} \Biggl[\frac{2k^2-\omega\mu_0\sigma\sqrt{k^2-\lambda^2}}{(\omega\mu_0\sigma)^2 (2\lambda)}\Biggr]   \nonumber\\
&\mbox{} +ik^3\int_0^\infty d\tau\,(1+i\tau)^2\Biggl[\frac{1}{2k^2-e^{-i\pi/4}\,\omega\mu_0\sigma k \sqrt{\tau}\sqrt{2+i\tau}}-\frac{1}{2k^2+e^{-i\pi/4}\,\omega\mu_0\sigma k \sqrt{\tau}\sqrt{2+i\tau}}\Biggr]\nonumber\\
&\mbox{}\qquad \times \sqrt{\frac{2}{\pi k\rho(1+i\tau)}}\,e^{i(k\rho-3\pi/4)}\,e^{-k\rho\tau}\Biggr\}~,
\end{align*}
where for the last integral we appropriately deformed the integration path in the upper $\lambda$-half-plane, set $\lambda=k(1+i\tau)$ and integrated on each side of the positive real $\tau$-axis.  The major contribution to integration comes from the vicinity of $\tau=0$ with width $\mathcal O((k\rho)^{-1})$; the corresponding integrand yields a term of the order of $\omega\mu_0 \sigma/k$. By enforcing the condition $|\omega\mu_0\sigma|\ll |k|$, we reduce the preceding formula for $E_{1\rho}^{({\rm vd})}$ to
\begin{align}\label{eq:E1r-lg-kr}
E_{1\rho}^{({\rm vd})}&\sim \frac{i\omega\mu_0 k}{4\pi}\Biggl\{\frac{k_m^3}{k^3}\,\sqrt{\frac{2\pi}{k_m\rho}} \,e^{i(k_m\rho-\pi/4)}\nonumber\\
&\mbox{} -i\,\frac{\omega\mu_0\sigma}{2k}\sqrt{2} e^{-i\pi/4}\sqrt{\frac{2}{\pi k\rho}}\,e^{i(k\rho-3\pi/4)}\int_0^\infty d\tau\,\sqrt{\tau}\,e^{-k\rho\tau}\Biggr\}\nonumber\\
&= \frac{i\omega\mu_0 k}{4\pi} \Biggl\{ \frac{k_m^3}{k^3}\,\sqrt{\frac{2\pi}{k_m\rho}} \,e^{i(k_m\rho-\pi/4)}+\frac{i\omega\mu_0\sigma}{2k}\,\frac{e^{ik\rho}}{k^2\rho^2}\Biggr\}~,
\end{align}
in agreement with~\eqref{eq:E1r-vd-ex-b} which comes from the exact series expansion for this component.

In the same vein, by~\eqref{eq:E1z-vd} and asymptotic formula~\eqref{eq:J-asymp} for $\nu=0$ we have
\begin{align}
E_{1z}^{({\rm vd})}&=-\frac{\omega\mu_0}{2\pi k^2}\int_0^\infty \dl\ J_0(\lambda\rho)\ \frac{\lambda^3}{\sqrt{k^2-\lambda^2}}\, \frac{k^2+\omega\mu_0\sigma\sqrt{k^2-\lambda^2}}{2k^2+\omega\mu_0\sigma\sqrt{k^2-\lambda^2}}\nonumber\\
&\mbox{}\stackrel{|k\rho|\gg 1}{\sim} -\frac{\omega\mu_0}{4\pi k^2}\int_{-\infty}^\infty \dl\ \frac{\lambda^3}{\sqrt{k^2-\lambda^2}}\ \frac{k^2+\omega\mu_0\sigma\sqrt{k^2-\lambda^2}}{2k^2+\omega\mu_0\sigma\sqrt{k^2-\lambda^2}}\sqrt{\frac{2}{\pi\lambda\rho}}\,e^{i(\lambda\rho-\pi/4)}~.
\end{align}
By singling out the contributions from two singular points, at $\lambda=k_m$ and $\lambda=k$, and additionally imposing $|\omega\mu_0\sigma|\ll |k|$, after some algebra we find
\begin{align}\label{eq:E1z-lg-kr}
E_{1z}^{({\rm vd})}&\sim  \frac{\omega\mu_0}{4\pi}  \Biggl\{2\pi i\frac{4k^4}{(\omega\mu_0\sigma)^3}\sqrt{\frac{2}{\pi k_m\rho}}e^{i(k_m\rho-\pi/4)}\nonumber\\
&\mbox{} -ik^2\int_0^\infty d\tau\,\sqrt{\frac{2}{\pi k\rho(1+i\tau)}}(1+i\tau)^3 e^{i(k\rho-\pi/4)}\,e^{-(k\rho)\tau}\Biggl[\frac{1}{-e^{-i\pi/4}\,k\sqrt{2+i\tau}\sqrt{\tau}}\nonumber\\
&\mbox{} \times \frac{k^2-e^{-i\pi/4}\,\omega\mu_0\sigma k \sqrt{\tau}\sqrt{2+i\tau}}{2k^2-e^{-i\pi/4}\,\omega\mu_0\sigma k \sqrt{\tau}\sqrt{2+i\tau}}
-\frac{1}{e^{-i\pi/4}\,k\sqrt{2+i\tau}\sqrt{\tau}}
\nonumber\\
&\mbox{} \times \frac{k^2+e^{-i\pi/4}\,\omega\mu_0\sigma k \sqrt{\tau}\sqrt{2+i\tau}}{2k^2+e^{-i\pi/4}\,\omega\mu_0\sigma k \sqrt{\tau}\sqrt{2+i\tau}}\Biggr]\nonumber\\
&\sim \frac{\omega\mu_0 k}{4\pi} \Biggl\{ 8\pi i\frac{k^3}{(\omega\mu_0\sigma)^3}\sqrt{\frac{2}{\pi k_m\rho}}\,e^{i(k_m\rho-\pi/4)}+\frac{i}{k\rho}\,e^{ik\rho}\Biggr\}~,
\end{align}
by expanding the integrand at $\tau=0$. The preceding result is
in agreement with~\eqref{eq:E1z-vd-ex} of the exact solution for $|\omega\mu_0\sigma|\ll |k|$ and $|k\rho|\gg 1$. By a similar computation, \eqref{eq:B-reg1-vd} yields
\begin{align}\label{eq:B1ph-lg-kr}
B_{1\phi}^{({\rm vd})}&= \frac{i\mu_0}{2\pi}\int_0^\infty \dl\ J_1(\lambda\rho)\ \frac{\lambda^2}{\sqrt{k^2-\lambda^2}}\ \frac{k^2+\omega\mu_0\sigma\sqrt{k^2-\lambda^2}}{2k^2+\omega\mu_0\sigma\sqrt{k^2-\lambda^2}}\nonumber\\
&\stackrel{|k\rho|\gg 1}{\sim} \frac{i\mu_0}{4\pi}\int_{-\infty}^\infty \dl\ \frac{\lambda^2}{\sqrt{k^2-\lambda^2}}\ \frac{k^2+\omega\mu_0\sigma\sqrt{k^2-\lambda^2}}{2k^2+\omega\mu_0\sigma\sqrt{k^2-\lambda^2}}\,\sqrt{\frac{2}{\pi\lambda\rho}}\,e^{i(\lambda\rho-3\pi/4)}\nonumber\\
&\stackrel{|\omega\mu_0\sigma|\ll |k|}{\sim} \mbox{} -\frac{i\mu_0 k^2}{4\pi}\Biggl\{ 4\pi \,\frac{k^2}{(\omega\mu_0\sigma)^2}\sqrt{\frac{2}{\pi k_m\rho}}\,e^{i(k_m\rho-3\pi/4)}+\frac{1}{k\rho}\,e^{ik\rho}\Biggr\}~,
\end{align}
which, alternatively,  results from~\eqref{eq:B1ph-vd-ex} of the exact solution if $|\omega\mu_0\sigma|\ll |k|$ and $|k\rho|\gg 1$.

\subsubsection{Field of horizontal dipole on film}
In this case, the contribution from the branch point at $\lambda=k$ should be affected by the proximity of the pole at $\lambda=k_e$  if $|(k_e-k)\rho|\le \mathcal O(1)$. This effect is taken into account via the integral of Lemma B.1, where $(k_e-k)\rho\sim \wp_0$ enters as a parameter. This situation is typical in radiowave propagation on the boundary separating media of significantly different wavenumbers.~\cite{KOW}

In view of~\eqref{eq:J} and~\eqref{eq:J-asymp}, the limit as $a\downarrow 0$ and $z\uparrow 0$ of~\eqref{eq:E2r-hd} yields
\begin{align}
E_{2\rho}^{({\rm hd})}&=-\frac{\omega\mu_0}{2\pi}\cos\phi \int_0^\infty \dl\,\Biggl\{\frac{1}{\rho}\,J_1(\lambda\rho)\ \frac{1}{2\sqrt{k^2-\lambda^2}+\omega\mu_0\sigma}+\lambda J_1'(\lambda\rho)\frac{\sqrt{k^2-\lambda^2}}{2k^2+\omega\mu_0\sigma\sqrt{k^2-\lambda^2}}\Biggr\}\nonumber\\
&\stackrel{|k\rho|\gg 1}{\sim} -\frac{\omega\mu_0}{4\pi}\cos\phi \Biggl\{\int_{-\infty}^\infty \dl\ \frac{1}{\rho}\sqrt{\frac{2}{\pi\lambda\rho}}\,e^{i(\lambda\rho-3\pi/4)}\ \frac{1}{2\sqrt{k^2-\lambda^2}+\omega\mu_0\sigma}\nonumber\\
&\hphantom{-\frac{\omega\mu_0}{4\pi}\cos\phi \Biggl\{} +\int_{-\infty}^\infty \dl\,i\lambda\sqrt{\frac{2}{\pi\lambda\rho}}\,e^{i(\lambda\rho-3\pi/4)}\,\frac{\sqrt{k^2-\lambda^2}}{2k^2+\omega\mu_0\sigma\sqrt{k^2-\lambda^2}}\Biggr\}~.
\end{align}
There is only one residue contribution, $E_{2\rho,k_m}^{({\rm hd})}$, to this component, which comes from the last integral and concerns the simple pole at $\lambda=k_m$:
\begin{equation*}
E_{2\rho,k_m}^{({\rm hd})}\sim -\frac{\omega\mu_0}{4\pi}\cos\phi \,2\pi i\Biggl\{ ik_m\sqrt{\frac{2}{\pi k_m\rho}} \, e^{i(k_m\rho-3\pi/4)}\,\sqrt{k^2-k_m^2}\lim_{k\to k_m}\Biggl[\frac{2k^2-\omega\mu_0\sigma \sqrt{k^2-\lambda^2}}{(\omega\mu_0\sigma)^2\,(2\lambda)}\Biggr]\Biggr\}~.
\end{equation*}
The contribution from the branch point at $\lambda=k$ is singled out as
\begin{align*}
E_{2\rho,k}^{({\rm hd})}&\sim -\frac{\omega\mu_0}{4\pi}\cos\phi\, e^{i(k\rho-3\pi/4)}\sqrt{\frac{2}{\pi k\rho}}\Biggl\{\frac{k_e-k}{\rho}
\int_{-\infty}^\infty d\tau'\, \biggl(1+\frac{k_e-k}{k}\tau'\biggr)^{-1/2}e^{i(k_e-k)\rho \tau'}\nonumber\\
&\mbox{} \times \frac{1}{2i\sqrt{2k}\sqrt{k_e-k}\sqrt{1+\frac{k_e-k}{2k}\tau'}\,\sqrt{\tau'}+\omega\mu_0\sigma}+(ik)^2 \int_{0}^\infty d\tau\, \sqrt{1+i\tau}\,e^{-(k\rho)\tau}\nonumber\\
&\mbox{} \quad \times \Biggl[\frac{-e^{-i\pi/4}\,k \sqrt{2+i\tau}\sqrt{\tau}}{2k^2-e^{-i\pi/4}\,\omega\mu_0\sigma k \sqrt{2+i\tau}\sqrt{\tau}}-\frac{e^{-i\pi/4}\,k \sqrt{2+i\tau} \sqrt{\tau}}{2k^2+e^{-i\pi/4}\,\omega\mu_0\sigma k \sqrt{2+i\tau}\sqrt{\tau}}\Biggr]\Biggr\}~.
\end{align*}
For the first integral above, we have set $\lambda=k+(k_e-k)\tau'$; whereas for the second integral we deformed the path in the upper $\lambda$-half-plane and then set $\lambda=k(1+i\tau)$ as before.
In each of the transformed integrals, the major contribution arises from a small neighborhood of the origin. The value of the first integral  crucially depends on which term in its denominator prevails when $(k_e-k)\rho\tau' =\mathcal O(1)$; by this interplay, the parameter $\wp_0$ emerges naturally. In contrast, in the second integral such interplay is not present; one may simply Taylor-expand the integrand in powers of $\omega\mu_0\sigma/k$ and keep the leading-order term. To subtract the primary field contribution in the first integral, we split each term, evaluated near $\tau'=0$, as
\begin{equation*}
\frac{1}{2i\sqrt{2k}\sqrt{k_e-k}\sqrt{\tau'}+\omega\mu_0\sigma}= \frac{1}{2i\sqrt{2k}\sqrt{k_e-k}\sqrt{\tau'}}- \frac{\omega\mu_0\sigma}{2i\sqrt{2k}\sqrt{k_e-k} \sqrt{\tau'}\big(2i\sqrt{2k}\sqrt{k_e-k}\sqrt{\tau'}+\omega\mu_0\sigma\big)}~
\end{equation*}
and carry out the integration without further approximations for these terms. Thus, we find
\begin{align*}
E_{2\rho,k}^{({\rm hd})}&\stackrel{|\omega\mu_0\sigma|\ll |k|}{\sim} -\frac{\omega\mu_0}{4\pi}\,\cos\phi \Biggl\{ -\frac{2}{k\rho^2}\,e^{ik\rho}-\frac{k_e-k}{\rho}\,\frac{\omega\mu_0\sigma}{2i\sqrt{2k}\sqrt{k_e-k}}\,\sqrt{\frac{2}{\pi k\rho}} \,e^{i(k\rho-3\pi/4)}\,\mathfrak I\Biggr\}~,
\end{align*}
where $\mathfrak I$ is given in Lemma B.1. Thus, by $E_{2\rho}^{({\rm hd})}\sim E_{2\rho,k_m}^{({\rm hd})}+E_{2\rho,k}^{({\rm hd})}$ we compute
\begin{equation}\label{eq:E2r-lg-kr}
E_{2\rho}^{({\rm hd})}\sim -\frac{\omega\mu_0}{4\pi}\cos\phi \Biggl\{\frac{8\,k^4}{(\omega\mu_0\sigma)^3} \sqrt{\frac{2\pi }{k_m\rho}} e^{i(k_m\rho-3\pi/4)}-\frac{2 e^{ik\rho}}{k\rho^2}+\frac{\omega\mu_0\sigma}{2k\rho}\sqrt{\frac{\pi}{k\rho}} e^{ik\rho}\mathcal F(\wp_0)\Biggr\}~,
\end{equation}
which is in agreement with~\eqref{eq:E2r-hd-ex-b} for $|k\rho|\gg 1$.

The remaining components can be approximately computed by the same methodology, although the algebraic details are different in each case. By~\eqref{eq:E2ph-hd}, we have
\begin{align}
E_{2\phi}^{({\rm hd})}&=\frac{\omega\mu_0}{2\pi}\sin\phi \int_0^\infty \dl\,\Biggl\{\lambda\,J_1'(\lambda\rho)\ \frac{1}{2\sqrt{k^2-\lambda^2}+\omega\mu_0\sigma}+\frac{1}{\rho}\, J_1(\lambda\rho)\frac{\sqrt{k^2-\lambda^2}}{2k^2+\omega\mu_0\sigma\sqrt{k^2-\lambda^2}}\Biggr\}\nonumber\\
&\stackrel{|k\rho|\gg 1}{\sim} \frac{\omega\mu_0}{4\pi}\sin\phi \Biggl\{\int_{-\infty}^\infty \dl\ i\lambda\,\sqrt{\frac{2}{\pi\lambda\rho}}\,e^{i(\lambda\rho-3\pi/4)}\ \frac{1}{2\sqrt{k^2-\lambda^2}+\omega\mu_0\sigma}\nonumber\\
&\hphantom{-\frac{\omega\mu_0}{4\pi}\cos\phi \Biggl\{} +\int_{-\infty}^\infty \dl\,\frac{1}{\rho}\,\sqrt{\frac{2}{\pi\lambda\rho}}\,e^{i(\lambda\rho-3\pi/4)}\,\frac{\sqrt{k^2-\lambda^2}}{2k^2+\omega\mu_0\sigma\sqrt{k^2-\lambda^2}}\Biggr\}~.
\end{align}
By separating the pole and branch-point contributions, we find
\begin{align}\label{eq:E2ph-lg-kr}
E_{2\phi}^{({\rm hd})}&\stackrel{|\omega\mu_0\sigma|\ll |k|}{\sim} \frac{\omega\mu_0}{4\pi}\sin\phi\Biggl\{ -\frac{8\pi i k^4}{(\omega\mu_0\sigma)^3}\frac{1}{k_m\rho}\sqrt{\frac{2}{\pi k_m\rho}} e^{i(k_m\rho-3\pi/4)}+i\Biggr(-\frac{1}{\rho}+\frac{1}{k^2\rho^3}\Biggr)e^{ik\rho}\nonumber\\
&\mbox{} \qquad -\frac{\omega\mu_0\sigma}{2\sqrt{2}}\,\sqrt{k_e-k}\sqrt{k}\,\sqrt{\frac{2}{\pi k\rho}}\,e^{i(k\rho-3\pi/4)}\,\mathfrak I\Biggr\}~,
\end{align}
which, by Lemma~B.1, is in agreement with~\eqref{eq:E2ph-hd-ex-b} if $|k\rho|\gg 1$. We note in passing that the $e^{ik\rho}/(k^2\rho^3)$ term in the first line of~\eqref{eq:E2ph-lg-kr} arises from the (TM-polarization) integral containing $(2k^2+\omega\mu_0\sigma\sqrt{k^2-\lambda^2})^{-1}$, is small compared to  $e^{ik\rho}/\rho$ in this far-field regime and, thus, should be dropped to leading order in $k\rho$.

By~\eqref{eq:E2z-hd}, the $z$-component of the electric field does not contain any TE polarization term and thus lacks the Fresnel integrals. By similar asymptotics, we reach the following result.
\begin{align}\label{eq:E2z-lg-kr}
E_{2z}^{({\rm hd})}&=  -\frac{i\omega\mu_0}{2\pi} \cos\phi \int_0^\infty \dl\,\lambda^2\,J_1(\lambda\rho)\ \frac{1}{2k^2+\omega\mu_0\sigma\sqrt{k^2-\lambda^2}}\nonumber\\
&\mbox{} \stackrel{|k\rho|\gg 1}{\sim} -\frac{i\omega\mu_0}{4\pi}\,\cos\phi\int_{-\infty}^\infty \dl\,\lambda^2\,\sqrt{\frac{2}{\pi\lambda\rho}}\,e^{i(\lambda\rho-3\pi/4)}\,\frac{1}{2k^2+\omega\mu_0\sigma\sqrt{k^2-\lambda^2}}\nonumber\\
&\stackrel{|\omega\mu_0\sigma|\ll |k|}{\sim} -\frac{i\omega\mu_0}{4\pi}\cos\phi\,\Biggl\{ 4\pi i k_m\frac{k^2}{(\omega\mu_0\sigma)^2}\sqrt{\frac{2}{\pi k_m\rho}}\,e^{i(k_m\rho-3\pi/4)}-\frac{i\omega\mu_0\sigma}{2}\frac{e^{ik\rho}}{k^2\rho^2}\Biggr\}~;
\end{align}
cf.~\eqref{eq:E2z-hd-ex} from the exact solution, recalling that $k_m\sim i 2k^2/(\omega\mu_0\sigma)$.

Next, we turn attention to the components of the magnetic field. By~\eqref{eq:B2r-hd}, we have
\begin{align}
B_{2\rho}^{({\rm hd})}&=\frac{\mu_0}{2\pi}\sin\phi \int_0^\infty \dl\,\Biggl\{\frac{1}{\rho}\,J_1(\lambda\rho)\ \frac{k^2}{2k^2+\omega\mu_0\sigma\sqrt{k^2-\lambda^2}}+\lambda J_1'(\lambda\rho)\frac{\sqrt{k^2-\lambda^2}}{2\sqrt{k^2-\lambda^2}+\omega\mu_0\sigma}\Biggr\}\nonumber\\
&\stackrel{|k\rho|\gg 1}{\sim} \frac{\mu_0}{4\pi}\sin\phi \Biggl\{\int_{-\infty}^\infty \dl\ \frac{1}{\rho}\sqrt{\frac{2}{\pi\lambda\rho}}\,e^{i(\lambda\rho-3\pi/4)}\ \frac{k^2}{2k^2+\omega\mu_0\sigma\sqrt{k^2-\lambda^2}}\nonumber\\
&\hphantom{-\frac{\omega\mu_0}{4\pi}\cos\phi \Biggl\{} -\int_{-\infty}^\infty \dl\,\frac{i\lambda}{2}\sqrt{\frac{2}{\pi\lambda\rho}}\,e^{i(\lambda\rho-3\pi/4)}\,\frac{\omega\mu_0\sigma}{2\sqrt{k^2-\lambda^2}+\omega\mu_0\sigma}\Biggr\}~;
\end{align}
in the third line we replaced $\sqrt{k^2-\lambda^2}/(2\sqrt{k^2-\lambda^2}+\omega\mu_0\sigma)$ by $1/2-(\omega\mu_0\sigma/2)/(2\sqrt{k^2-\lambda^2}+\omega\mu_0\sigma)$ and integrated out the constant. For the first integral, we deform the path in the upper $\lambda$-half-plane, picking up the residue from $\lambda=k_m$, and then evaluate the contribution from the vicinity of $\lambda=k$. For the second integral, we set $\lambda\mapsto \tau$ with  $\tau=(\lambda-k)/(k_e-k)$ and evaluate the contribution from the vicinity of $\tau=0$, ending up with Fresnel integrals via Lemma B.1. Accordingly, we derive the expression
\begin{align}\label{eq:B2r-lg-kr}
B_{2\rho}^{({\rm hd})}&\sim \frac{\mu_0}{4\pi}\sin\phi \Biggl\{\frac{4\pi i \,k^4}{(\omega\mu_0\sigma)^2}\frac{1}{k_m\rho}\sqrt{\frac{2}{\pi k_m\rho}}\,e^{i(k_m\rho-3\pi/4)}+\frac{i\omega\mu_0\sigma}{2}\Biggl(\frac{1}{\rho}-\frac{1}{k^2\rho^3}\Biggr)e^{ik\rho}\nonumber\\
&\hphantom{\frac{\mu_0}{4\pi}\sin\phi \Biggl\{ } -i\frac{(\omega\mu_0\sigma)^2}{4}\sqrt{\frac{\pi}{k\rho}}\,e^{ik\rho}\,\mathcal F(\wp_0)\Biggr\}~;
\end{align}
the $e^{ik\rho}/(k^2\rho^3)$ term in the parenthesis of the first line comes from the TM-polarization integral and is neglected. The ensuing asymptotic result can also be obtained from~\eqref{eq:B2r-hd-ex-b} in the far-field regime, $|k\rho|\gg 1$.

In this vein, by~\eqref{eq:B2ph-hd} we compute the following expression [cf.~\eqref{eq:B2ph-hd-ex-b}].
\begin{align}\label{eq:B2ph-lg-kr}
B_{2\phi}^{({\rm hd})}&=\frac{\mu_0}{2\pi}\cos\phi \int_0^\infty \dl\,\Biggl\{\lambda\,J_1'(\lambda\rho)\ \frac{k^2}{2 k^2+\omega\mu_0\sigma\sqrt{k^2-\lambda^2}}+\frac{1}{\rho}\, J_1(\lambda\rho)\frac{\sqrt{k^2-\lambda^2}}{2\sqrt{k^2-\lambda^2}+\omega\mu_0\sigma}\Biggr\}\nonumber\\
&\stackrel{|k\rho|\gg 1}{\sim} \frac{\mu_0}{4\pi}\cos\phi \Biggl\{\int_{-\infty}^\infty \dl\ i\lambda\,\sqrt{\frac{2}{\pi\lambda\rho}}\,e^{i(\lambda\rho-3\pi/4)}\ \frac{k^2}{2k^2 +\omega\mu_0\sigma\sqrt{k^2-\lambda^2}}\nonumber\\
&\hphantom{-\frac{\omega\mu_0}{4\pi}\cos\phi \Biggl\{} -\int_{-\infty}^\infty \dl\,\frac{1}{2\rho}\,\sqrt{\frac{2}{\pi\lambda\rho}}\,e^{i(\lambda\rho-3\pi/4)}\,\frac{\omega\mu_0\sigma}{2\sqrt{k^2-\lambda^2} +\omega\mu_0\sigma}\Biggr\}\nonumber\\
&\stackrel{|\omega\mu_0\sigma|\ll |k|}{\sim} \frac{\mu_0}{4\pi}\cos\phi \Biggl\{ -\frac{4\pi\,k^4}{(\omega\mu_0\sigma)^2}\,\sqrt{\frac{2}{\pi k_m\rho}}\,e^{i(k_m\rho-3\pi/4)}+\omega\mu_0\sigma\, \frac{e^{ik\rho}}{k\rho^2}\nonumber\\
&\mbox{} \hphantom{\frac{\mu_0}{4\pi}\cos\phi \Biggl\{  } \qquad -\frac{(\omega\mu_0\sigma)^2}{4}\frac{1}{k\rho}\sqrt{\frac{\pi}{k\rho}}\,e^{ik\rho}\,\mathcal F(\wp_0)\Biggr\}~.
\end{align}
Here, the branch-point contribution for each polarization (TE and TM) integral carries a free-space term proportional to $[\omega\mu_0 \sigma/(k\rho^2)]e^{ik\rho}$, besides the Fresnel integral contribution.

We conclude this subsection with the approximate computation of the $z$-component of the magnetic field by similar means. This component expresses only TE polarization and, thus, lacks any pole contribution (by Remark~3). By~\eqref{eq:B2z-hd} and similar algebraic manipulations as above, we find
\begin{align}\label{eq:B2z-lg-kr}
B_{2z}^{({\rm hd})}&=  \frac{i\mu_0}{2\pi} \sin\phi \int_0^\infty \dl\,\lambda^2\,J_1(\lambda\rho)\ \frac{1}{2 \sqrt{k^2-\lambda^2}+\omega\mu_0\sigma}\nonumber\\
&\mbox{} \stackrel{|k\rho|\gg 1}{\sim} \frac{i\mu_0}{4\pi}\,\sin\phi\int_{-\infty}^\infty \dl\,\lambda^2\,\sqrt{\frac{2}{\pi\lambda\rho}}\,e^{i(\lambda\rho-3\pi/4)}\,\frac{1}{2\sqrt{k^2-\lambda^2}+\omega\mu_0\sigma}\nonumber\\
&\stackrel{|\omega\mu_0\sigma|\ll |k|}{\sim} \frac{i\mu_0 k^2}{4\pi}\sin\phi\,\Biggl\{ -\frac{1}{k\rho}\,e^{ik\rho}+\frac{\omega\mu_0\sigma}{2k}\,\sqrt{\frac{\pi}{k\rho}}\,e^{ik\rho}\,\mathcal F(\wp_0)\Biggr\}~,
\end{align}
in agreement with~\eqref{eq:B2z-hd-ex-c}.

\subsection{Near (static) field: $|k_m\rho|\ll 1$ with $|\omega\mu_0\sigma|\ll |k|$}
In this case, the major contribution in each Sommerfeld-type integral comes from $\lambda=\mathcal O(1/\rho)\gg |k_m|\gg |k|$. Thus, we make the approximations
\begin{equation*}
\sqrt{k^2-\lambda^2}\sim i\lambda~,\quad 2k^2+\omega\mu_0\sigma\sqrt{k^2-\lambda^2}\sim i\lambda\,\omega\mu_0\sigma~,\quad 2\sqrt{k^2-\lambda^2}+\omega\mu_0\sigma\sim 2i\lambda~,
\end{equation*}
where $\lambda>0$; and exactly evaluate the ensuing integrals containing $J_\nu(\lambda\rho)$.

\subsubsection{Vertical dipole}
In the limit $a\downarrow 0$ with $0<z<a$,~\eqref{eq:E1r-vd} yields
\begin{equation}
E_{1\rho}^{({\rm vd})}\stackrel{|k_m\rho|\ll 1}{\sim} \frac{i\omega\mu_0}{4\pi k^2}\frac{i2k^2}{\omega\mu_0\sigma}\int_0^\infty \dl\,\lambda\,J_1(\lambda\rho)=-\frac{\omega\mu_0}{2\pi}\frac{1}{\omega\mu_0\sigma}\frac{1}{\rho^2}~,
\end{equation}
which agrees with~\eqref{eq:E1r-vd-ex-b} in the regime where $|k\rho|\ll |k_m\rho|\ll 1$ and $|\omega\mu_0\sigma|\ll |k|$. In this regime, the factor $e^{ik\rho}$ as well as terms containing the Struve and Neumann functions must be expanded out in powers of $k\rho$ and $k_m\rho$;~\cite{Bateman-II} after several mutual cancellations, it is found that the term $(k_m\rho)^{-2}$ in the second line of~\eqref{eq:E1r-vd-ex-b} provides the leading-order contribution. Likewise, \eqref{eq:E1z-vd} furnishes the integral
\begin{equation}
E_{1z}^{({\rm vd})}\sim \frac{i\omega\mu_0}{2\pi k^2}\int_0^\infty \dl\,\lambda^2\,J_0(\lambda\rho)=\frac{i\omega\mu_0}{2\pi k^2}\frac{1}{\rho}\int_0^\infty \dl\,\lambda\,\frac{d}{d\lambda}[\lambda J_1(\lambda)]=-\frac{i\omega\mu_0}{2\pi k^2} \frac{1}{\rho^3}~,
\end{equation}
via integration by parts; this result is in agreement with~\eqref{eq:E1z-vd-ex} where the leading-order contribution comes from the last term. The remaining component is provided by~\eqref{eq:B-reg1-vd}, which is reduced to the following expression [cf.~\eqref{eq:B1ph-vd-ex}].
\begin{equation}
B_{1\phi}^{({\rm vd})}\sim \frac{\mu_0}{2\pi}\int_0^\infty \,\lambda\,J_1(\lambda\rho)=\frac{\mu_0}{2\pi}\,\frac{1}{\rho^2}~.
\end{equation}

\subsubsection{Horizontal dipole}
In the static  field of the horizontal dipole, the two types of polarization may provide contributions of different orders of magnitude. By~\eqref{eq:E2r-hd}, we have
\begin{align}
E_{2\rho}^{({\rm hd})}&\sim -\frac{\omega\mu_0}{4\pi}\cos\phi \int_0^\infty \dl\,\lambda\,\Biggl\{\frac{1}{2i\lambda}\big[J_0(\lambda\rho)+J_2(\lambda\rho)\big]+\frac{1}{\omega\mu_0\sigma}\big[J_0(\lambda\rho)-J_2(\lambda\rho)\big]\Biggr\}\nonumber\\
&\sim -\frac{\omega\mu_0}{4\pi}\,\cos\phi\, \frac{2}{\omega\mu_0\sigma}\int_0^\infty \dl\,\frac{\partial}{\partial \rho} J_1(\lambda\rho)=\frac{\omega\mu_0}{2\pi}\,\cos\phi\,\frac{1}{\omega\mu_0\sigma}\,\frac{1}{\rho^2}~,
\end{align}
with dominance of the term pertaining to TM polarization. This result agrees with~\eqref{eq:E2r-hd-ex-b},  in which $H_1^{(1)\prime}(k_m\rho)$ and $Y_1'(k_m\rho)$ provide the leading-order contribution after cancellation of the $\mathcal O(1/\rho^3)$ terms. By~\eqref{eq:E2ph-hd}, the $\phi$-component of the electric field is
\begin{align}
E_{2\phi}^{({\rm hd})}&\sim \frac{\omega\mu_0}{4\pi}\sin\phi \int_0^\infty \dl\,\lambda\,\Biggl\{\frac{1}{2i\lambda}\big[J_0(\lambda\rho)-J_2(\lambda\rho)\big]+\frac{1}{\omega\mu_0\sigma}\big[J_0(\lambda\rho)+J_2(\lambda\rho)\big]\Biggr\}\nonumber\\
&\sim \frac{\omega\mu_0}{4\pi}\,\sin\phi\, \frac{1}{\omega\mu_0\sigma}\frac{2}{\rho}\int_0^\infty \dl\,J_1(\lambda\rho)=\frac{\omega\mu_0}{2\pi}\,\sin\phi\,\frac{1}{\omega\mu_0\sigma}\,\frac{1}{\rho^2}~,
\end{align}
in agreement with~\eqref{eq:E2ph-hd-ex-b} where the leading-order term comes from the expansion of $(k_m\rho)^{-1} H_1^{(1)}(k_m\rho)$ and $(k_m\rho)^{-1} Y_1(k_m\rho)$ for $|k_m\rho|\ll 1$ under $|\omega\mu_0\sigma|\ll |k|$. The remaining component of the electric field is evaluated on the basis of~\eqref{eq:E2z-hd}, viz.,
\begin{equation}
E_{2z}\sim -\frac{i\omega\mu_0}{2\pi}\,\cos\phi\,\frac{1}{i\omega\mu_0\sigma}\int_0^\infty \dl\,\lambda\,J_1(\lambda\rho)=-\frac{\omega\mu_0}{2\pi}\,\cos\phi\,\frac{1}{\omega\mu_0\sigma}\,\frac{1}{\rho^2}~.
\end{equation}
This result smoothly connects with~\eqref{eq:E2z-hd-ex-b} where the leading-order contribution comes from the $(k_m\rho)^{-2}$ term of the second line.

We proceed to the computation of the magnetic field. By~\eqref{eq:B2r-hd}, we write
\begin{align}
B_{2\rho}^{({\rm hd})}&\sim \frac{\mu_0}{4\pi}\,\sin\phi \int_0^\infty \dl\,\lambda  \Biggl\{\big[J_0(\lambda\rho)+J_2(\lambda\rho)\big]\frac{k^2}{i\lambda\,\omega\mu_0\sigma}+\big[J_0(\lambda\rho)-J_2(\lambda\rho)\big]\frac{1}{2}\Biggr\}\nonumber\\
&\sim \frac{\mu_0}{4\pi}\,\sin\phi \int_0^\infty \dl\,\frac{\partial}{\partial\rho}J_1(\lambda\rho)=-\frac{\mu_0}{4\pi}\,\sin\phi\,\frac{1}{\rho^2}~.
\end{align}
This formula also results from~\eqref{eq:B2r-hd-ex-b}; the leading-order term comes from the expansion of $(k_m\rho)^{-1} H_1^{(1)}(k_m\rho)$ and $(k_m\rho)^{-1} Y_1(k_m\rho)$. By~\eqref{eq:B2ph-hd}, we obtain
\begin{align}
B_{2\phi}^{({\rm hd})}&\sim \frac{\mu_0}{4\pi}\,\cos\phi\,\int_0^\infty \dl\,\lambda \Biggl\{ \big[J_0(\lambda\rho)-J_2(\lambda\rho)\big]\frac{k^2}{i\lambda\,\omega\mu_0\sigma}+\big[J_0(\lambda\rho)+J_2(\lambda\rho)\big] \frac{1}{2}\Biggr\}\nonumber\\
&\sim \frac{\mu_0}{4\pi}\,\cos\phi\,\frac{1}{\rho}\int_0^\infty \dl\,J_1(\lambda\rho)=\frac{\mu_0}{4\pi}\,\frac{\cos\phi}{\rho^2}~.
\end{align}
Now compare this approximation to~\eqref{eq:B2ph-hd-ex-b}; the latter reduces to the above expression by expansion in powers of $k\rho$ and $k_m\rho$. The remaining component of the magnetic field  is
\begin{equation}
B_{2z}^{({\rm hd})}\sim \frac{i\mu_0}{2\pi}\,\sin\phi \int_0^\infty \dl\, \lambda^2\,J_1(\lambda\rho)\,\frac{1}{2i\lambda}=\frac{\mu_0}{4\pi}\,\frac{\sin\phi}{\rho^2}~.
\end{equation}
This formula manifestly agrees with~\eqref{eq:B2z-hd-ex-b} if $|k\rho|\ll 1$ and $|\omega\mu_0\sigma|\ll |k|$.


\begin{thebibliography}{50}

\bibitem{Torres-book}L.~E.~F. Foa Torres, S. Roche, and J.-C. Charlier, {\em Introduction to Graphene-Based Nanomaterials: From Electronic Structure to Quantum Transport} (Cambridge University Press, Cambridge, UK, 2014).

\bibitem{Zhang-book}X.~C. Zhang and J. Xu, {\em Introduction to THz Wave Photonics} (Springer, Berlin, 2010).

\bibitem{Maier-book}S.~A. Maier, {\em Plasmonics: Fundamentals and Applications} (Springer, New York, 2007).


\bibitem{Vakil11}A. Vakil and N. Engheta, ``Transformation optics using graphene,'' Science {\bf 332}, 1291--1294 (2011).

\bibitem{Ju11}L. Ju, B.~S. Geng, J. Horng, C. Girit, M. Martin, Z. Hao, H.~A. Bechtel, X.~G. Liang, A. Zettl, Y.~R. Shen, and F. Wang, ``Graphene plasmonics for tunable terahertz metamaterials,'' Nature Nanotechnol.\ {\bf 6},  630--634 (2011).

\bibitem{Cheng13}J. Cheng, W.~L. Wang, H. Mosallaei, and E. Kaxiras, ``Surface plasmon engineering in graphene functionalized with
organic molecules: A multiscale theoretical investigation,'' Nano Lett.\ {\bf 14}, 50--56 (2014).

\bibitem{Hanson08}G.~W. Hanson, ``Dyadic Green's functions and guided surface waves for a surface conductivity model of graphene,'' J.\ Appl.\ Phys.\ {\bf 103}, 064302 (2008); Erratum, {\em ibid.}\ {\bf 113}, 029902 (2013).

\bibitem{Hanson11}G.~W. Hanson, A.~B. Yakovlev, and A. Mafi, ``Excitation of discrete and continuous spectrum for a surface conductivity model of graphene,'' J.\ Appl.\ Phys.\ {\bf 110}, 114305
(2011).

\bibitem{Lovat15}G. Lovat and R. Araneo, ``Semi-analytical representation of the two-dimensional time-domain Green's function of a graphene sheet in the intraband regime,'' IEEE Trans.\ Nanotech.\ {\bf 14}, 681--688 (2015).

\bibitem{Raether-book}H. Raether, {\em Surface Plasmons on Smooth and Rough Surfaces and on Gratings} (Springer, Berlin, 1986).

\bibitem{Bludov13}Yu.~V. Bludov, A. Ferreira, N.~M.~R. Peres, and M.~I. Vasilevskiy,
``A primer on surface plasmon-polaritons in graphene,'' Int.\ J.\ Modern Phys.\ B {\bf 27}, 1341001 (2013).

\bibitem{Zhang12}J. Zhang, L. Zhang, and W. Xu. ``Surface plasmon polaritons: Physics and applications'', J.\ Phys.\ D: Appl.\ Physics {\bf 45}, 113001 (2012).

\bibitem{Huidobro10}P.~A. Huidobro, M.~L. Nesterov, L. Mart\'{\i}n-Moreno and F.~J. Garc\'{\i}a-Vidal, ``Transformation optics for plasmonics,'' Nano Lett.\ {\bf 10}, 1985--1990 (2010).

\bibitem{Renger10}J. Renger, M. Kadic, G. Dupont, S.~S. Acimovic, S. Guenneau, R. Quidant, and S. Enoch, ``Hidden progress: Broadband plasmonic invisibility,'' Opt.\ Express {\bf 18}, 15757--15768 (2010).

\bibitem{Green12}M.~A. Green and S. Pillai, ``Harnessing plasmonics for solar cells'', Nature Photon. {\bf 6}, 130--132 (2012).

\bibitem{Luo04}X. Luo and T. Ishihara, ``Surface plasmon resonant interference nanolithography technique,'' Appl.\ Phys.\ Lett.\ {\bf 84}, 4780--4782 (2004).

\bibitem{Gonzalez14}P. Alonso-Gonz\'{a}lez, A.~Y. Nikitin, F. Golmar, A. Centeno,  A. Pesquera, S. V\'{e}lez, J. Chen, G. Navickaite, F. Koppens, A. Zurutuza,
F. Casanova,  L. E. Hueso, and R. Hillenbrand, ``Controlling graphene plasmons with resonant metal antennas and spatial conductivity patterns'', Science {\bf 344}, 1369--1373 (2014).

\bibitem{Sommerfeld1899}A. Sommerfeld, ``\"Über die Fortpflanzung electrodynamischer Wellen l\"angs eines Drahtes,'' Ann.\ Phys.\ und Chemie {\bf 67}, 233--290 (1899).

\bibitem{Sommerfeld1909}A. Sommerfeld, ``\"Uber die Ausbreitung der Wellen in der drahtlosen Telegraphie,'' Ann.\ Phys.\ (Leipzig) {\bf 28}, 665--736 (1909).

\bibitem{Zenneck1907}J. Zenneck, ``\"Über die Fortpflanzung ebener elektromagnetischer Wellen l\"angs einer ebenen Leiterfl\"ache und ihre Beziehung zur drahtlosen Telegraphie,'' Ann.\ der Phys.\ {\bf 23}, 846--866 (1907).

\bibitem{vanderPol31}B. van der Pol, ``\"Uber die Ausbreitung elektromagnetischer Wellen,'' Zeitschrift f\"ur Hochfrequenztechnik {\bf 37}, 152--156 (1931).

\bibitem{Fok33}V. Fock, ``Z\"ur Berechnung des elektromagnetischen Wechselstromfeldes bei ebener Begrenzung,'' Ann.\ Phys.\ (Leipzig) {\bf 17}, 401--420 (1933).

\bibitem{Norton36}K.~A. Norton, ``The propagation of radio waves over the surface of the earth and in the upper atmosphere,'' Proc.\ IRE {\bf 24}, 1367--1387 (1936).

\bibitem{Banos}A. Ba$\tilde{\rm n}$os, {\em Dipole Radiation in the Presence of a Conducting Half-Space} (Pergamon, Oxford, 1966).

\bibitem{Wait}J.~R. Wait, {\em Electromagnetic Waves in Stratified Media} (Pergamon, New York, 1970).

\bibitem{KOW}R.~W.~P. King, M. Owens, and T.~T. Wu, {\em Lateral Electromagnetic Waves: Theory and Applications to Communications, Geophysical Exploration, and Remote Sensing}, (Springer-Verlag, New York, 1992).

\bibitem{Nikitin11}A. Yu. Nikitin, F. Guinea, F.~J. Garcia-Vidal, and L. Martin-Moreno, ``Fields radiated by a nanoemitter in a graphene sheet,'' Phys.\ Rev.\ B {\bf 84}, 195446 (2011).

\bibitem{Lebedev-book}N.~N. Lebedev, {\em Worked Problems in Applied Mathematics} (Dover, New York, 1979), pp.\ 165--168.

\bibitem{Muller}C. M\"uller, {\em Foundations of the Mathematical Theory of Electromagnetic Waves} (Springer-Verlag, Berlin, 1969), pp.~10, 81.

\bibitem{Margetis01}D. Margetis and T.~T. Wu, ``Exactly calculable field components of electric dipoles in planar boundary,'' J.\ Math.\ Phys.\ {\bf 42}, 713--745 (2001).

\bibitem{Bateman-II}Bateman Manuscript Project, {\em Higher Transcendental Functions}, edited by A. Erd\'{e}lyi (McGraw-Hill, New York, 1953), Vol.~II, Chap.~VII.

\bibitem{Hardy}G.~H. Hardy, {\em Divergent Series} (Chelsea, New York, 1991), Chap.~I.

\bibitem{Falkovsky07}L.~A. Falkovsky and S.~S. Pershoguba, ``Optical far-infrared properties of a graphene monolayer and multilayer,'' Phys.\ Rev.\ B {\bf 76}, 153410 (2007).


\bibitem{Gan12}C.~H. Gan, H.~S. Chu, and E.~P. Li, ``Synthesis of highly confined surface plasmon modes with doped graphene sheets in the mid-infrared
and terahertz frequencies,'' Phys.\ Rev.\ B {\bf 85}, 125431 (2012).

\bibitem{Watson}G.~N. Watson, {\em A Treatise on the Theory of Bessel Functions} (Cambridge University Press, New York, 1995), pp.\ 434,\,435.

%\bibitem{Jauch}J.~M. Jauch and F. Rohrlich, {\em The Theory of Photons and Electrons: The Relativistic Quantum Field Theory of Charged %Particles with Spin One-half} (Springer-Verlag, 2nd Edition, Berlin, 1980).

\bibitem{Fikioris-book}G. Fikioris, {\em Mellin Transform Method for Integral Evaluation: Introduction and Applications to Electromagnetics} (Morgan and Claypool, San Rafael, CA, 2007), pp.\ 35--38.

\bibitem{Bateman-I}Bateman Manuscript Project, {\em Higher Transcendental Functions}, edited by A. Erd\'{e}lyi (McGraw-Hill, New York, 1953), Vol.~I, p.~267.

\bibitem{Sommerfeld-pde}A. Sommerfeld, {\em Partial Differential Equations in Physics} (Academic, New York, 1949).

%\bibitem{Xu-12}H.~J. Xu, W.~B. Lu, W. Zhu, Z.~G. Dong, and T.~J. Cui,
%``Efficient manipulation of surface plasmon polariton waves in graphene,'' Appl.\ %Phys.\ Lett.\ {\bf 100}, 243110 (2012).

\bibitem{Low14}T. Low, R. Rold\'{a}n, H. Wang, F. Xia, P. Avouris, L. Mart\'{\i}n Moreno, and Francisco Guinea, ``Plasmons and screening in monolayer and multilayer black phosphorus,'' Phys.\ Rev.\ Lett.\ {\bf 113}, 106802 (2014).

\end{thebibliography}
\end{document}